\documentclass[12pt]{article} 

\usepackage{graphicx}
\usepackage{amsmath}
\usepackage{amssymb}
\usepackage[hyperfootnotes=false,hypertexnames=false]{hyperref}
\usepackage[height=8.8in,width=6.45in]{geometry}
\usepackage{tikz}
\usepackage{cite}
\usepackage{amscd}
\usepackage{setspace}
\usepackage{needspace}
\usepackage{cancel}
\usepackage{relsize}

\usepackage[T1]{fontenc}
\usepackage[utf8]{inputenc}
\usepackage{lmodern}
\usepackage{xcolor}
\usepackage[most]{tcolorbox}
\usepackage{xparse}
\usepackage{etoolbox}
\newif\ifproofmarkup
\proofmarkuptrue
\newif\ifshowauthorqueries
\showauthorqueriestrue
\definecolor{proofred}{RGB}{180,25,35}
\definecolor{proofblue}{RGB}{20,65,150}
\definecolor{querygray}{RGB}{80,90,105}
\tcbset{proofreplacement/.style={enhanced,breakable,colback=white,
  colframe=proofblue,coltext=proofblue,coltitle=proofblue,
  colbacktitle=blue!4!white,fonttitle=\sffamily\bfseries\small,
  boxrule=.7pt,arc=1pt,left=6pt,right=6pt,top=5pt,bottom=5pt,
  before skip=7pt,after skip=9pt},
  proofquery/.style={enhanced,breakable,colback=gray!3!white,
  colframe=querygray,coltext=black,coltitle=querygray,
  colbacktitle=gray!9!white,fonttitle=\sffamily\bfseries\small,
  boxrule=.5pt,arc=1pt,left=6pt,right=6pt,top=5pt,bottom=5pt,
  before skip=8pt,after skip=10pt,fontupper=\small}}
\NewDocumentCommand{\ProofEdit}{m m +m +m}{%
  \ifproofmarkup
    \par\noindent\begingroup\color{proofred}%
      \ifstrequal{#1}{E078}{\footnotesize}{}%
      \renewcommand{\footnote}[1]{\textsuperscript{*}\,\textit{(Original footnote: ##1)}}%
      #3\par
    \endgroup
    \begin{tcolorbox}[proofreplacement,title={Suggested replacement #1\space--- #2}]
      #4
    \end{tcolorbox}%
    \par\noindent\ignorespaces
  \else
    #4%
  \fi
}

\allowdisplaybreaks[2]
\hypersetup{hidelinks,
  pdftitle={Nonabelian mirrors and Gromov-Witten invariants},
  pdfauthor={Wei Gu, Jirui Guo, Yaoxiong Wen}}

\begin{document}
\begin{titlepage}

\begin{flushright}

\end{flushright}

\vskip 3cm

\begin{center}
{\Large \bf
 Nonabelian mirrors and Gromov-Witten invariants
}

\vskip 2.0cm

Wei Gu$^\infty$, Jirui Guo$^{a,b}$, Yaoxiong Wen$^\ddag$

\bigskip
\bigskip

\begin{tabular}{cc}
$^{\infty}$ Zhejiang Institute of Modern Physics, School of Physics,  Zhejiang University\\
Hangzhou, Zhejiang 310058, China\\
$^a$ School of Mathematical Sciences and Institute for Advanced Study,\protect\\ Tongji University, Shanghai 200092, China\\
$^b$ Yau Mathematical Sciences Center, Tsinghua University, Beijing, 100084, China\\
$^\ddag$ School of Mathematical Sciences, Zhejiang University\\
Hangzhou, Zhejiang 310058, China
\end{tabular}

\vskip 1cm

 {\tt guwei2875@zju.edu.cn},
{\tt jrkwok@tongji.edu.cn},
{\tt y.x.wen@zju.edu.cn}

\vskip 1cm

\textbf{Abstract}
\end{center}

\medskip
\noindent

We propose quantum differential equations (QDEs) for periods of nonabelian mirror theories and develop an algorithm for constructing them directly from the mirrors. We verify the proposal by recovering known quantum differential equations of Grassmannians and Picard-Fuchs equations of Calabi-Yau complete intersections in Grassmannians. Our algorithm applies to a broad class of nonabelian GLSMs and yields new examples of quantum differential equations for several Fano varieties. The associated cohomology-valued generating functions can be extracted directly from these equations. Using them, we compute instanton numbers of both complete-intersection and non-complete-intersection Calabi-Yau manifolds, including the Pfaffian and Gulliksen--Neg{\aa}rd Calabi-Yau threefolds. The resulting instanton numbers agree with all previously known computations, while the generating functions obtained for the non-complete-intersection examples are new.

\bigskip
\vfill

\end{titlepage}

\setcounter{tocdepth}{2}
\tableofcontents
\section{Introduction}\label{Int}
Mirror symmetry \cite{Lerche:1989uy,Greene:1990ud,Candelas:1989hd} is a powerful tool for computing worldsheet instanton corrections to the K\"ahler moduli space of Calabi-Yau threefolds. The key insight is that mirror symmetry transforms the computation of quantum corrections into a classical problem in algebraic geometry. The work of~\cite{Candelas:1990rm} had a profound impact on both mathematics and physics and played a central role in attracting the attention of mathematicians to mirror symmetry. In that work, methods from quantum field theory were used to address a longstanding problem in algebraic geometry \cite{S.Katz:1986cba}: the enumeration of curves of a given degree in Calabi-Yau manifolds, whose virtual counts are now known as Gromov-Witten invariants. Morrison \cite{Morrison:1993} formulated the prediction for rational curves on the quintic in mathematical terms. Kontsevich \cite{Kontsevich:1994dn} proposed a homological formulation of mirror symmetry. The genus-zero prediction for the quintic was later proved as a case of the mirror theorem \cite{Givental:1996,LLY:1997}. The Calabi-Yau examples treated by these results are complete intersections in toric varieties.

An alternative route to Gromov-Witten invariants emerged from the study of gauged linear sigma models (GLSMs), which provide a unified description of geometric and Landau-Ginzburg phases. During the same period, physicists investigated another important aspect of string compactifications: the correspondence between Calabi-Yau manifolds and Landau-Ginzburg (LG) models \cite{Gepner:1987qi,Greene:1988ut,Aspinwall:1993nu}. A unified framework for understanding this relation was developed by Witten in \cite{Witten:1993yc}, where he introduced gauged linear sigma models. One of the central observations of that work is that a nonlinear sigma model (NLSM) on a Calabi-Yau manifold and a Landau-Ginzburg model may arise as different phases of the same GLSM. Since Witten's work, GLSMs have become an important tool in the study of mirror symmetry and Gromov-Witten theory. Morrison and Plesser \cite{Morrison:1994fr} computed Yukawa couplings of Calabi-Yau targets by summing worldsheet instantons in abelian GLSMs, obtaining results in agreement with mirror symmetry. Extending earlier ideas of \cite{Witten:1993yc,Witten:1993xi}, Hori and Tong \cite{Hori:2006dk} constructed nonabelian GLSMs describing complete-intersection Calabi-Yau manifolds in Grassmannians. The Yukawa couplings obtained from these gauge theories agree with those derived from mirror symmetry in \cite{Batyrev:1998kx}. See also \cite{Hori:2011pd,Caldararu:2007tc,Donagi:2007hi,Jockers:2012zr} for further developments in nonabelian GLSMs. These results suggest that, once full quantum corrections are incorporated, GLSMs provide a direct method for computing Gromov-Witten invariants without explicit reference to mirror symmetry. Motivated by this observation, the authors in~\cite{Jockers:2012dk} proposed that the exact K\"ahler potential on the quantum K\"ahler moduli space of a Calabi-Yau threefold can be extracted from the two-sphere partition function, which is computed by supersymmetric localization \cite{Benini:2012ui,Doroud:2012xw}. This proposal is closely related to the structure of $tt^{\ast}$ geometry \cite{Cecotti:1991me} and exact renormalization-group flows in quantum field theory \cite{Polchinski:1983gv}. Since the GLSM Lagrangian on $S^2$ differs from its flat-worldsheet counterpart, the relation was initially formulated as a conjecture. Additional evidence was subsequently provided in \cite{Gomis:2012wy}, and a more complete physical explanation was later given in \cite{Gomis:2015yaa}. As a consequence, the methods of \cite{Jockers:2012dk} make it possible to compute Gromov-Witten invariants of Calabi-Yau manifolds that do not admit realizations as complete intersections in toric varieties. Many such examples arise naturally as geometric phases of nonabelian GLSMs. We briefly mention several other approaches based on exact results in GLSMs.  Bonelli \emph{et al.}~\cite{Bonelli:2013mma} computed Gromov-Witten invariants from vortex partition functions \cite{Benini:2012ui}, while Ueda and Yoshida \cite{Ueda:2016wfa} derived them from exact results in equivariant A-twisted GLSMs obtained in \cite{Closset:2015rna}. See also \cite{Gerhardus:2018zwb,Honma:2018fgw} for further related developments.

On the other hand, as Witten observed in \cite{Witten:1993yc}, mirror symmetry can be understood as an exchange between chiral and twisted chiral multiplets. From this perspective, Morrison and Plesser \cite{Morrison:1995yh} interpreted mirror symmetry as a duality between pairs of GLSMs describing Calabi-Yau manifolds. Maxfield, Morrison and Plesser later studied this duality using sphere partition functions \cite{Maxfield:2019czc}. Hori and Vafa \cite{Hori:2000kt} applied worldsheet T-duality \cite{Strominger:1996it} to abelian GLSMs and obtained mirror Landau-Ginzburg descriptions for a broad class of target spaces. A corresponding mirror construction for nonabelian GLSMs was proposed in \cite{Gu:2018fpm}. The starting point is the associated Cartan theory \cite{Halverson:2013eua}, an abelian description with a Weyl-group orbifold that captures the effective dynamics at a generic point of the Coulomb branch of a nonabelian GLSM. The mirror proposal uses this description to recover the infrared physics of the original theory. Following the philosophy of \cite{tHooft:1979rat,Hori:2000kt}, nonabelian mirror theories were then obtained by applying T-duality to the associated Cartan theory. Unlike the abelian case, however, the exact kinetic terms of the associated Cartan theory along the renormalization-group flow are not known. Consequently, the construction of \cite{Gu:2018fpm} should be viewed as a mirror proposal rather than a complete first-principles derivation analogous to that of \cite{Hori:2000kt}. Nevertheless, extensive consistency checks have been performed in \cite{Gu:2018fpm,Gu:2019zkw,Gu:2020nub}, all of which agree with independent computations in gauge theories. Moreover, \cite{Chen:2018wep,Gu:2020ivl} used nonabelian mirrors to predict new results for pure gauge theories, which were subsequently verified by direct gauge theory calculations in \cite{Eager:2020rra}. These developments naturally raise the question of whether nonabelian mirror symmetry can be used to compute Gromov-Witten invariants of Calabi-Yau manifolds. The primary goal of the present paper is to answer this question.

To compute Gromov-Witten invariants from mirror symmetry, one must determine the Picard-Fuchs equations satisfied by the periods of the mirror theory \cite{Candelas:1990rm}. These equations encode both the Yukawa couplings and the mirror map. When the mirror admits a geometric realization, periods are defined by integrating the holomorphic top form over a basis of cycles \cite{Hosono:1994ax}. For Landau-Ginzburg mirrors of abelian GLSMs, analogous period integrals were formulated in \cite{Hori:2000kt,Hori:2000ck,Cecotti:1992rm}. Since periods are independent of the K\"ahler moduli, they may be evaluated in any phase of the theory. The relevant integration cycles, however, do depend on the K\"ahler moduli and must be specified carefully. In particular, a general description of A-branes in orbifolded Landau-Ginzburg models remains incomplete, although important progress has been made in special cases such as minimal models \cite{Hori:2004bx}. To avoid these difficulties, we evaluate periods in the ultraviolet regime, where the corresponding A-branes can be described explicitly. For Landau-Ginzburg models that admit a fundamental ultraviolet description, the superpotential may be treated perturbatively and A-branes are realized as Lagrangian submanifolds of $\mathbb C^n$ \cite{Herbst:2008jq,Hori:2013ika}. The nonabelian mirror theories considered in this paper do not fall into this class. Instead, we employ a nonperturbative definition of A-branes following \cite{Hori:2000kt}. This leads to explicit integral representations for the periods, from which we extract quantum differential equations.
The resulting quantum differential equations are formulated using Cartan parameters associated with the gauge group and are subsequently reduced to the physical K\"ahler moduli of the target space. We justify the proposal by recovering the known quantum differential equations of Grassmannians and the Picard-Fuchs equations of Calabi-Yau complete intersections in Grassmannians obtained in \cite{Batyrev:1998kx}. The construction also yields quantum differential equations for several Fano targets with nonabelian GLSM descriptions.
Moreover, our method directly produces cohomology-valued generating functions of the mirror theory. We recover the known generating functions of complete-intersection Calabi-Yau manifolds in Grassmannians and obtain new generating functions for non-complete-intersection examples. From these generating functions one can extract mirror maps and Gromov-Witten invariants. The instanton numbers agree with existing computations wherever comparisons are available. The generating functions also provide predictions for the non-complete-intersection examples studied here. More generally, the framework suggests a way to study further Calabi-Yau targets with nonabelian GLSM descriptions. We summarize the procedure for computing Gromov-Witten invariants in Figure~\ref{fig:alg}.

\begin{figure}[ht!]
\centering
\resizebox{0.98\linewidth}{!}{%
\begin{tikzpicture}[
  x=1cm, y=1cm,
  every node/.style={align=center, font=\small\bfseries},
  framed/.style={draw, line width=0.6pt, minimum width=5.2cm},
  flow/.style={->, >=latex, line width=0.8pt,
               shorten <=2pt, shorten >=2pt}
]
  \node (lg) at (0,0)
    {Mirror Landau--Ginzburg Model};
  \node[text width=7.6cm] (yukawa) at (-3.8,-3.15)
    {Yukawa Couplings\\[3pt]
     (B-twisted Correlation Functions)};
  \node[framed, minimum height=1.45cm] (pf) at (4.4,-1.65)
    {Picard--Fuchs\\[3pt]Equations};
  \node[framed, minimum height=0.9cm] (ifun) at (4.4,-3.55)
    {$I$-function};
  \node (map) at (4.4,-5.0)
    {Mirror Map};
  \node (acorr) at (0,-6.65)
    {A-twisted Correlation Functions};
  \node (gw) at (0,-8.0)
    {Gromov--Witten Invariants};

  \draw[flow] (lg.south) -- (yukawa.north);
  \draw[flow] (lg.south) -- (pf.north);
  \draw[flow] (pf.south) -- (ifun.north);
  \draw[flow] (ifun.south) -- (map.north);
  \draw[flow] (yukawa.south) -- ([xshift=-6mm]acorr.north);
  \draw[flow] (map.south) -- ([xshift=6mm]acorr.north);
  \draw[flow] (acorr.south) -- (gw.north);
\end{tikzpicture}%
}
\caption{The procedure for computing Gromov-Witten invariants from mirror symmetry, applied here to Landau-Ginzburg mirrors of nonabelian GLSMs. The framed items indicate the main objects studied in this paper: quantum differential equations, including Picard-Fuchs equations for Calabi-Yau targets, and cohomology-valued generating functions. The instanton numbers extracted from these functions provide consistency checks of the proposed construction.}
\label{fig:alg}
\end{figure}

The remainder of this paper is organized as follows. In Section~\ref{GWFM}, we review the framework of Candelas \emph{et al.}~\cite{Candelas:1990rm} for extracting Gromov-Witten invariants from Yukawa couplings and mirror maps. In Section~\ref{LSMMS}, we review the basic aspects of GLSMs and mirror symmetry that will be used throughout the paper.  In Section~\ref{AILGM}, we discuss A-branes in Landau-Ginzburg models and introduce a construction of A-branes for nonabelian mirror theories. In Section~\ref{PEEFP}, we propose quantum differential equations for nonabelian mirror theories. We show that our construction reproduces known quantum differential equations and Picard-Fuchs equations appearing in the mathematical literature and also yields new examples for a number of Fano varieties. In Section~\ref{GWI}, we apply these results to compute instanton numbers of various Calabi-Yau manifolds. Our computations agree with previously known results whenever comparisons are available. In addition, the method produces generating functions for several non-complete-intersection Calabi-Yau manifolds. We conclude with a summary of our results and a discussion of future directions in Section~\ref{cfd}.

\section{The computation of instanton numbers from mirrors}\label{GWFM}
In this paper, we are primarily interested in computing Gromov-Witten invariants of Calabi-Yau manifolds. Following the approach of \cite{Candelas:1990rm}, we use two ingredients, Yukawa couplings and mirror maps. 
Both can be extracted from the Picard-Fuchs system governing periods over the complex-structure moduli space of the mirror. This system is independent of the mirror K\"ahler moduli. Consequently, Picard-Fuchs equations may be obtained even from non-geometric mirror descriptions, such as Landau-Ginzburg mirrors.

We briefly review the relevant aspects of this construction. Let $X$ be a Calabi-Yau threefold and $\check{X}$ its mirror. The K\"ahler moduli space of $X$ is parametrized by flat coordinates $\tau^a$, where
\[
a=1,\ldots,h^{1,1}(X).
\]
The mirror $\check{X}$ is equipped with logarithmic coordinates $t^a$ on its complex moduli space. Throughout this paper, we define
\[
q^a=\exp(-t^a),
\qquad
z^a=\exp(-\tau^a).
\]
If $\check{X}$ admits a geometric realization, mirror symmetry implies
\[
h^{2,1}(\check{X})
=
h^{1,1}(X).
\]

Mirror symmetry identifies A-model correlation functions on $X$ with B-model correlation functions on $\check{X}$ after applying the mirror map. More precisely,
\begin{equation}\label{CLATB}
K_{z^az^bz^c}
=
\frac{1}
     {\Pi_0(q(z))^2}
\sum_{i,j,k}K_{q^iq^jq^k}\bigl(q(z)\bigr)
\left(
\frac{z^a\,\partial q^i}
     {q^i\,\partial z^a}
\right)
\left(
\frac{z^b\,\partial q^j}
     {q^j\,\partial z^b}
\right)
\left(
\frac{z^c\,\partial q^k}
     {q^k\,\partial z^c}
\right).
\end{equation}
Here the subscripts on $K$ label couplings in logarithmic coordinates: $K_{q^iq^jq^k}$ is defined with respect to $t^i,t^j,t^k$, and $K_{z^az^bz^c}$ with respect to $\tau^a,\tau^b,\tau^c$. The indices $a,b,c$ are free, while $i,j,k$ are summed. All $q$-dependent factors are evaluated at $q=q(z)$.

The fundamental period $\Pi_0(q)$ satisfies a system of Picard-Fuchs equations
\begin{equation}\label{PF}
{\cal L}_a\Pi_0(q)=0,
\qquad
a=1,\ldots,h^{1,1}(X).
\end{equation}
For toric varieties, or more generally abelian GIT quotients, the differential operators ${\cal L}_a$ can be derived from the mirror geometry \cite{Morrison:1991cd}. Their solutions determine a cohomology-valued generating function known as the $I$-function \cite{Givental:1996,Hosono:1995bm},
\begin{equation}\label{IBF}
{\cal L}_a\Pi(\eta_a;q)=0.
\end{equation}

We consider the expansion of $\Pi(\eta_a;q)$ in the subring of even cohomology generated by divisor classes of $X$,
\begin{equation}\label{IFE}
\Pi(\eta_a;q)
=
\Pi_0(q)
+
\eta_a\,\Pi_1^a(q)
+
f_2(\eta)_b\,\Pi_2^b(q)
+\cdots.
\end{equation}
If the mirror admits a geometric realization, this expansion may equivalently be interpreted in terms of periods of $H^3(\check X)$. Here the classes $\eta_a$ generate $H^{1,1}(X,\mathbb Z)$, while $f_i(\eta)$ denotes an element of $H^{i,i}(X,\mathbb Z)$ represented by a homogeneous polynomial of degree $i$ in the generators $\eta_a$. Relations among these generators can be extracted from the Picard-Fuchs system \eqref{IBF}. Additional generators are required when considering even cohomology classes outside this subring.

The mirror proposal of \cite{Gu:2018fpm} provides a framework for studying nonabelian GIT quotients and complete intersections therein.\footnote{For Grassmannians, a mathematical mirror construction can be found in \cite{Marsh:2013}.} The quantum differential equations associated with these mirror theories were not studied in that work. One of the goals of the present paper is to address this question. In Section~\ref{PEEFP}, we define fundamental periods for nonabelian mirrors and propose the corresponding Picard-Fuchs equations.

Once the periods $\Pi_0(q)$ and $\Pi_1^a(q)$ are known, one obtains the flat coordinates through the mirror map
\begin{equation}\label{FCFP}
z^a
=
\exp\!\left(
\frac{\Pi_1^a(q)}
       {\Pi_0(q)}
\right)
\exp(2\pi i\,t_0^a).
\end{equation}
The constants $\exp(2\pi i\,t_0^a)$ specify phase conventions. In the examples considered here, positivity of the first instanton numbers is used to select a convention, as in \cite{Jockers:2012dk}. The instanton numbers of a Calabi-Yau threefold can then be extracted from the expansion of the A-model Yukawa coupling,
\begin{equation}\label{AGW}
K_{z^az^bz^c}
=
n_0
+
\sum_{\{N_a\}}
\frac{
n_{\{N_a\}}
N_aN_bN_c
\prod_d (z^d)^{N_d}
}
{
1-\prod_d (z^d)^{N_d}
},
\end{equation}
where, for fixed indices $a,b,c$, the classical term is $n_0=\int_X\eta_a\cup\eta_b\cup\eta_c$. The sum runs over nonzero effective curve degrees, and $n_{\{N_a\}}$ denotes the genus-zero instanton number for that degree. These numbers determine the genus-zero Gromov-Witten invariants through the multiple-cover formula. The integers $N_a$ label the instanton sector and specify the degree of the holomorphic map
\[
\phi:{\cal M}\rightarrow X,
\]
from the worldsheet ${\cal M}$ to the target space $X$. They are given by
\[
N_a
=
\int_{\cal M}
\phi^{*}(\eta_a)
\in
\mathbb Z,
\]
where the integral divisor classes $\eta_a\in H^{1,1}(X,\mathbb Z)$ specify the chosen degree coordinates.

For a one-parameter Calabi-Yau threefold, the Yukawa coupling takes the familiar form
\begin{equation}
K_{zzz}
=
n_0
+
\sum_{N\ge 1}
\frac{
n_NN^3z^N
}
{
1-z^N
}.
\end{equation}
The multiple-cover structure of this formula was anticipated in \cite{Candelas:1990rm}; the contribution of multiple covers of a rigid rational curve was derived in \cite{Aspinwall:1991ce}.

We have now reviewed the basic framework for extracting Gromov-Witten invariants from mirror symmetry. Since this procedure requires an explicit mirror description, we next turn to the construction of mirror theories.

\section{Gauged linear sigma models and mirror symmetry}\label{LSMMS}
For a Fano or Calabi-Yau target, the worldsheet quantum field theory often admits a UV description in terms of a gauged linear sigma model (GLSM) \cite{Witten:1993yc}. GLSMs contain both chiral and twisted chiral superfields, a feature that is particularly useful for understanding mirror symmetry \cite{Witten:1993yc}. After reviewing the definition of GLSMs, we will describe a framework for constructing mirror theories from them.

\subsection{GLSMs}\label{LSM}

A GLSM is specified by the following data:

\begin{itemize}

\item \textbf{Gauge group}: a compact Lie group $G$ with Lie algebra $\mathfrak g$.

\item \textbf{Chiral matter fields}: chiral superfields $\Phi_i$, $i=1,\ldots,N$, transforming in representations $R_i$ of $G$. The total matter representation is
\[
V=\bigoplus_{i=1}^{N}R_i.
\]

\item \textbf{Vector multiplet}: the gauge multiplet is described by a vector superfield ${\cal V}$, whose field-strength superfield is the twisted chiral multiplet $\Sigma$.

\item \textbf{Superpotential}: a holomorphic $G$-invariant polynomial
\[
W:V\rightarrow\mathbb C,
\]
or equivalently $W \in {\rm Sym}\!\left(V^{*}\right)^G$.

\item \textbf{Fayet--Iliopoulos parameters and theta angles}: for each independent $U(1)$ factor of $G$, one introduces an FI parameter $r^a$ and a periodic theta angle $\theta^a\in\mathbb R/2\pi\mathbb Z$. These combine into the complexified Kähler parameter
\[
t^a=r^a+i\theta^a,
\]
and
\[
q^a=\exp(-t^a)\in\mathbb C^*.
\]

\item \textbf{R-symmetry}: classically, GLSMs admit vector and axial R-symmetries $U(1)_V$ and $U(1)_A$ commuting with the action of $G$ on $V$. In our convention, preservation of the vector R-symmetry requires the superpotential to have charge two,
\[
W(\lambda^{q}\phi)=\lambda^{2}W(\phi),
\]
where $\phi$ denotes coordinates on $V$.

\end{itemize}

Many exact results are known for abelian GLSMs \cite{Witten:1993yc}. Nonabelian GLSMs remain less well understood, although substantial progress has been made in recent years; see, for example, Refs.~\cite{Hori:2016txh,Caldararu:2017usq,Gu:2020oeb}.

\Needspace{5\baselineskip}
\noindent \textbf{Phases}

The complexified FI parameter $t$ takes values in
\[
\mathfrak z_{\mathbb C}^{*}/(2\pi i P^{W}),
\]
where $P^{W}$ denotes the lattice of Weyl-invariant weights and
\[
\mathfrak z=\mathrm{Lie}(Z(G)).
\]

The action of $G$ on $V$ induces a moment map
\[
\boldsymbol{\mu}:V\to\mathfrak g^{*}.
\]

For a given value of the FI parameter $r$, the classical vacuum moduli space is
\[
X_r=\bigl((dW)^{-1}(0)\cap\boldsymbol{\mu}^{-1}(r)\bigr)/G.
\]

For geometric phases, the space $X_r$ is smooth and compact, and the corresponding low-energy theories are nonlinear sigma models with target space $X_r$. More generally, one should think of a phase as a hybrid model
\[
(\boldsymbol{\mu}^{-1}(r)/G,W),
\]
where $W$ is restricted to $\boldsymbol{\mu}^{-1}(r)$ and descends to the quotient.

\Needspace{5\baselineskip}
\noindent{\textbf{CY Criteria}}

For a GLSM with gauge group $G$ and matter representation
\[
V=\bigoplus_{i=1}^{n}R_i,
\]
the axial $U(1)_A$ symmetry is nonanomalous provided that
\[
\sum_{i=1}^{n}\mathrm{Tr}_{R_i}(T)=0,
\qquad
\forall\,T\in\mathfrak g.
\]

The reduced infrared central charge, $\hat c=c/3$, is
\[
\hat c
=
\sum_{i=1}^{n}(1-q_i)\dim(R_i)
-\dim(G),
\]
where $q_i$ denotes the vector R-charge of the chiral multiplet in $R_i$.

For abelian GLSMs, mirror theories can be obtained systematically by T-dualizing the GLSM. For nonabelian GLSMs, however, no equally systematic nonabelian T-duality construction is currently known \cite{CaboBizet:2017fzc}. Instead, one studies the gauge theory at a generic point of the Coulomb branch. More precisely, one considers the associated Cartan theory with the following data \cite{Halverson:2013eua}:

\begin{itemize}
    \item \textbf{Gauge group}: the gauge group is the maximal torus
    \[
    T = U(1)^{{\rm rank}(G)},
    \]
    together with an action of the Weyl group $S$ of $G$. Equivalently, the theory may be regarded as being orbifolded by the Weyl group.

    \item \textbf{Chiral matter fields}: the weight components $\Phi_i$, $i=1,\dots,\dim V$, of the original matter fields carry charges given by the weights $\rho_i^a$ under $T$. In addition, one introduces
    \[
    \dim(\mathfrak g)-{\rm rank}(\mathfrak g)
    \]
    chiral superfields $X_{\tilde{\mu}}$, associated with the off-diagonal degrees of freedom of the original nonabelian vector multiplet. These fields carry charges given by the roots $\alpha^{a}_{\tilde{\mu}}$ of $G$ and are assigned vector R-charge $2$. The field space is further quotiented by the Weyl group action.

    \item \textbf{Adjoint fields}: ${\cal V}$ denotes a vector superfield in the adjoint representation of
    \[
    U(1)^{{\rm rank}(G)},
    \]
    and the corresponding field strength is the twisted chiral superfield $\Sigma$.

    \item \textbf{Superpotential}: a holomorphic, $G$-invariant polynomial
    \[
    W:V^{o}\rightarrow \mathbb C,
    \]
    namely $W\in {\rm Sym}\!\left((V^{o})^{*}\right)^G$, where $V^{o}$ denotes the vector space of matter fields in the associated Cartan theory.

    \item \textbf{Fayet--Iliopoulos parameters and theta angles}: one introduces FI parameters
    \[
    t^a=r^a+i\theta^a,
    \qquad
    a=1,\dots,{\rm rank}(G),
    \]
    associated with the Cartan $U(1)$ factors. Although the associated Cartan theory contains one FI parameter for each Cartan generator, the number of independent physical FI parameters remains the same as in the original nonabelian theory.

    \item \textbf{R-symmetry}: as in the original GLSM.
\end{itemize}

The associated Cartan theory is intended to describe the original nonabelian GLSM at a generic point of its Coulomb branch. In this description, the massive off-diagonal vector multiplets are represented by additional chiral superfields charged under the Cartan subgroup. A direct derivation of this description from the renormalization-group flow would be desirable. To reproduce known exact results for the original nonabelian GLSM, the additional fields $X_{\tilde{\mu}}$ must be assigned vector R-charge $2$ \cite{Halverson:2013eua}. In anomaly and central-charge matching, these fields account for the negative contribution of the off-diagonal vector multiplets. This assignment is also supported by exact GLSM computations \cite{Halverson:2013eua}. The associated Cartan theory thus provides an effective description of Coulomb-branch observables and serves as the starting point for the nonabelian mirror construction.

\subsection{Mirror constructions}\label{MC}
Having reviewed gauged linear sigma models, we now specify the data of their mirror theories. The mirror fields are twisted chiral in the original GLSM supersymmetry convention and chiral after the mirror exchange of the two R-symmetries; we use the latter terminology when discussing the B-twisted mirror.

\medskip

\Needspace{5\baselineskip}
\noindent{\textbf{Abelian mirrors}}

\medskip

For an abelian GLSM with gauge group $U(1)^k$ and no superpotential, the mirror theory is described by the following data:
\begin{itemize}
\item \textbf{Mirror of the chiral matter fields}: twisted chiral fields
\[
Y_i=\varrho_i+i\vartheta_i,
\qquad i=1,\dots,N,
\]
where $\varrho_i\in\mathbb{R}$ and $\vartheta_i$ is a periodic coordinate on $S^1$ with period $2\pi$, together with
\[
\Sigma_a\in\mathbb{C},
\qquad a=1,\dots,k.
\]
The field space is
\[
(\mathbb{C}^{\ast})^N\times\mathbb{C}^k.
\]

\item \textbf{Complex moduli parameters}: the complexified FI parameters
\[
t^a=r^a+i\theta^a.
\]

\item \textbf{Superpotential}:
\[
W
=
\sum_a
\Sigma_a
\Bigl(
\sum_i Q_i^aY_i
-t^a(\mu)
\Bigr)
+
\mu
\sum_i e^{-Y_i},
\]
where $\mu$ denotes the worldsheet renormalization scale.
\end{itemize}

The mirror theory is therefore a Landau-Ginzburg (LG) model. For a noncompact Calabi-Yau target space,\footnote{This should not be confused with a noncompact Calabi-Yau equipped with a nontrivial superpotential. Such theories are hybrid models rather than ordinary nonlinear sigma models.} the K\"ahler potential is irrelevant under the renormalization-group flow of the LG theory \cite{Vafa:1988uu}. Consequently, we do not specify the K\"ahler potential and instead work with the flat K\"ahler metric.

The construction above originates from the work of Hori and Vafa \cite{Hori:2000kt}. They first studied the mirror of the abelian Higgs model and then obtained mirrors of general toric varieties by considering a larger gauge theory in which additional $U(1)$ factors are gauged and subsequently taking an appropriate weak-coupling limit. This procedure was referred to as \emph{localization}; see \cite[Section 3.2.1]{Hori:2000kt} for details. The mirror of the abelian Higgs model is the LG theory on
\[
\mathbb{C}^{\ast}\times\mathbb{C}
\]
with superpotential
\[
W
=
\Sigma\bigl(Y-t(\mu)\bigr)
+
\mu e^{-Y}.
\]
The linear coupling $\Sigma(Y-t)$ follows directly from worldsheet T-duality. The exponential term $\mu e^{-Y}$ encodes the contributions of vortex sectors in the original gauge theory and is required for agreement with the exact dynamics \cite{Gu:2017nye}.  A related phenomenon occurs in three-dimensional mirror symmetry, where infrared descriptions are naturally formulated in terms of gauge-invariant operators rather than elementary ultraviolet fields \cite{Tong:2005un,Intriligator:2013lca}.

\medskip

For GLSMs with superpotentials, Hori and Vafa obtained mirror descriptions indirectly. For $q_i=0$, the fundamental mirror variable is $Y_i$; the $X_i$-coordinate formulas below apply to $q_i\ne0$, with $2/q_i$ required to be a positive integer. Motivated by exact results, we shall instead adopt the following direct description of the mirror theory:
\begin{itemize}
\item \textbf{Chiral matter fields}:
\[
X_i
=
\exp\!\left(
-\frac{q_i}{2}Y_i
\right),
\]
where $q_i$ is the vector R-charge of the corresponding matter field in the original GLSM, together with
\[
\Sigma_a\in\mathbb{C},
\qquad a=1,\dots,k.
\]
The field space is
\[
\prod_{i=1}^{N}
\frac{\mathbb{C}}{\mathbb{Z}_{2/q_i}}
\times
\mathbb{C}^k.
\]

\item \textbf{Complex moduli parameters}:
\[
t^a=r^a+i\theta^a.
\]

\item \textbf{Superpotential}:
\[
W
=
\sum_a
\Sigma_a
\left(
-
\sum_i
\frac{2Q_i^a}{q_i}
\log X_i
-
t^a(\mu)
\right)
+
\mu
\sum_i
X_i^{2/q_i}.
\]
\end{itemize}

The mirror superpotential has vector R-charge two in the mirror supersymmetry convention, corresponding to axial R-charge two in the original GLSM convention. In the $X_i$ coordinates, the mirror superpotential contains logarithmic singularities. These singularities indicate that the mirror should be interpreted as an effective field theory rather than a fundamental UV-complete description. Consequently, the standard perturbative canonical quantization is not directly applicable. Nevertheless, the path integral remains well defined, and the theory flows to the expected orbifold superconformal field theory in the infrared.

One advantage of this formulation is that, near the large-complex-structure limit, A-branes can be described directly in the UV without explicitly resolving the orbifold structure appearing in the infrared theory. We shall return to this point in Section~\ref{AILGM}.

\Needspace{5\baselineskip}
\noindent{\textbf{Nonabelian mirrors}}

For a nonabelian GLSM, the mirror theory is given by a Weyl-group orbifold of a Landau-Ginzburg model. The Landau-Ginzburg model contains the following chiral superfields:
\begin{itemize}
    \item $\dim V=\sum_{m=1}^{N}\dim(R_m)$ chiral superfields
    \[
    X_i=\exp\!\left(-\frac{q_i}{2}Y_i\right),
    \]
    or $Y_i$ if $q_i=0$, one for each weight of $V$ counted with multiplicity, where $q_i$ is the vector R-charge of the corresponding original matter multiplet;

    \item ${\rm rank}(\mathfrak g)$ chiral superfields $\Sigma_a$, associated with a choice of Cartan subalgebra of the Lie algebra $\mathfrak g$ of $G$;

    \item $\dim(\mathfrak g)-{\rm rank}(\mathfrak g)$ chiral superfields $X_{\tilde{\mu}}$, associated with the roots of $G$ and encoding the contributions of the off-diagonal vector multiplets.
\end{itemize}

The superpotential is
\begin{align}\label{MLGS}
W
=&
\sum_{a=1}^{{\rm rank}(\mathfrak g)}
\Sigma_a
\left(
\sum_m
\sum_{\rho\in\Lambda_{\mathfrak R_m}}
\rho^a\,Y_{\rho^m}
-
\sum_{\tilde{\mu}=1}^{\dim(\mathfrak g)-{\rm rank}(\mathfrak g)}
\alpha_{\tilde{\mu}}^a
\ln X_{\tilde{\mu}}
-
t^a
\right)
\nonumber\\
&
+
\sum_m
\sum_{\rho\in\Lambda_{\mathfrak R_m}}
e^{-Y_{\rho^m}}
+
\sum_{\tilde{\mu}=1}^{\dim(\mathfrak g)-{\rm rank}(\mathfrak g)}
X_{\tilde{\mu}} .
\end{align}

The Weyl-group orbifold acts by mapping weights to weights,
\begin{equation}
Y_i \mapsto Y_j,
\qquad
\sum_a \Sigma_a \rho_i^a
\mapsto
\sum_a \Sigma_a \rho_j^a ,
\end{equation}
roots to roots,
\begin{equation}
X_{\tilde{\mu}}
\mapsto
X_{\tilde{\nu}},
\qquad
\sum_a \Sigma_a \alpha_{\tilde{\mu}}^a
\mapsto
\sum_a \Sigma_a \alpha_{\tilde{\nu}}^a ,
\end{equation}
and acts on the formal Cartan parameters $t^a$ in the representation
dual to its action on the Cartan fields $\Sigma_a$, so that
\[
\sum_{a=1}^{{\rm rank}(\mathfrak g)} \Sigma_a t^a
\]
is invariant. The physical FI parameters are restricted to the
Weyl-invariant locus, with theta angles understood modulo their
periodicities. The superpotential is invariant under this combined
action.

If one ignores the superpotential, the field space is
\[
\left(\mathbb C^\ast\right)^{\sum_i \dim(R_i)}
\times
\mathbb C^{\dim(\mathfrak g)}
\Big/ S ,
\]

which contains fixed loci of the Weyl-group action. The mirror construction of \cite{Gu:2018fpm} starts from the associated Cartan theory, itself an effective description of the nonabelian GLSM at a generic point of the Coulomb branch. Accordingly, we regard the mirror Landau-Ginzburg model as an effective theory, without asserting that it gives a UV-complete dual description.

The logarithmic singularities of the superpotential also obstruct a straightforward free-field description. They do not, however, prevent the proposed theory from serving as a mirror: the mirror is required to reproduce the infrared physics of the nonlinear sigma model on the original target space.

It remains desirable to obtain a genuine nonabelian T-duality construction. Such a construction would provide a direct physical derivation of nonabelian mirror symmetry from a UV-complete theory. Alternatively, a detailed understanding of the renormalization-group flow of the K\"ahler potential in the associated Cartan theory could also furnish a first-principles derivation of the nonabelian mirror proposal of \cite{Gu:2018fpm}.

\subsection{Yukawa couplings}\label{YC}
In this paper, we compute Gromov-Witten invariants of Calabi-Yau manifolds following the procedure described in Section~\ref{GWFM}. As explained there, the computation requires knowledge of the Yukawa couplings. We therefore briefly review the relevant results for GLSMs.

For abelian theories, Yukawa couplings can be extracted from Picard-Fuchs equations, as originally studied in \cite{Candelas:1990rm,Morrison:1993}.\footnote{In Section~\ref{PEEFP} we will define Picard-Fuchs equations for nonabelian mirrors. In principle, the corresponding Yukawa couplings can be derived from these equations, but we shall not pursue this approach here.} They may also be obtained from instanton sums in the GLSM description \cite{Morrison:1994fr}. In this paper, we instead employ a formula derived from supersymmetric localization, which applies equally well to nonabelian GLSMs.

Building on localization techniques developed in supersymmetric gauge theories \cite{Pestun:2007rz,Pestun:2016zxk}, Closset {\it et al.}~\cite{Closset:2015rna} computed A-twisted correlation functions of GLSMs. The genus-zero correlation functions are given by
\begin{equation}\label{ACL}
\langle {\cal O}(\sigma) \rangle_{g=0}
=
\frac{1}{|S|}
\sum_{\mathbf{k}}\left(\prod_{a=1}^{\mathrm{rank}(G)}(q^a)^{k^a}\right)
\oint_{{\rm JK}(r^{UV})}
\left(
\prod_{a=1}^{{\rm rank}(G)}
\frac{d\widehat{\sigma}_a}{2\pi i}
\right)
{\cal Z}^{\rm 1\text{-}loop}_{\mathbf{k}}(\widehat{\sigma})
\,{\cal O}(\widehat{\sigma}),
\end{equation}
where $\mathbf{k}=(k^1,\ldots,k^{\mathrm{rank}(G)})$ runs over the allowed magnetic fluxes, $\widehat{\sigma}_a$ are complex coordinates on the Cartan subalgebra of the gauge group, and ${\rm JK}(r^{UV})$ denotes the Jeffrey--Kirwan residue contour determined by the ultraviolet FI parameter $r^{UV}$. We denote the root system of $G$ by $\Delta(G)$.

The one-loop determinant factorizes into matter and vector multiplet contributions,
\begin{equation}
{\cal Z}^{\rm 1\text{-}loop}
=
{\cal Z}^{\rm 1\text{-}loop}_{\rm matter}
\,{\cal Z}^{\rm 1\text{-}loop}_{\rm gauge},
\end{equation}
where
\begin{equation}
{\cal Z}^{\rm 1\text{-}loop}_{\rm matter}
=
\prod_{\rho\in\mathfrak{R}}
\left[
\frac{1}{\rho(\widehat{\sigma})}
\right]^{\rho(\mathbf{k})-q_\rho+1},
\qquad
{\cal Z}^{\rm 1\text{-}loop}_{\rm gauge}
=
(-1)^{\sum_{\alpha>0}\alpha(\mathbf{k})}
\prod_{\alpha\in\Delta(G)}
\alpha(\widehat{\sigma}).
\end{equation}

Using the proposed Landau-Ginzburg mirrors reviewed in Section~\ref{MC}, one may also compute these correlation functions in the B-twisted mirror theory. At genus zero, B-model correlation functions localize to the critical points of the superpotential. Since only the bosonic zero modes contribute, we denote the lowest scalar components of the chiral superfields by the corresponding lowercase letters. The correlation functions can then be expressed as residue integrals \cite{Gu:2018fpm,Vafa:1990mu},
\begin{equation}\label{MBCF}
\langle {\cal O} \rangle_{g=0}
=
\frac{1}{|S|}
\oint
\frac{
\displaystyle
\bigwedge_a d\sigma_a
\wedge
\bigwedge_{\widetilde{\mu}} dx_{\widetilde{\mu}}
\wedge
\bigwedge_{i,a} dy_i^a
\wedge
\bigwedge_p dx_p
}
{
\displaystyle
\prod_a \partial_{\sigma_a}W
\cdot
\prod_{\widetilde{\mu}}
\partial_{x_{\widetilde{\mu}}}W
\cdot
\prod_{i,a}
\partial_{y_i^a}W
\cdot
\prod_p
\partial_{x_p}W
}
\,{\cal O}.
\end{equation}

Mirror symmetry identifies the A-twisted correlation functions of the GLSM with the B-twisted correlation functions of the mirror Landau-Ginzburg model. Consequently, Yukawa couplings may be computed equivalently from either side of the mirror correspondence. In the examples studied in this paper, we will primarily use the mirror description, since the residue formula \eqref{MBCF} provides a particularly efficient method for extracting Yukawa couplings and, subsequently, Gromov-Witten invariants.

\Needspace{5\baselineskip}
\noindent{\textbf{Calabi-Yau complete intersections in Grassmannians}}.

Hori and Tong constructed GLSMs describing Calabi-Yau complete intersections in Grassmannians \cite{Hori:2006dk}. The gauge group is $U(k)$, with $N$ fundamental chiral multiplets $\Phi_i$ and $m$ chiral multiplets $P^\beta$ transforming in the representation $\det^{-Q_\beta}$ of the gauge group, where $Q_\beta>0$. The Calabi-Yau condition is
\[
N=\sum_{\beta=1}^{m}Q_\beta.
\]
To obtain a compact Calabi-Yau target space, one introduces the superpotential
\[
W=\sum_{\beta=1}^{m}P^\beta G_\beta(B),
\]
where $G_\beta$ is a homogeneous polynomial of degree $Q_\beta$ in the baryonic (Pl\"ucker) coordinates
\begin{equation}
B_{i_1\cdots i_k}
=
\epsilon_{a_1\cdots a_k}
\Phi^{a_1}_{i_1}
\cdots
\Phi^{a_k}_{i_k}.
\end{equation}

In the geometric phase, the fields $P^\beta$ are assigned vector R-charge $2$, while the fields $\Phi_i$ have vector R-charge $0$. The mirror Landau-Ginzburg model then follows directly from the GLSM data. Using either the localization formula \eqref{ACL} or the mirror residue formula \eqref{MBCF}, one obtains the residue expression
\begin{equation}\label{CLABG}
\langle{\cal O}\rangle
=
\left(
\prod_{\beta=1}^{m}Q_\beta
\right)
{\rm Res}
\frac{
\left[
\prod_{\widetilde{\mu}<\widetilde{\nu}}
(z_{\widetilde{\mu}}-z_{\widetilde{\nu}})^2
\right]
\left(
\sum_{a=1}^{k}z_a
\right)^m
{\cal O}(z)
}
{
\left[
\prod_{a=1}^{k-1}
(z_a^N-1)
\right]
\left[
1
+
(-1)^{k-N}
q
\prod_{\beta=1}^{m}
Q_\beta^{Q_\beta}
\left(
\sum_{a=1}^{k}z_a
\right)^N
\right]
},
\end{equation}
where
\[
z_a=\frac{\sigma_a}{\sigma_k}.
\]

Here $z_k=1$ by definition. Applying \eqref{CLABG} to suitable gauge-invariant observables yields the Yukawa couplings for a number of Calabi-Yau threefolds \cite{Closset:2015rna}. The observables are holomorphic functions of the gauge-invariant coordinates
\[
u_j(\sigma)\equiv {\rm Tr}(\sigma^j).
\]

\medskip

\begin{itemize}
    \item $X_4\subset G(2,4)$.

The correlation functions are
\begin{equation}\label{CLGR24}
\langle u_1(\sigma)^3\rangle
=
\frac{8}{1-2^{10}q},
\qquad
\langle u_2(\sigma)u_1(\sigma)\rangle
=
0.
\end{equation}

The singular locus is located at
\[
q=2^{-10},
\]
which also follows from the effective twisted superpotential \cite{Hori:2006dk}.

 \item $X_{2^2,1}\subset G(2,5)$.

The Yukawa couplings are
\begin{align}
\langle u_1(\sigma)^3\rangle
&=
\frac{20}
     {1+11(2^4q)-(2^4q)^2},
\nonumber\\
\langle u_1(\sigma)u_2(\sigma)\rangle
&=
\frac{4(1-32q)}
     {1+11(2^4q)-(2^4q)^2}.
\label{CLGR252}
\end{align}

 \item $X_{3,1^2}\subset G(2,5)$.

The Yukawa couplings are
\begin{align}
\langle u_1(\sigma)^3\rangle
&=
\frac{15}
     {1+11(3^3q)-(3^3q)^2},
\nonumber\\
\langle u_1(\sigma)u_2(\sigma)\rangle
&=
\frac{3(1-54q)}
     {1+11(3^3q)-(3^3q)^2}.
\label{CLGR25}
\end{align}

 \item $X_{2,1^4}\subset G(2,6)$.

The Yukawa couplings are
\begin{align}
\langle u_1(\sigma)^3\rangle
&=
\frac{28}
     {(1+2^2q)(1-27(2^2q))},
\nonumber\\
\langle u_1(\sigma)u_2(\sigma)\rangle
&=
\frac{8(1+18q)}
     {(1+2^2q)(1-27(2^2q))}.
\label{CLGR26}
\end{align}

 \item $X_{1^6}\subset G(3,6)$.

The Yukawa couplings are
\begin{align}
\langle u_1(\sigma)^3\rangle
&=
\frac{42}
     {(1-q)(1-64q)},
\nonumber\\
\langle u_1(\sigma)u_2(\sigma)\rangle
&=
0,
\nonumber\\
\langle u_3(\sigma)\rangle
&=
-\frac{6(1-8q)}
       {(1-q)(1-64q)}.
\label{CLGR36}
\end{align}

 \item $X_{1^7}\subset G(2,7)$.

The Yukawa couplings are
\begin{align}
\langle u_1(\sigma)^3\rangle
&=
\frac{14(3+q)}
     {1+57q-289q^2-q^3},
\nonumber\\
\langle u_1(\sigma)u_2(\sigma)\rangle
&=
\frac{14(1-9q)}
     {1+57q-289q^2-q^3}.
\label{CLGR27}
\end{align}
\end{itemize}

The poles of these Yukawa couplings correspond to discriminant loci in the K\"ahler moduli space. For the examples considered here, these loci are conifold singularities. 

\section{A-branes in Landau-Ginzburg models}\label{AILGM}

We now study A-type boundary conditions in Landau-Ginzburg models \cite{Hori:2000ck}. We denote the corresponding A-branes by $\gamma_i$.

For a B-twisted bulk theory with A-type boundaries, one can define the following pairing between B-model observables ${\cal O}_a(X)$ and A-branes:
\begin{equation}\label{PER}
  \Pi_{a,i}=\left\langle\gamma_{i}\mid{\cal O}_{a}(X)\right\rangle.
\end{equation}

In the $tt^{\ast}$ framework, these periods satisfy the flatness equations
\begin{equation}\nonumber
  \nabla_{a}\Pi_{b}=\left(D_{a}+C_{a}\right)\Pi_{b}=0 \qquad \overline{\nabla}_{a}\overline{\Pi}_{b}=\left(\overline{D}_{a}+\overline{C}_{a}\right)\overline{\Pi}_{b}=0,
\end{equation}
where $C_{a}$ denotes multiplication by the chiral-ring observable corresponding to ${\cal O}_{a}(X)$, and $D_{a}$ is the covariant derivative defined in \cite{Cecotti:1991me}. 

Localization in the B-twisted bulk suggests that only constant maps from the worldsheet with boundary to the LG target contribute to the period path integral. The resulting finite-dimensional integral is
\begin{equation}\label{PLG}
  \Pi_{a,i}=\int_{\gamma_{i}}{\cal O}_{a}(X)\exp\left(-W(X)\right)\Omega,
\end{equation}
where $\Omega$ is the holomorphic top form on the noncompact field space of the Landau-Ginzburg theory. For the identity insertion ${\cal O}_{a}(X)=1$, the expression becomes the Landau-Ginzburg analogue of the central charge associated with the A-brane $\gamma_i$:
\begin{equation}\label{0PER}
  \Pi_{1,i}=\int_{\gamma_{i}}\exp\left(-W(X)\right)\Omega.
\end{equation}
Let $t^a$ be the coordinates on the complex moduli space. We then have
\begin{equation}\label{DP}
  \partial_{t^{a}}\Pi_{1,i}=\Pi_{a,i}=\int_{\gamma_{i}}{\cal O}_{a}(X)\exp\left(-W(X)\right)\Omega,
\end{equation}
\begin{equation}\label{DDP}
  \partial_{t^{a}}\partial_{t^{b}}\Pi_{1,i}=\partial_{t^{a}}\Pi_{b,i}=C^{c}_{ab}\Pi_{c,i},
\end{equation}
where ${\cal O}_{a}(X)$ is the chiral operator associated with the deformation parametrized by $t^a$. The periods thus encode the structure constants of the chiral ring. To make the period integrals explicit, we first discuss the relevant A-branes.

We consider Landau-Ginzburg models with a noncompact Calabi-Yau target and a holomorphic superpotential $W$. When a UV description is available in which $W$ can be treated perturbatively, A-branes may first be described in the corresponding free theory. We begin with the targets $\mathbb C^n$ and $(\mathbb C^\star)^n$.

\noindent\textbf{Special Lagrangians.} For a Calabi-Yau target space $Y$, a special Lagrangian submanifold is an oriented\footnote{An orientation is specified by an equivalence class of nowhere-vanishing real top forms, where two forms are equivalent if they differ by a positive function.} Lagrangian submanifold $\gamma\subset Y$ satisfying
\begin{equation}\label{SPC}
  \operatorname{Im}\!\left(e^{-i\pi\xi(\gamma)}\Omega\big|_{\gamma}\right)=0
\end{equation}
for some real constant $\xi\left(\gamma\right)$.

$\bullet$ Consider the following linear A-branes in $\mathbb C^n$. Rotating the coordinate axes of $\mathbb R^n\subset\mathbb C^n$ by angles $\theta_i$ gives the embedding
\begin{equation}\nonumber
  \gamma^{\overrightarrow{\theta}}=\left\{\left(e^{i\theta_{1}}x_{1},e^{i\theta_{2}}x_{2},\ldots,e^{i\theta_{n}}x_{n}\right):x_{i}\in \mathbb{R}\right\}.
\end{equation}

The pullback of the holomorphic $n$-form $\Omega=dz_1\wedge\cdots\wedge dz_n$ to this submanifold is
\begin{equation}\nonumber
  \Omega=e^{i\left(\theta_{1}+\cdots+\theta_{n}\right)}dx_{1}\wedge dx_{2}\wedge\cdots\wedge dx_{n}.
\end{equation}
The special Lagrangian condition \eqref{SPC} is satisfied by choosing
\begin{equation}\nonumber
  \xi\left(\gamma\right)=\frac{1}{\pi}\sum^{n}_{i=1}\theta_{i}.
\end{equation}

We denote by $\gamma^{\overrightarrow{0}}$ the plane with $\theta_1=\cdots=\theta_n=0$, equipped with the orientation determined by $dx_1\wedge\cdots\wedge dx_n$. For this plane, we choose $\xi(\gamma)=0$.

$\bullet$ On $(\mathbb C^\star)^n$, an analogous family of A-branes is described in logarithmic coordinates $Y_i=\rho_i+i\theta_i$, where $\rho_i\in\mathbb R$ and $\theta_i\sim\theta_i+2\pi$, so that $e^{-Y_i}\in\mathbb C^\star$. For fixed angles $\theta_i$, we set
\begin{equation}\label{PLGCS}
  \gamma^{\overrightarrow{\theta}}=\left\{\left(\rho_{1}+i\theta_{1},\rho_{2}+i\theta_{2},\cdots,\rho_{n}+i\theta_{n}\right):\rho_{i}\in \mathbb{R}\right\}.
\end{equation}
With $\Omega=dY_1\wedge\cdots\wedge dY_n$ and the orientation determined by $d\rho_1\wedge\cdots\wedge d\rho_n$, we may choose $\xi(\gamma)=0$.

When singularities of the superpotential obstruct a perturbative UV description, a nonperturbative description of A-branes is needed. We therefore use the supersymmetric boundary conditions described in \cite{Hori:2000kt,Hori:2000ck}, together with the non-renormalization properties of the LG superpotential.

$\bullet$ Consider an LG model with target $(\mathbb C^\star)^m\times\mathbb C^n$ and superpotential $W$. For the A-type boundary condition used here, the image of the Lagrangian submanifold under $W$ lies on a line parallel to the real axis. Equivalently,
\begin{equation}\label{WC}
{\rm{Im}} W={\rm constant}.
\end{equation}
When the critical values have distinct imaginary parts, one can choose rays in the $W$-plane that start at the critical values and run parallel to the positive real axis. The corresponding A-branes are discussed in \cite{Hori:2000ck}. The condition \eqref{WC} can be modified by introducing a suitable boundary interaction. Our LG models contain $\Sigma$ fields in addition to matter fields. Following \cite{Hori:2013ika,Herbst:2008jq}, we choose a Lagrangian integration cycle in the $\Sigma$-field space so that the periods reduce to contour integrals, while retaining \eqref{WC} for the remaining field directions. Independence of the chosen contour is understood to hold only under admissible deformations that preserve convergence and do not cross singularities.

\Needspace{5\baselineskip}
\noindent\textbf{The mirror of the GLSM for $\mathbb{CP}^{N-1}$.} This mirror is an LG model with target $(\mathbb C^\star)^N\times\mathbb C$ and superpotential
 \begin{equation}\nonumber
   W=\Sigma\left(\sum^{N}_{i=1}Y_{i}-t\right)+\sum^{N}_{i=1}e^{-Y_{i}}.
 \end{equation}

The critical points are obtained by solving the vacuum equations
\begin{equation}\nonumber
  \partial_{Y_{i}}W=\partial_\Sigma W=0,
\end{equation}
which gives
\begin{equation}\nonumber
  \Sigma=X_{i}=e^{-Y_{i}},\qquad \Sigma^{N}=X_{i}^{N}=q.
\end{equation}

This motivates the following choice of Lagrangian integration cycle:
\begin{equation}\nonumber
  \gamma=\left\{\left(L, \rho_{1}+i\left(\frac{\theta}{N}+\frac{2\pi k}{N}\right),\cdots,\rho_{N}+i\left(\frac{\theta}{N}+\frac{2\pi k}{N}\right)\right): \rho_{i}\in\left(-\infty,\frac{r}{N}\right)\right\}.
\end{equation}
We leave the contour $L$ in the $\Sigma$-field space implicit, subject to the conditions discussed in \cite{Hori:2013ika}. In particular, $L$ should be an admissible deformation of the real locus that avoids singularities and ensures convergence of the hemisphere partition function. In the large-volume limit, the additional constraint imposed by the superpotential selects a subclass of the cycles in \eqref{PLGCS}, as expected from the perturbative description.

\noindent\textbf{The mirror of the GLSM for $Gr(k,N)$.} The proposed mirror is the $S_k$ orbifold of an LG model with field space $(\mathbb C^\star)^{Nk}\times\mathbb C^{k^2}$ and superpotential
 \begin{equation}\nonumber
   W=\Sigma_{a}\left(\sum^{N}_{i=1}Y^{a}_{i}+\alpha^{a}_{\tilde{\mu}\tilde{\nu}}\log X_{\tilde{\mu}\tilde{\nu}}-t\right)+\sum^{N}_{i=1}\sum^{k}_{a=1}e^{-Y^{a}_{i}}+\sum_{\tilde{\mu}\tilde{\nu}}X_{\tilde{\mu}\tilde{\nu}}.
 \end{equation}
Solving the vacuum equations
\begin{equation}\nonumber
  \partial_{Y^{a}_{i}}W=\partial_{X_{\tilde{\mu}\tilde{\nu}}}W=\partial_{\Sigma_a}W=0
\end{equation}
gives the critical locus
\begin{equation}\nonumber
  \Sigma_{a}=X^{a}_{i}=e^{-Y^{a}_{i}},\qquad X_{\tilde{\mu}\tilde{\nu}}=\Sigma_{\tilde{\mu}}-\Sigma_{\tilde{\nu}}\neq 0,\qquad \Sigma_{a}^{N}=\left(X^{a}_{i}\right)^{N}=\left(-1\right)^{k-1}q.
\end{equation}

These critical-point equations motivate the following candidate Lagrangian integration cycle:
\begin{align}\label{LSGR}
  \gamma=\Bigg\{\Bigg. & L,\Bigg(\Bigg. \rho^{a}_{1}+i\left(\frac{\theta}{N}+\frac{2\pi k^{a}+\pi\left(k-1\right)}{N}\right),\cdots,\rho^{a}_{N}+i\left(\frac{\theta}{N}+\frac{2\pi k^{a}+\pi\left(k-1\right)}{N}\right),
   \\\nonumber
   &\bigcup_{\tilde{\mu}\neq\tilde{\nu}} \left(\left(e^{-i\left(\frac{\theta}{N}+\frac{2\pi k^{\tilde{\mu}}+\pi\left(k-1\right)}{N}\right)}-e^{-i\left(\frac{\theta}{N}+\frac{2\pi k^{\tilde{\nu}}+\pi\left(k-1\right)}{N}\right)}\right)e^{-\frac{r}{N}}+\rho\right)\Bigg.\Bigg):
    \\\nonumber
    &\rho^{a}_{i}\in\left(-\infty,\frac{r}{N}\right),\quad \rho\in [0,+\infty),\quad k^{a}\neq k^{b}\quad \rm{if}\quad \textit{a}\neq \textit{b} \Bigg.\Bigg\}.
\end{align}

As above, we leave the contour $L$ in the $\Sigma_a$ directions
implicit and regard \eqref{LSGR} as a candidate integration cycle
on the covering Landau--Ginzburg model. Each Weyl image of
$\gamma$ is equipped with the orientation induced by the group
action. For the holomorphic integration form
\[
\Omega=
\left(\bigwedge_a d\Sigma_a\right)
\wedge
\left(\bigwedge_{i,a}\frac{dX_i^a}{X_i^a}\right)
\wedge
\left(\bigwedge_{\mu\neq\nu}dX_{\mu\nu}\right),
\]
with a fixed ordering of the differentials, the Weyl action gives
\[
w^*\Omega=(\operatorname{sgn}w)^N\Omega,
\qquad w\in S_k.
\]
Since the superpotential is Weyl invariant, the corresponding
period integrals satisfy
\[
\int_{w_*\gamma}e^{-W}\Omega
=
(\operatorname{sgn}w)^N
\int_\gamma e^{-W}\Omega.
\]

In this proposed construction, the fiber over $\gamma$ is specified by the identification above, rather than by the torus fibration appearing in the geometric picture of \cite{Strominger:1996it}. The root fields $X_{\tilde\mu\tilde\nu}$ are massive and are to be integrated out when passing to the mirror of the geometric target. In the large-volume limit, the configuration \eqref{LSGR} is intended to be compatible with this fiber identification.

Integrating out the $\Sigma_a$ fields also requires care. In the physical path integral, both $\Sigma_a$ and $\overline\Sigma_a$ are integrated out, and the full measure must respect the Weyl-group action. The resulting effective theory may have a holomorphic top form that descends only up to a sign. The applicability of the B-twist in this situation is discussed in \cite{Sharpe:2006qd}; the corresponding low-energy A-branes must carry compatible orbifold data. Similar questions arise when extending this construction to other Fano targets and Calabi-Yau manifolds. 

\section{Quantum differential equations for periods}\label{PEEFP}

In the previous section, we defined A-branes in the Landau-Ginzburg models under consideration. These branes provide the starting point for computing periods and deriving their quantum differential equations. We carry out this procedure below.

\subsection{Quantum differential equations of nonabelian mirrors}\label{PFNM}

We first consider quantum differential equations for mirrors of nonabelian GIT quotients, and then extend the discussion to mirrors of complete intersections in these spaces.

\subsubsection{Nonabelian GIT quotient spaces}\label{NGQS}
As described in Section~\ref{MC}, the mirror is a Landau-Ginzburg model with field space $\left(\left(\mathbb{C}^{\star}\right)^{n}\times \mathbb{C}^{r+k}\right)/S$, where $S$ is the Weyl group of $G$. Its superpotential is
\begin{align}\label{MLGS-GIT}
  W = & \sum^{k}_{a=1}\Sigma_{a}\left(\sum^{n}_{i=1}\rho_{i}^{a}Y_{i}-\sum^{r}_{\tilde{\mu}=1}\alpha^{a}_{\widetilde{\mu}}\ln X_{\tilde{\mu}}-t^{a}\right) \\\nonumber
  & +\sum_{i}\exp\left(-Y_{i}\right)+\sum^{r}_{\tilde{\mu}=1}X_{\tilde{\mu}}.
\end{align}
The physical FI parameters correspond to the abelian factors of $G$. To formulate the differential equations, we first introduce one formal parameter for each Cartan generator. We solve these equations for the generating functions and subsequently restrict the parameters to the physical K\"ahler moduli. The Weyl-group action must be tracked throughout this procedure, with physical quantities obtained after restriction.

The period is
\begin{align}\label{PMNG}
  \Pi=&\int \prod^{k}_{a=1}d\Sigma_{a}\prod^{n}_{i=1}dY_{i}\prod^{r}_{\tilde{\mu}=1}dX_{\tilde{\mu}}
  \\\nonumber
  &\exp\left(-\sum^{k}_{a=1}\Sigma_{a}\left(\sum^{n}_{i=1}\rho_{i}^{a}Y_{i}-\sum^{r}_{\tilde{\mu}=1}\alpha^{a}_{\widetilde{\mu}}\ln X_{\tilde{\mu}}-t^{a}\right)-\sum_{i}\exp\left(-Y_{i}\right)-\sum^{r}_{\tilde{\mu}=1}X_{\tilde{\mu}}\right).
\end{align}
Each $Y$ field is a logarithmic coordinate valued in $\mathbb{C}/(2\pi i\mathbb{Z})$, so its global symmetry acts additively as a shift,
\[
Y\mapsto Y-\log\mu .
\]
By contrast, each $X=e^{-Y}$ is an ordinary LG chiral field valued in $\mathbb{C}$, and the same symmetry acts multiplicatively as a scaling,
\[
X\mapsto \mu X .
\]
Following Appendices~\ref{APFMT} and \ref{PFEMS}, we introduce coefficients $\mu_i$ for $e^{-Y_i}$ and $\nu_\alpha$ for $X_\alpha$, keeping $t^a$ fixed. The rescaling $X_\alpha\mapsto\nu_\alpha X_\alpha$ supplies the root Jacobian in the deformed period
\begin{align}\label{PMNG2}
  \Pi(\mu,\nu;t)={}&\int\prod_a d\Sigma_a\prod_i dY_i\prod_{\alpha\in\Delta}(\nu_\alpha\,dX_\alpha)
  \\\nonumber
  &\times\exp\!\left[-\sum_a\Sigma_a\!\left(\sum_i\rho_i^aY_i-\sum_{\alpha\in\Delta}\alpha^a\log X_\alpha-t^a\right)
       -\sum_i\mu_i e^{-Y_i}-\sum_{\alpha\in\Delta}\nu_\alpha X_\alpha\right].
\end{align}
Here $\Delta$ is the full root system. To differentiate only the exponential, set $Z=\Pi/\prod_{\alpha\in\Delta}\nu_\alpha$. Write $(x_J,c_J,\lambda_J)$ for $(e^{-Y_i},\rho_i,\mu_i)$ or $(X_\alpha,\alpha,\nu_\alpha)$. On the Cartan constraint, $\prod_J x_J^{c_J^a}=q^a$, where $q^a=e^{-t^a}$. Each coefficient derivative inserts $-x_J$, so
\begin{equation}\label{PMNG3}
  \prod_{c_J^a>0}\partial_{\lambda_J}^{\,c_J^a} Z
   =(-1)^{\sum_i\rho_i^a}q^a
     \prod_{c_J^a<0}\partial_{\lambda_J}^{\,-c_J^a} Z .
\end{equation}
The roots occur in opposite pairs, so $\sum_{\alpha\in\Delta}\alpha^a=0$. To integrate them out, choose a set $\Delta_+$ of positive roots. Write $S^a=\sum_{\alpha\in\Delta_+}\alpha^a$, $T^a=t^a-i\pi S^a$, and $\theta_a=-\partial_{t^a}$. With the root cycles specified below, the period at $\mu_i=\nu_\alpha=1$ is, up to an overall normalization,
\begin{equation}\label{AblP}
  \Pi(t)=\prod_{\alpha\in\Delta_+}[-\alpha(\theta)]\, F(t),
  \qquad F(t)=\Pi^{\rm Ab}(T),\qquad \alpha(\theta)=\sum_b\alpha^b\theta_b .
\end{equation}
The abelian period $F$ satisfies Eq.~\eqref{PFMT2} evaluated at $T$. Setting $D_i=\sum_b\rho_i^b\theta_b$ and $s_a=\sum_{\rho_i^a>0}\rho_i^a$, we obtain
\begin{equation}\label{ABLP2}
  \left[\prod_{\rho_i^a>0}\prod_{\ell=0}^{\rho_i^a-1}(D_i-\ell)\right]F
  =(-1)^{s_a+S^a}q^a
   \left[\prod_{\rho_i^a<0}\prod_{\ell=0}^{-\rho_i^a-1}(-D_i+\ell)\right]F .
\end{equation}
We use an integral Cartan basis. The two signs on the right come from the matter fields and from $e^{-T^a}=(-1)^{S^a}q^a$, respectively. In the large-volume regime the remaining Cartan and matter integrations are taken over $L\times\mathbb R^n$, with the theta-angle dependence of these contours left implicit. Candidate cycles for the Grassmannian mirror are discussed in Section~\ref{AILGM}.

We now evaluate the root integral at fixed $\Sigma_a$. Write
\[
z_\alpha=\alpha(\Sigma)
=\sum_{a=1}^{k}\alpha^a\Sigma_a.
\]
For each pair of roots $\alpha$ and $-\alpha$, the corresponding
fields contribute
\[
W_\alpha
=
X_\alpha+X_{-\alpha}
-z_\alpha\log X_\alpha
+z_\alpha\log X_{-\alpha}
\]
to the superpotential. The critical equations give
\[
X_\alpha=z_\alpha,
\qquad
X_{-\alpha}=-z_\alpha,
\]
as in the root-field elimination of \cite{Gu:2018fpm}.

For fixed real $\Sigma$ with $z_\alpha>0$, we choose
the positive real axis for $X_\alpha$ and the Hankel
contour $\mathcal H$ for $X_{-\alpha}$, with
$\log(-z_\alpha)=\log z_\alpha+i\pi$.
The Hankel contour is oriented clockwise from $\arg X=2\pi$
to $\arg X=0$ and represents the Lefschetz thimble
through $X_{-\alpha}=-z_\alpha$.\footnote{
Writing $X_{-\alpha}=re^{i\varphi}$, the condition
\[
\operatorname{Im}
\left(X_{-\alpha}+z_\alpha\log X_{-\alpha}\right)
=\pi z_\alpha
\]
gives
\[
r=\frac{z_\alpha(\pi-\varphi)}{\sin\varphi},
\qquad 0<\varphi<2\pi.
\]
The contour passes through $X_{-\alpha}=-z_\alpha$
at $\varphi=\pi$.
}
The root integration cycle is
\[
\gamma_{\mathrm{root}}
=
\prod_{\alpha\in\Delta_+}
\left(
\mathbb R_{>0}^{(X_\alpha)}
\times
\mathcal H^{(X_{-\alpha})}
\right),
\]
with orientation
\[
\bigwedge_{\alpha\in\Delta_+}
\left(dX_\alpha\wedge dX_{-\alpha}\right).
\]

For
\[
-1<\operatorname{Re}z_\alpha<1,
\]
the positive-ray integral converges, and the small-circle
contribution to the Hankel integral vanishes as the circle
shrinks. Comparing the two logarithmic branches gives
\begin{equation}\label{WIC}
\begin{aligned}
\int_0^\infty dX_\alpha\,
X_\alpha^{z_\alpha}e^{-X_\alpha}
&=\Gamma(1+z_\alpha),\\
\int_{\mathcal H}dX_{-\alpha}\,
X_{-\alpha}^{-z_\alpha}e^{-X_{-\alpha}}
&=
\bigl(1-e^{-2\pi i z_\alpha}\bigr)
\Gamma(1-z_\alpha).
\end{aligned}
\end{equation}
Taking the product over positive roots, we find
\begin{equation}\label{WIGR}
\mathcal I_{\mathrm{root}}(\Sigma)
=
\prod_{\alpha\in\Delta_+}
\bigl(1-e^{-2\pi i z_\alpha}\bigr)
\Gamma(1-z_\alpha)\Gamma(1+z_\alpha).
\end{equation}

The reflection formula gives
\[
\bigl(1-e^{-2\pi i z}\bigr)
\Gamma(1-z)\Gamma(1+z)
=
\bigl(1-e^{-2\pi i z}\bigr)
\frac{\pi z}{\sin\pi z}
=
2\pi i z\,e^{-i\pi z}.
\]
The root integrals therefore reduce to
\begin{equation}\label{WIGR2}
\mathcal I_{\mathrm{root}}(\Sigma)
=
(2\pi i)^{|\Delta_+|}
\exp\!\left(
-i\pi\sum_{\alpha\in\Delta_+}\alpha(\Sigma)
\right)
\prod_{\alpha\in\Delta_+}\alpha(\Sigma).
\end{equation}
The right-hand side is entire in $\Sigma$. We use this
continuation beyond the initial convergence strip when
performing the remaining integrations.

For $z_\alpha\neq0$, the root saddle satisfies
\[
W_\alpha(z_\alpha,-z_\alpha)=i\pi z_\alpha,
\qquad
\det W_\alpha''(z_\alpha,-z_\alpha)
=-\frac{1}{z_\alpha^2},
\]
where $W_\alpha''$ is the Hessian with respect to
$X_\alpha$ and $X_{-\alpha}$. Its Gaussian integral,
with the orientation chosen above, is
$2\pi i z_\alpha e^{-i\pi z_\alpha}$, in agreement
with the paired root integral.

Absorbing $(2\pi i)^{|\Delta_+|}$ into the normalization,
the period with these root cycles becomes
\begin{equation}\label{PMNG4}
\begin{aligned}
\Pi={}&
\int_{L\times\mathbb R^n}
\left(\prod_{a=1}^{k}d\Sigma_a\right)
\left(\prod_{i=1}^{n}dY_i\right)
\prod_{\alpha\in\Delta_+}\alpha(\Sigma)
\\
&\quad\times
\exp\!\left[
-\sum_{a=1}^{k}\Sigma_a
\left(
\sum_{i=1}^{n}\rho_i^aY_i
-t^a
+i\pi\sum_{\alpha\in\Delta_+}\alpha^a
\right)
-\sum_{i=1}^{n}e^{-Y_i}
\right].
\end{aligned}
\end{equation}
The FI parameters are thus shifted to
\[
t_{\mathrm{eff}}^a
=
t^a-i\pi\sum_{\alpha\in\Delta_+}\alpha^a.
\]
Since $\theta_a e^{\sum_b t^b\Sigma_b}=-\Sigma_a e^{\sum_b t^b\Sigma_b}$, Eq.~\eqref{PMNG4} gives
\begin{equation}\label{PMNG5}
  \Pi(t)=\prod_{\alpha\in\Delta_+}[-\alpha(\theta)]\,\Pi^{\rm Ab}(T),
\end{equation}
where the abelian period on the specified large-volume cycle is
\begin{equation}\label{PMNG6}
  \Pi^{\rm Ab}(T)=\int_{L\times\mathbb R^n}\!
   \prod_a d\Sigma_a\prod_i dY_i\,
   \exp\!\left[-\sum_a\Sigma_a\!\left(\sum_i\rho_i^aY_i-T^a\right)-\sum_i e^{-Y_i}\right].
\end{equation}
For the Grassmannian this gives the alternating differential operator discussed in the appendix of \cite{Hori:2000kt}.

\Needspace{5\baselineskip}
\noindent \textbf{Example: Grassmannian}

The gauged linear sigma model for $Gr(k,N)$ is a $U(k)$ gauge theory with $N$ fundamental chiral superfields. The dual Landau-Ginzburg model is $\left(\left(\mathbb{C}^{\star}\right)^{kN}\times\mathbb{C}^{k^{2}}\right)/S_{k}$ with superpotential
\begin{equation}\label{MGR}
  W=\sum^{k}_{a=1}\Sigma_{a}\left(\sum^{N}_{p=1}Y^{a}_{p}-\sum_{\widetilde{\mu}}\alpha^{a}_{\widetilde{\mu}}\log X_{\widetilde{\mu}}-t^{a}\right)+\sum_{pa}\exp\left(-Y^{a}_{p}\right)+\sum^{k(k-1)}_{\widetilde{\mu}=1}X_{\widetilde{\mu}}.
\end{equation}
For $\Delta_+=\{e_a-e_b:a<b\}$, we have $S^a=k+1-2a$. There are $N$ fields of charge $+1$ in each Cartan row. The auxiliary abelian period therefore satisfies
\begin{equation}\label{MPGR}
  \theta_a^N F(t)=(-1)^{N+k-1}q^a F(t).
\end{equation}
The Grassmannian-mirror period is obtained as
\begin{equation}\label{MPGR2}
  \Pi(t)=\left.\prod_{1\leq a<b\leq k}(-\theta_a+\theta_b)\,F(t^1,\ldots,t^k)\right|_{t^1=\cdots=t^k=t}.
\end{equation}

On the unshifted toric branch, the unit period of $\mathbb{CP}^{N-1}$ is $\sum_{d\geq0}(-1)^{Nd}q^d/(d!)^N$. In the large-volume regime, $\Pi^{Ab}$ is the period of the mirror of the abelian GLSM with target $(\mathbb{CP}^{N-1})^k$. The differential operator in Eq.~\eqref{MPGR2} was discussed in the appendix of \cite{Hori:2000kt}. As we will see, applying this operator and restricting to the physical parameter selects the antisymmetric part of the abelian cohomology expansion. After the appropriate normalization, the surviving terms are indexed by a basis of the cohomology of the Grassmannian.

 \subsubsection{Nonabelian Calabi-Yau manifolds} \label{NCYM}
Our main application is the computation of Gromov-Witten invariants of Calabi-Yau manifolds, but the same approach also applies to differential equations for other complete intersections. The construction parallels that in the abelian case.

The mirror is an LG model with target space $\left(\left(\mathbb{C}^{\star}\right)^{n}\times\mathbb{C}^{m}\times \mathbb{C}^{r+k}\right)/S$, where $S$ is the Weyl group of $G$, and superpotential
 \begin{equation}\label{NCYMW}
   W=\sum^{k}_{a=1}\Sigma_{a}\left(\sum^{n}_{i=1}\rho^{a}_{i}Y_{i}-\sum^{m}_{j=1}\widetilde{\rho}^{a}_{j}\log X_{j}-\sum^{r}_{\widetilde{\mu}=1}\alpha^{a}_{\widetilde{\mu}}\log X_{\widetilde{\mu}}-t^{a}\right)+\sum^{n}_{i=1}\exp\left(-Y_{i}\right)+\sum^{m}_{j=1}X_{j}+\sum^{r}_{\widetilde{\mu}=1}X_{\widetilde{\mu}}.
 \end{equation}
We retain all $k$ Cartan parameters until the final restriction to the physical FI parameters. Let $\widetilde D_j=\sum_b\widetilde\rho_j^b\theta_b$, and take $\widetilde\rho_j^a\leq0$ in the geometric basis used here. As in Appendix~\ref{PFEMS}, the R-charge 2 fields insert $\prod_j\widetilde D_j$ into the abelian local-model period. Denoting the resulting period on the shifted branch by $F_{\rm c}(t)$, we obtain
\begin{equation}\label{NMSPAB}
   \Pi(t)=\prod_{\alpha\in\Delta_+}[-\alpha(\theta)]\,F_{\rm c}(t).
\end{equation}
With $s_a$ and $D_i$ as in Eq.~\eqref{ABLP2}, the closed equations for $F_{\rm c}$ are
\begin{equation}\label{NMSPF}
\begin{aligned}
  \left[\prod_{\rho_i^a>0}\prod_{\ell=0}^{\rho_i^a-1}(D_i-\ell)\right]F_{\rm c}
  ={}&(-1)^{s_a+S^a}q^a
     \left[\prod_{\rho_i^a<0}\prod_{\ell=0}^{-\rho_i^a-1}(-D_i+\ell)\right]\\
     &\times\left[\prod_j\prod_{\ell=1}^{-\widetilde\rho_j^a}(-\widetilde D_j+\ell)\right]F_{\rm c}.
\end{aligned}
\end{equation}
The product for each R-charge 2 field starts at $\ell=1$ because $\widetilde D_j q^a=q^a(\widetilde D_j+\widetilde\rho_j^a)$. Empty products are one. We apply Eq.~\eqref{NMSPAB} before restricting the parameters to the physical K\"ahler moduli.

We now illustrate the construction with a concrete example.

\Needspace{5\baselineskip}
\noindent \textbf{Example: $Gr(2,4)[4]$}

The GLSM for the Calabi-Yau hypersurface in $Gr(2,4)$ is a $U(2)$ gauge theory with four fundamental chiral superfields and one chiral superfield in the $\det^{-4}$ representation. Its mirror is an LG model with target space $\bigl((\mathbb{C}^{\star})^{8}\times\mathbb{C}^{4}\times\mathbb{C}\bigr)/S_{2}$ and superpotential
\begin{align}\label{MSGR24-W}
  W&=\Sigma_{1}\left(\sum^{4}_{i=1}Y^{1}_{i}+4\log X_{p}+\log X_{1}-\log X_{2}-t^{1}\right)
   \\\nonumber
   &+\Sigma_{2}\left(\sum^{4}_{i=1}Y^{2}_{i}+4\log X_{p}-\log X_{1}+\log X_{2}-t^{2}\right)
   \\\nonumber
   &+\sum^{4}_{i=1}\exp\left(-Y^{1}_{i}\right)+\sum^{4}_{i=1}\exp\left(-Y^{2}_{i}\right)+X_{p}+X_{1}+X_{2}.
\end{align}

In this case $F_{\rm c}$ is proportional to $(\theta_1+\theta_2)f$, where $f$ is the abelian local-model period in Eq.~\eqref{PFAbGR244}. The compact equations are
\begin{equation}\label{MSGR24}
  \theta_1^4 F_{\rm c}=-q^1\prod_{\ell=1}^{4}\bigl(4(\theta_1+\theta_2)+\ell\bigr)F_{\rm c},
\end{equation}
\begin{equation}\label{MSGR242}
  \theta_2^4 F_{\rm c}=-q^2\prod_{\ell=1}^{4}\bigl(4(\theta_1+\theta_2)+\ell\bigr)F_{\rm c}.
\end{equation}
The mirror period of $Gr(2,4)[4]$ is obtained as
\begin{equation}\nonumber
  \Pi(t)=\left.(\partial_{t^1}-\partial_{t^2})F_{\rm c}(t^1,t^2)\right|_{t^1=t^2=t}.
\end{equation}

\subsection{Matter fields with vector R-charge 2 and mirrors}\label{VR2FM}
Section~\ref{MC} explained how the vector R-charge of a matter field determines the corresponding fundamental variable in the mirror. We now examine this relation in more detail, in preparation for the Gromov-Witten computations in Section~\ref{GWI}.

\Needspace{5\baselineskip}
\noindent{\textbf{GLSM for the quintic versus GLSM for
${\cal O}(-5)$}}.
In Appendix~\ref{PFEMS}, it is shown that the periods in the
mirrors of these two GLSMs satisfy the relation
$\widetilde{\Pi}=5\theta\Pi$, where $\widetilde{\Pi}$ and $\Pi$
are the periods of the quintic and the line bundle
$\mathcal{O}_{\mathbb{CP}^4}(-5)$, respectively. 

We now study these two Landau-Ginzburg models in more detail,
largely following the discussion in
\cite[pp.~476--477]{Hori:2003ic}. The mirror of the GLSM for the quintic is a Landau-Ginzburg
model with ambient field space
$(\mathbb{C}^{\ast})^{5}\times\mathbb{C}^{2}$ and superpotential
\begin{equation}\nonumber
W
=
\Sigma
\left(
\sum_{i=1}^{5}Y_i+5\log X_p-t
\right)
+
\sum_{i=1}^{5}e^{-Y_i}
+
X_p.
\end{equation}
The logarithmic expression is understood on the locus
$X_p\neq0$.

The mirror of the NLSM on the quintic is the Landau-Ginzburg
SCFT with target space $\mathbb{C}^5/G_{\rm LG}$, where
\[
G_{\rm LG}
=
\left\{
(\omega_1,\ldots,\omega_5):
\omega_i^5=1,\quad
\prod_{i=1}^{5}\omega_i=1
\right\}
\cong(\mathbb{Z}_5)^4
\]
acts by $X_i\mapsto\omega_iX_i$. The superpotential is
\begin{equation}\nonumber
W
=
X_1^5+\cdots+X_5^5
+
q^{-\frac15}X_1X_2X_3X_4X_5.
\end{equation}
We write
\[
\psi=q^{-1/5},
\qquad
M=X_1X_2X_3X_4X_5,
\]
and work away from $\psi^5=-5^5$.

This is the low-energy effective theory of the mirror
Landau-Ginzburg model of the GLSM for the quintic.
In a projective description, the diagonal $\mathbb{Z}_5$
is absorbed into projective scaling, leaving
$G_{\rm LG}/\mathbb{Z}_5\cong(\mathbb{Z}_5)^3$.
One can obtain the ring relations of the untwisted sector
by computing
\begin{equation}\nonumber
\partial_{X_i}W=0,
\qquad i=1,\ldots,5,
\end{equation}
which gives
\begin{equation}\nonumber
5X_i^4+\psi\prod_{j\neq i}X_j=0,
\qquad i=1,\ldots,5.
\end{equation}
A basis of the orbifold-invariant Jacobian ring in the
untwisted sector is
\[
1,\qquad M,\qquad M^2,\qquad M^3.
\]
These operators correspond to the even cohomology classes
of the quintic, which lie in $H^{p,p}$ for $p=0,1,2,3$,
as expected from mirror symmetry. The corresponding ground
states are represented by
\begin{equation}\label{GSQ}
M^p|\Omega\rangle_{\rm c},
\qquad p=0,1,2,3,
\end{equation}
where the ground state $|\Omega\rangle_{\rm c}$ is the
compact-theory vacuum, i.e.\ the state corresponding to
the unit class of the quintic.

The mirror of the GLSM for ${\cal O}(-5)$ is a
Landau-Ginzburg model with target space
$(\mathbb{C}^{\ast})^{6}\times\mathbb{C}$ and superpotential
\begin{equation}\nonumber
W
=
\Sigma
\left(
\sum_{i=1}^{5}Y_i-5Y_p-t
\right)
+
\sum_{i=1}^{5}e^{-Y_i}
+
e^{-Y_p}.
\end{equation}
Eliminating $\Sigma$ imposes
\[
\sum_{i=1}^{5}Y_i-5Y_p=t.
\]
With
\[
X_i=e^{-Y_i/5},
\]
the superpotential becomes
\[
W=\sum_{i=1}^{5}X_i^5+\psi M.
\]
In these coordinates, the mirror is an LG model on $T/G$,
where $T=(\mathbb C^*)^5$ and
\[
G=G_{\rm LG}
=
\left\{
(\omega_1,\ldots,\omega_5):
\omega_i^5=1,\quad
\prod_{i=1}^{5}\omega_i=1
\right\}
\simeq(\mathbb Z_5)^4.
\]
The group acts by $X_i\mapsto\omega_iX_i$.

For the period integrals, we use the algebraic twisted
de Rham cohomology
\[
\mathcal H_{\rm nc}
=
H^5\!\left(
\Omega^\bullet_{\rm alg}(T),
d-dW\wedge
\right)^G,
\]
with logarithmic volume form
\[
\Omega_{\rm nc}
=
\frac{dX_1}{X_1}\wedge\cdots\wedge\frac{dX_5}{X_5}.
\]
For $\psi^5\neq-5^5$, this space has the basis\footnote{%
\begin{samepage}
The cohomology can be computed using polynomial logarithmic
forms on $\mathbb C^5$ along the divisor $D=\{M=0\}$.
This logarithmic complex computes the same algebraic
twisted de Rham cohomology as the complex on $T$.
Its degree filtration has associated Koszul complex
for the sequence
\[
X_i\partial_{X_i}W=5X_i^5+\psi M,
\qquad i=1,\ldots,5.
\]
For $\psi^5\neq-5^5$, these polynomials have only the
origin as a common zero and form a regular sequence.
The $G$-invariant associated graded cohomology is therefore
represented by
\[
\left(
\frac{\mathbb C[X_1,\ldots,X_5]}
     {(5X_i^5+\psi M)_{i=1}^{5}}
\right)^G
\simeq
\frac{\mathbb C[M]}{(M^5)}.
\]
Its five monomial representatives lift to the five
twisted de Rham classes displayed above.
This associated graded description does not impose
$[M^5\Omega_{\rm nc}]=0$ in the twisted de Rham cohomology.
\end{samepage}
}
\[
[\Omega_{\rm nc}],
\quad
[M\Omega_{\rm nc}],
\quad
[M^2\Omega_{\rm nc}],
\quad
[M^3\Omega_{\rm nc}],
\quad
[M^4\Omega_{\rm nc}].
\]

The number of classes agrees with
\[
\dim H^{\rm even}\!\left(
\operatorname{Tot}\mathcal O_{\mathbb P^4}(-5),
\mathbb C
\right)=5.
\]
In the state notation used below, we write
\begin{equation}\label{GST}
M^p|\Omega\rangle_{\rm nc}
\ \longleftrightarrow\
[M^p\Omega_{\rm nc}],
\qquad p=0,\ldots,4.
\end{equation}
These symbols denote the cohomology classes entering
the period integrals.

Here $|\Omega\rangle_{\rm nc}$ denotes the formal state
associated with the unit cohomology class of the total space
of ${\cal O}(-5)\to\mathbb{CP}^4$ under the mirror correspondence.
In the large-volume sigma-model description, this state is
Poincar\'e dual to the full noncompact total space and is
non-normalizable.

One can explain the difference between these cohomological
sectors on the gauge theory side following the discussion
in \cite{Hori:2003ic}. We restrict to the ambient
even-cohomology sector generated by $\Sigma$.
Consider the GLSM for the noncompact Calabi-Yau
${\cal O}(-5)$. The cohomological states in this sector can
be represented by
\[
\mathbf{1},\Sigma,\ldots,\Sigma^4.
\]
These states are in one-to-one correspondence with the
cohomology classes of ${\cal O}(-5)$ by mapping $\Sigma$
to the pullback of the hyperplane class
$H\in H^2(\mathbb{CP}^4)$.

Our convention is that the insertion associated with
$\theta$ is represented by the formal class $\Sigma$.
In the displayed GLSM mirror integral, this insertion is
$-\Sigma_{\rm GLSM}$, since
\[
\partial_t e^{-W}
=
\Sigma_{\rm GLSM}e^{-W},
\qquad
\theta e^{-W}
=
-\Sigma_{\rm GLSM}e^{-W}.
\]

The GLSM for the quintic is a compact theory that can be
understood by turning on a perturbative superpotential in
the GLSM for the noncompact Calabi-Yau ${\cal O}(-5)$.
The states corresponding to normalizable forms in the
noncompact theory survive in the compact theory.
In the sector considered here, these are represented by
$\Sigma,\ldots,\Sigma^4$, while $\mathbf{1}$ corresponds
to the cohomology class dual to the total space of
${\cal O}(-5)$ and is non-normalizable.

The analysis of the states in the gauge theory suggests
the following correspondence, up to normalization:
\begin{equation}\label{mbgl}
|\Omega\rangle_{\rm nc}
\longleftrightarrow
\mathbf{1},
\qquad
|\Omega\rangle_{\rm c}
\longleftrightarrow
\Sigma|\Omega\rangle_{\rm nc}.
\end{equation}
These identifications refer to the large-volume
cohomological correspondence between the two theories.

The relation between the vacua can also be read from the
period integrals, as discussed in Appendix~\ref{PFEMS}.
Let
\[
\Omega_0
=
dX_1\wedge\cdots\wedge dX_5
=
M\Omega_{\rm nc}.
\]
The restriction $j:T\hookrightarrow\mathbb C^5$ maps
the compact period classes to
\[
j^*[M^p\Omega_0]
=
[M^{p+1}\Omega_{\rm nc}],
\qquad p=0,\ldots,3.
\]
Thus the four compact period classes map to the subspace
spanned by
\[
[M\Omega_{\rm nc}],
\quad
[M^2\Omega_{\rm nc}],
\quad
[M^3\Omega_{\rm nc}],
\quad
[M^4\Omega_{\rm nc}].
\]

The normalization $\widetilde\Pi=5\theta\Pi$ corresponds
to the compact period form
\[
\widetilde\Omega_{\rm c}
=
\psi\,\Omega_0.
\]
Indeed, for compatible integration cycles,
\begin{align*}
5\theta\Pi
&=
-\psi\frac{\partial}{\partial\psi}
\int_{\gamma_{\rm nc}}e^{-W}\Omega_{\rm nc}
\\
&=
\psi\int_{\gamma_{\rm nc}}
e^{-W}M\Omega_{\rm nc}
\\
&=
\psi\int_{\gamma_{\rm c}}e^{-W}\Omega_0
=
\widetilde\Pi.
\end{align*}
In the large-volume regime, the A-brane constructions
in the two models give the period relation above.
With compatible normalizations, the corresponding
large-volume state identification is
\begin{equation}\label{mbgl2}
|\Omega\rangle_{\rm c}
\longleftrightarrow
5\Sigma|\Omega\rangle_{\rm nc}.
\end{equation}
Here the right-hand side represents the
$-5\Sigma_{\rm GLSM}$ insertion in the GLSM mirror integral.

In Section~\ref{GWFM}, we expanded the generating function
in a basis of even cohomology beginning with the unit class.
The cohomology-valued generating functions below use the
coefficient rings
\[
R_{\rm nc}
=
\mathbb C[\Sigma]/(\Sigma^5),
\qquad
R_{\rm c}
=
\mathbb C[\Sigma]/(\Sigma^4).
\]
These relations refer to the A-model cohomology coefficients,
rather than to multiplication of representatives in
$\mathcal H_{\rm nc}$.

Write
\[
\Pi_{{\cal O}(-5)}(\Sigma;t)
=
\sum_{j=0}^{4}\Sigma^j f_j(t),
\]
with the unit component normalized so that
\[
\theta f_0(t)=0.
\]
Then $\theta\Pi_{{\cal O}(-5)}$ lies in the ideal
$(\Sigma)\subset R_{\rm nc}$.
In this normalization, the generating function used to
compute the quintic Gromov-Witten invariants is
\begin{equation}\label{PGNQ}
\Pi_{\rm quintic}(\Sigma;t)
=
\frac{
\theta\Pi_{{\cal O}(-5)}(\Sigma;t)
}{\Sigma}.
\end{equation}
Division by $\Sigma$ here denotes the inverse of the
linear map
\[
R_{\rm c}
\longrightarrow
(\Sigma)\subset R_{\rm nc},
\qquad
f\longmapsto\Sigma f.
\]
Explicitly,
\[
\frac{\theta\Pi_{{\cal O}(-5)}(\Sigma;t)}{\Sigma}
=
\theta f_1(t)
+
\Sigma\theta f_2(t)
+
\Sigma^2\theta f_3(t)
+
\Sigma^3\theta f_4(t)
\qquad \bmod\ \Sigma^4.
\]

Now, as discussed in Eq.~\eqref{IFE}, we can expand
$\Pi_{\rm quintic}(\Sigma;t)$ as
\begin{equation}\nonumber
\begin{aligned}
\Pi_{\rm quintic}(\Sigma;t)
={}&
\widetilde\Pi_0(q)
+
\Sigma\widetilde\Pi_1(q)
+
\Sigma^2\widetilde\Pi_2(q)
+
\Sigma^3\widetilde\Pi_3(q),
\\[-2pt]
&\hspace{7cm}\bmod\ \Sigma^4.
\end{aligned}
\end{equation}
Equivalently, one may use
\[
5\Sigma\Pi_{\rm quintic}
=
5\theta\Pi_{{\cal O}(-5)},
\]
provided the expansion basis is
\[
5\Sigma,\quad
5\Sigma^2,\quad
5\Sigma^3,\quad
5\Sigma^4,
\]
and the relation $\Sigma^5=0$ is imposed.
The same normalization procedure extends to other
complete intersections.

\noindent{\textbf{Roots in mirrors}}. The fields associated with roots in the Cartan theory have vector R-charge 2. Their fundamental mirror variables are $X_{\widetilde{\mu}}=\exp(-Z_{\widetilde{\mu}})$, analogous to the fundamental variable $X_{p}$ for the $p$-field in the mirror of the GLSM for the quintic. However, in the case of the quintic, the $p$-field has gauge charge $-5$ and vector R-charge 2, and the GLSM superpotential is $W=pG_5$, where $G_5$ is a homogeneous polynomial of degree five. In the geometric phase, the equation $G_5=0$ defines the quintic in $\mathbb{CP}^{4}$. Mathematically, quantum Serre duality relates the twisted theories associated with ${\cal O}(-5)$ and ${\cal O}(5)$, while quantum Lefschetz relates the latter theory to the genus-zero Gromov-Witten theory of the quintic \cite{Coates:2001}. The $p$-field is a fiber coordinate on ${\cal O}(-5)$; restricted to the quintic, this is the dual of its normal bundle. One may expect that the R-charge 2 chiral superfields from the roots in the nonabelian GLSM have an analogous interpretation in terms of normal bundles over the nonabelian target in the associated abelian ambient space. We leave a detailed discussion of this point of view to future work.

For the cohomology-valued generating function, $\theta_a e^{-\sum_b t^b\Sigma_b}=\Sigma_a e^{-\sum_b t^b\Sigma_b}$. With the scalar period normalized as in Eq.~\eqref{PMNG5}, we choose the cohomology-valued normalization
\begin{equation}\label{NBCR}
  \widetilde\Pi(\Sigma;t)
  =\frac{(-1)^{|\Delta_+|}\Pi(\Sigma;t)}{\prod_{\alpha\in\Delta_+}\alpha(\Sigma)}
  =\frac{\prod_{\alpha\in\Delta_+}\alpha(\theta)}{\prod_{\alpha\in\Delta_+}\alpha(\Sigma)}
    \,\Pi^{\rm Ab}(\Sigma;T).
\end{equation}
For the Grassmannian generating functions below, we use the common logarithmic lift $T^a_{\rm sym}=T^a+2\pi i(k-a)=t^a+i(k-1)\pi$. It gives the same $e^{-T^a}$ for each lift. For logarithmic periods we use the branch obtained by continuing from $T^a$ to $T^a_{\rm sym}$ along these imaginary shifts at large volume, transporting the cycles along the path. We then take the Weyl-invariant part and restrict to the physical parameters. The Pfaffian phase uses the continuation specified in Section~\ref{PFCYGW}.

\subsubsection{Fano varieties}

\Needspace{5\baselineskip}
\noindent $\mathbf{Gr(2,4).}$
We begin with $Gr(2,4)$ and then consider general Grassmannians. Following Eq.~\eqref{AblP}, the generating function is
\begin{equation}\label{PG24}
  \Pi_{Gr(2,4)}\left(\Sigma_{1},\Sigma_{2}; t\right)=\left.\frac{-\partial_{t^{1}}+\partial_{t^{2}}}{\Sigma_{1}-\Sigma_{2}}\right |_{t^{1}=t^{2}=t}\Pi_{\mathbb{CP}^{3}\times\mathbb{CP}^{3}}\left(\Sigma_{1}, \Sigma_{2}; t^{1}+i\pi, t^{2}+i\pi\right)
\end{equation}
for the mirror of $Gr(2,4)$. We use $\Pi_{\mathbb{CP}^{3}\times\mathbb{CP}^{3}}$ to denote the generating function of the mirror of $\mathbb{CP}^{3}\times\mathbb{CP}^{3}$ and $\Pi_{Gr(2,4)}$ to denote the generating function of the mirror of $Gr(2,4)$.

The mirror of the GLSM for $\mathbb{CP}^{3}\times\mathbb{CP}^{3}$ is the Landau-Ginzburg model $\left(\mathbb{C}^{\star}\right)^{8}\times\mathbb{C}^{2}$ with the superpotential
\begin{equation}\nonumber
  W=\Sigma_{1}\left(\sum^{4}_{i=1}Y^{1}_{i}-t^{1}-i\pi\right)+\Sigma_{2}\left(\sum^{4}_{i=1}Y^{2}_{i}-t^{2}-i\pi\right)+\sum^{4}_{i=1}\exp\left(-Y^{1}_{i}\right)+\sum^{4}_{i=1}\exp\left(-Y^{2}_{i}\right).
\end{equation}
The chiral ring relations can be obtained from the vacuum equations
\begin{equation}\nonumber
  \partial_{\Sigma_{a}}W=\partial_{Y^{a}_{i}}W=0,\qquad a=1,2 \quad {\rm and} \quad i=1,\cdots,4,
\end{equation}
 giving rise to
 \begin{equation}\nonumber
   \Sigma_{1}=X^{1}_{i}=\exp\left(-Y^{1}_{i}\right),\quad \Sigma^{4}_{1}=-q^{1};\qquad \Sigma_{2}=X^{2}_{i}=\exp\left(-Y^{2}_{i}\right),\quad  \Sigma^{4}_{2}=-q^{2}.
 \end{equation}
A basis of the chiral ring consists of 16 operators:
\begin{equation}\nonumber
  \Sigma^{i_{1}}_{1}\Sigma^{i_{2}}_{2},\quad i_{1},i_{2}=0,\cdots,3.
\end{equation}
Following Witten~\cite{Witten:1993xi}, we identify $\Sigma_a$ with the hyperplane class $H_a$ of the corresponding projective-space factor. The basis monomials of the mirror Jacobian ring thus correspond to the cohomology classes $H_1^{i_1}H_2^{i_2}$, with $i_1,i_2=0,\ldots,3$, of $\mathbb{CP}^{3}\times\mathbb{CP}^{3}$. We expand the generating function in this basis:
\begin{equation}\label{ABEP}
\Pi_{\mathbb{CP}^{3}\times\mathbb{CP}^{3}}\left(\Sigma_{1}, \Sigma_{2}; t^{1}+i\pi, t^{2}+i\pi\right)=\Sigma^{i_{1}}_{1}\Sigma^{i_{2}}_{2}\Pi_{\mathbb{CP}^{3}\times\mathbb{CP}^{3},i_{1}i_{2}}\left(t^{1}+i\pi, t^{2}+i\pi\right).
\end{equation}
After restriction to $t^1=t^2=t$, the operator $\partial_{t^1}-\partial_{t^2}$ annihilates functions symmetric in $t^1,t^2$. The symmetric combinations of the 16 abelian periods span a ten-dimensional subspace. They occur as the coefficients in Eq.~\eqref{ABEP} of symmetric polynomials in
\begin{equation}\nonumber
  \mathbb{C}\left[\Sigma_{1},\Sigma_{2}\right]/\langle\Sigma^{4}_{1}, \Sigma^{4}_{2}\rangle.
\end{equation}
The six independent antisymmetric combinations multiply antisymmetric polynomials of the form
\begin{equation}\nonumber
  \left(\Sigma_{1}-\Sigma_{2}\right)S_{\lambda}\left(\Sigma_{1},\Sigma_{2}\right).
\end{equation}
Here $\lambda$ runs over the six Young diagrams contained in a $2\times2$ rectangle, and $S_\lambda$ denotes the corresponding Schur polynomial. These Schur polynomials are
\begin{equation}\nonumber
  1,~ \Sigma_{1}+\Sigma_{2},~ \Sigma^{2}_{1}+\Sigma_{1}\Sigma_{2}+\Sigma^{2}_{2},~ \Sigma_{1}\Sigma_{2},~ \Sigma_{1}\Sigma_{2}\left(\Sigma_{1}+\Sigma_{2}\right),~ \Sigma^{2}_{1}\Sigma^{2}_{2}.
\end{equation}
Now it is easy to see from (\ref{PG24}) that
\begin{equation}\label{PGR24}
  \Pi_{Gr(2,4)}\left(\Sigma_{1}, \Sigma_{2}; t\right)=\sum_{\lambda}S_{\lambda}\left(\Sigma_{1},\Sigma_{2}\right)\Pi_{\lambda}\left(t\right).
\end{equation}
The coefficient functions $\Pi_\lambda(t)$ span the image of the operator $(\partial_{t^1}-\partial_{t^2})|_{t^1=t^2=t}$. Up to the common phase shift and normalization used above, they have the Wronskian form
\begin{equation}\nonumber
  \partial_{t}\Pi_{\mathbb{CP}^{3}, i}(t)\Pi_{\mathbb{CP}^{3}, j}(t)-\Pi_{\mathbb{CP}^{3}, i}(t)\partial_{t}\Pi_{\mathbb{CP}^{3}, j}(t),\quad {\rm for}\quad i<j, \quad i,j=0,\cdots 3,
\end{equation}
where $\Pi_{\mathbb{CP}^{3},i}(t)$ denotes the coefficient of $\Sigma^i$ in the expansion
\begin{equation}\nonumber
  \Pi_{\mathbb{CP}^{3}}(\Sigma; t)=\Pi_{\mathbb{CP}^{3}, 0}(t)+\Sigma\Pi_{\mathbb{CP}^{3}, 1}(t)+\Sigma^{2}\Pi_{\mathbb{CP}^{3}, 2}(t)+\Sigma^{3}\Pi_{\mathbb{CP}^{3}, 3}(t), \quad \mathbf{{\rm mod}}\quad \Sigma^{4}.
\end{equation}

In Section~\ref{PFEIPP}, we show that Eq.~\eqref{PGR24} satisfies the known one-parameter quantum differential equation for the mirror of $Gr(2,4)$.

\Needspace{5\baselineskip}
\noindent $\mathbf{Gr(k,N).}$
The construction for a general Grassmannian parallels that for $Gr(2,4)$. Equation~\eqref{NBCR} gives
\begingroup\small 
\begin{align}\label{PGkN}
  \Pi_{Gr(k,N)}\left(\Sigma_{1},\cdots, \Sigma_{k};t\right)=&\prod_{1\leq a<b\leq k}\left.\frac{-\partial_{t^{a}}+\partial_{t^{b}}}{\Sigma_{a}-\Sigma_{b}}\right |_{t^{a}=t}\cdot\\\nonumber
  &\Pi_{(\mathbb{CP}^{N-1})^{k}}\left(\Sigma_{1},\cdots ,\Sigma_{k}; t^{1}+i(k-1)\pi,\cdots, t^{k}+i(k-1)\pi\right).
\end{align}
\endgroup
For the cohomology-valued expansion, we use the classical cohomology ring of $(\mathbb{CP}^{N-1})^k$. It has dimension $N^k$ and the presentation
\begin{equation}\nonumber
  \mathbb{C}\left[\Sigma_{1},\cdots , \Sigma_{k}\right]/\langle \Sigma^{N}_{1},\cdots, \Sigma^{N}_{k}\rangle.
\end{equation}
The alternating differential operator $\prod_{a<b}(\partial_{t^a}-\partial_{t^b})$, followed by restriction to $t^1=\cdots=t^k=t$, selects the alternating part of the abelian period expansion. The corresponding polynomials in the $\Sigma_a$ are alternating as well. Write $\varsigma=(\varsigma_1,\ldots,\varsigma_k)$ with $N>\varsigma_1>\cdots>\varsigma_k\geq0$. A basis of alternating polynomials is given by the determinants
\begin{equation}\label{RDAB}
  J_{\varsigma}\left(\Sigma_{1},\cdots, \Sigma_{k}\right)=\det\left(\Sigma_{a}^{\varsigma_{b}}\right)_{a,b}, \qquad 1\leq a,b\leq k.
\end{equation}
The special case with $\delta=\left(k-1, k-2,\cdots,0\right)$ is the Vandermonde determinant
\begin{equation}\label{VD}
  J_{\delta}\left(\Sigma_{1},\cdots, \Sigma_{k}\right)=\det\left(\Sigma_{a}^{\delta_{b}}\right)_{a,b}=\prod_{1\leq a<b\leq k}\left(\Sigma_{a}-\Sigma_{b}\right).
\end{equation}
Writing $\varsigma=\delta+\lambda$ identifies $\lambda$ with a partition contained in the $k\times(N-k)$ rectangle. The alternant formula for Schur polynomials gives
\begin{equation}\label{ScGr}
  J_{\varsigma}\left(\Sigma_{1},\cdots, \Sigma_{k}\right)=J_{\delta}\left(\Sigma_{1},\cdots, \Sigma_{k}\right)S_{\lambda}\left(\Sigma_{1},\cdots, \Sigma_{k}\right).
\end{equation}

The generating function for the mirror of $Gr(k,N)$
therefore has the Schur expansion
\begin{equation}\label{PGrE}
  \Pi_{Gr(k,N)}(\Sigma_1,\cdots,\Sigma_k;t)
  =
  \sum_{\lambda}
  S_{\lambda}(\Sigma_1,\cdots,\Sigma_k)
  \Pi_{\lambda}(t),
\end{equation}
where $\lambda$ runs over partitions contained in the
$k\times(N-k)$ rectangle, and
\begin{equation}\nonumber
  \Pi_{\lambda}(t)
  =
  \det\left[
  \partial_t^{a-1}
  \Pi_{\lambda_b+k-b}(t)
  \right]_{a,b=1}^{k}.
\end{equation}

This agrees with the description of Grassmannian cohomology
in terms of Schur polynomials in the Chern roots $\Sigma_a$
of the dual tautological bundle
~\cite{Witten:1993xi,Griffiths:1978,Martin:2000}.
For more general targets, including flag varieties
~\cite{Gu:2020}, cohomology has a basis of Schubert classes.
Applying the same approach to the corresponding generating
functions should permit comparisons with recent calculations
based on other methods
~\cite{Closset:2025akk,Closset:2026bnk}.

\noindent{\textbf{Symplectic Grassmannian}}. The GLSM for $SG(k,2n)$ studied in \cite{Gu:2020oeb,Gu:2020zpg} is a $U(k)$ gauge theory with $2n$ chirals $\Phi^{a}_{\pm i}$ in the fundamental representation $V$ $\left(a\in\{1,\cdots,k\}, i\in\{1,\cdots,n\}\right)$, and one chiral superfield $q_{ab}$ in the representation $\wedge^{2}V^{\ast}$, with superpotential
\begin{equation}\nonumber
  W=\sum_{\alpha,\beta}q_{ab}\Phi^{a}_{\alpha}\Phi^{b}_{\beta}\omega^{\alpha\beta}=\sum^{n}_{i=1}q_{ab}\Phi^{a}_{i}\Phi^{b}_{-i}.
\end{equation}
The superpotential breaks the global $U(2n)$ symmetry to $USp(2n)$, whose complexification is $Sp(2n,\mathbb C)$. Further details of this GLSM and its proposed nonabelian mirror can be found in \cite{Gu:2020oeb}, following the construction of \cite{Gu:2018fpm}. The mirror contains chiral superfields $\Sigma_a$, $Y_{ia}$, $U_{bc}$, and $X_{\mu\nu}$, with $i\in\{\pm1,\ldots,\pm n\}$ and $a,b,c,\mu,\nu\in\{1,\ldots,k\}$. Its superpotential is
\begin{align}\nonumber
  W&=\sum_{a}\Sigma_{a}\left(\sum_{i}Y_{ia}-\sum_{b>c}\rho^{a}_{bc}\ln U_{bc}-\sum_{\mu\neq\nu}\alpha^{a}_{\mu\nu}\ln X_{\mu\nu}-t_{\rm GLSM}^{a}\right)\\\nonumber
  &+\sum_{i,a}\exp\left(-Y_{ia}\right)+\sum_{b>c} U_{bc}+\sum_{\mu\neq\nu}X_{\mu\nu},
\end{align}
where $\rho^{a}_{bc}=-\delta^{a}_{b}-\delta^{a}_{c}$. We consider the general symplectic Grassmannians $SG(k,2n)$ with $k\leq n$, which contain Lagrangian Grassmannians as a special case when $k = n$.

In the generating function below we use $t^a=t_{\rm GLSM}^a$, with $q^a=e^{-t^a}$. The $2n$ fundamental fields give an even matter sign in each Cartan equation. The sign $(-1)^{(k-1)\sum_a d_a}$ comes from the root shift, and we absorb the degree-independent factor $e^{-i(k-1)\pi\sum_a\Sigma_a}$ into the normalization. Following Eq.~\eqref{NBCR}, we write the proposed generating function as
\begin{align}\label{PSG}
  \Pi_{SG(k,2n)}&\left(\Sigma_{a};t\right)=\prod_{1\leq a<b\leq k}\left.\frac{-\partial_{t^{a}}+\partial_{t^{b}}}{\Sigma_{a}-\Sigma_{b}}\right|_{t^{a}=t}\cdot \Pi^{Ab}\left(\Sigma_{a};t^{a}\right),
  \\
 \Pi^{Ab}\left(\Sigma_{a};t^{a}\right)= &\sum_{d_{1},\ldots,d_{k}=0}^{\infty}\frac{\prod_{1\leq b< c\leq k}\prod^{d_{b}+d_{c}}_{j=1}\left(\Sigma_{b}+\Sigma_{c}+j\right)}{\prod^{k}_{a=1}\prod^{d_{a}}_{l=1}\left(\Sigma_{a}+l\right)^{2n}}\cdot\\ \nonumber
&\left(-1\right)^{(k-1)\left(d_{1}+\cdots+d_{k}\right)}\exp\left(-t^{1}\left(d_{1}+\Sigma_{1}\right)-\cdots-t^{k}\left(d_{k}+\Sigma_{k}\right)\right).
\end{align}
Here $\Pi^{Ab}$ is a cohomology-valued series with $\Sigma_a^{2n}=0$ for each $a$. The special case of the Lagrangian Grassmannian $LG(n,2n)$ has been studied in the mathematical literature~\cite{Bertram:2004}.

\subsubsection{Calabi-Yau manifolds $Gr(k, N)\left[Q_{1},\cdots,Q_{m}\right]$}

Most of the examples below were computed by a different method in \cite{Batyrev:1998kx}. Revisiting them serves two purposes. First, the sign conventions for $q$ in the Yukawa couplings sometimes differ between the physics and mathematics literature~\cite{Closset:2015rna,Batyrev:1998kx}; we check that the extracted instanton numbers agree after accounting for these conventions. Second, our period formulas have combinatorial expressions different from those in \cite{Batyrev:1998kx}. The agreement of the resulting periods provides a further consistency check.

The Calabi-Yau condition is $\sum_{\beta=1}^{m}Q_\beta=N$. Following Section~\ref{NCYM}, our proposed mirror of the GLSM for $Gr(k,N)[Q_1,\ldots,Q_m]$ is an LG model with target space $\bigl[(\mathbb C^\ast)^{kN}\times\mathbb C^{k^2-k}\times\mathbb C^k\times\mathbb C^m\bigr]/S_k$, where $S_k$ acts diagonally on the gauge indices. Its superpotential is
\begin{align}\nonumber
  W=&\sum^{k}_{a=1}\Sigma_{a}\left(\sum^{N}_{i=1}Y^{a}_{i}+\sum^{m}_{\beta=1}Q_{\beta}\log X_{\beta}+\sum_{\widetilde{\mu}\neq\widetilde{\nu}}\alpha^{a}_{\widetilde{\mu}\widetilde{\nu}}\log X_{\widetilde{\mu}\widetilde{\nu}}-t^{a}\right)
  \\ \nonumber
  &+\sum^{N}_{i=1}\sum^{k}_{a=1}\exp\left(-Y^{a}_{i}\right)+\sum_{\widetilde{\mu}\neq\widetilde{\nu}} X_{\widetilde{\mu}\widetilde{\nu}}+\sum^{m}_{\beta=1}X_{\beta},
\end{align}
where $\widetilde{\mu},\widetilde{\nu}=1,\ldots,k$ and $\alpha^{a}_{\widetilde{\mu}\widetilde{\nu}}=-\delta^{a}_{\widetilde{\mu}}+\delta^{a}_{\widetilde{\nu}}$. Here the ordered-pair coefficient of $+\log X_{\widetilde{\mu}\widetilde{\nu}}$ is the negative of the root multiplying $-\log X_\alpha$ in Eq.~\eqref{MLGS-GIT}. Following Eq. (\ref{AblP}), we then have
\begin{align}\label{SIGRP}
 \Pi_{Gr(k, N)\left[Q_{1},\cdots,Q_{m}\right]} &\left(\Sigma_{1},\cdots ,\Sigma_{k}; t\right)=\prod_{1\leq a<b\leq k}\left.\frac{-\partial_{t^{a}}+\partial_{t^{b}}}{\Sigma_{a}-\Sigma_{b}}\right |_{t^{a}=t}\cdot\\\nonumber
  &\Pi_{{\cal M}}\left(\Sigma_{1},\cdots ,\Sigma_{k}, t^{1}+i(k-1)\pi,\cdots, t^{k}+i(k-1)\pi\right).
\end{align}
Here ${\cal M}$ is a formal label; we do not assign it a geometric interpretation. The auxiliary function $\Pi_{\cal M}$ is related to the period of the mirror of the abelian GIT quotient
\begin{equation}\nonumber
  {\cal T}=\operatorname{Tot}\!\left(\bigoplus_{\beta=1}^{m}{\cal O}(-Q_{\beta},\ldots,-Q_{\beta})\right)\quad \text{over}\quad (\mathbb{CP}^{N-1})^k.
\end{equation}
 More specifically, following the previous discussion on the effect of R-charge 2 matter fields in the mirror, we expect
 \begin{align}\nonumber
   &\Pi_{{\cal M}}\left(\Sigma_{1},\cdots ,\Sigma_{k}; t^{1}+i(k-1)\pi,\cdots, t^{k}+i(k-1)\pi\right)
   \\\nonumber
   &=\frac{\prod^{m}_{\beta=1}\left(Q_{\beta}\sum^{k}_{a=1}\theta_{a}\right)}{\prod_{\beta=1}^{m}\bigl(Q_{\beta}\sum_{a=1}^{k}\Sigma_{a}\bigr)}\Pi_{{\cal T}}\left(\Sigma_{1},\cdots ,\Sigma_{k}, t^{1}+i(k-1)\pi,\cdots, t^{k}+i(k-1)\pi\right).
 \end{align}
A direct physical derivation of the generating function $\Pi_{\cal T}$ would require precise control of the brane factors for A-cycles in LG mirrors. A general definition of these factors remains unavailable.\footnote{We thank Mauricio Romo for pointing this out to us.} We therefore motivate the generating function using the Gamma-function structure of mirror period integrals, and the relevant hemisphere partition functions in B-twisted LG models~\cite{Hori:2019}. On the gauge-theory side, B-brane factors have been studied in detail in \cite{Hori:2013ika}; see also \cite{Herbst:2008jq,Kapustin:2002bi,Walcher:2004tx}.\footnote{The authors of \cite{Knapp:2020oba,Erkinger:2020cjt} used GLSM hemisphere partition functions to obtain generating functions in LG phases and extract FJRW invariants~\cite{Fan:2007}.} Thus, $\Pi_{\cal T}$ can also be obtained from the gauge theory; see \cite[Section~8.2.2]{Hori:2013ika}. These generating functions are also well established in the mathematical literature~\cite{Coates:2001,Iritani:2009}. In the conventions used here, we obtain
\begingroup\small 
\begin{align}\label{MP}
  &\Pi_{{\cal T}}\left(\Sigma_{1},\cdots ,\Sigma_{k}; t^{1}+i(k-1)\pi,\cdots, t^{k}+i(k-1)\pi\right)
  \\\nonumber
  &\equiv\sum^{\infty}_{d_{1},\cdots,d_{k}= 0}\frac{\prod^{\sum^{k}_{a=1}Q_{1}d_{a}-1}_{i_{1}=0}\left(-Q_{1}\sum^{k}_{a=1}\Sigma_{a}-i_{1}\right)\times\cdots\times\prod^{\sum^{k}_{a=1}Q_{m}d_{a}-1}_{i_{m}=0}\left(-Q_{m}\sum^{k}_{a=1}\Sigma_{a}-i_{m}\right)}{\prod^{d_{1}}_{l_{1}=1}\left(\Sigma_{1}+l_{1}\right)^{N}\times\cdots\times\prod^{d_{k}}_{l_{k}=1}\left(\Sigma_{k}+l_{k}\right)^{N}}
  \\\nonumber
  &\cdot\exp\left(-\sum^{k}_{a=1}\left(\Sigma_{a}+d_{a}\right)\left(t^{a}+i(k-1)\pi\right)\right) \qquad \left({\rm mod}\quad \Sigma_{a}^{N},\quad a\in\left\{1,\cdots,k\right\}\right).
\end{align}
\endgroup
Empty products in Eq.~\eqref{MP} are understood to be one. In degree $d=\sum_a d_a$, the numerator contributes $(-1)^{d\sum_\beta Q_\beta}=(-1)^{Nd}$ and the root shift contributes $(-1)^{(k-1)d}$.
For a Calabi-Yau threefold, so that $k(N-k)-m=3$, we can now expand the generating function of the mirror in the part of even cohomology generated by the hyperplane class as
\begin{align}\nonumber
  \Pi_{Gr(k, N)\left[Q_{1},\cdots,Q_{m}\right]}& \left(\Sigma_{1},\cdots ,\Sigma_{k}; t\right)=\Pi_{Gr(k, N)\left[Q_{1},\cdots,Q_{m}\right],0} \left(t\right)+\left(\sum^{k}_{a=1}\Sigma_{a}\right)\cdot\Pi_{Gr(k, N)\left[Q_{1},\cdots,Q_{m}\right],1} \left(t\right)
  \\\nonumber
  &+\sum^{3}_{i=2}\left(\sum^{k}_{a=1}\Sigma_{a}\right)^{i}\cdot\Pi_{Gr(k, N)\left[Q_{1},\cdots,Q_{m}\right],i} \left(t\right), \qquad {\rm mod} \quad \left(\sum^{k}_{a=1}\Sigma_{a}\right)^{4}=0.
\end{align}
The components $\Pi_{Gr(k,N)[Q_1,\ldots,Q_m],\lambda}(t)$ can be computed from Eq.~\eqref{MP}. Dividing by the fundamental period gives the normalized generating function, which is identified with the ${\cal J}$-function after expressing it in the flat coordinate:
\begin{equation}\label{JF}
  {\cal J}_{Gr(k, N)\left[Q_{1},\cdots,Q_{m}\right]} \left(\Sigma_{1},\cdots ,\Sigma_{k}; t\right)\equiv\frac{\Pi_{Gr(k, N)\left[Q_{1},\cdots,Q_{m}\right]} \left(\Sigma_{1},\cdots ,\Sigma_{k}; t\right)}{\Pi_{Gr(k, N)\left[Q_{1},\cdots,Q_{m}\right],0}}.
\end{equation}
Then the flat coordinate on the K\"{a}hler moduli space of the Calabi-Yau threefold $Gr(k, N)\left[Q_{1},\cdots,Q_{m}\right]$ can be defined as
\begin{equation}\label{FCK}
  \tau\equiv-\frac{\Pi_{Gr(k, N)\left[Q_{1},\cdots,Q_{m}\right],1} \left(t\right)}{\Pi_{Gr(k, N)\left[Q_{1},\cdots,Q_{m}\right],0} \left(t\right)}-2\pi i t_{0}.
\end{equation}
For the examples considered here, we fix the remaining phase $t_0$ by requiring the first instanton number to be positive. A similar procedure applies to non-complete-intersection Calabi-Yau manifolds. In Section~\ref{GWI}, we use this framework to compute instanton numbers for several Calabi-Yau targets of nonabelian GLSMs.

\subsection{Quantum differential equations in physical parameters}\label{PFEIPP}
We now restrict the differential equations formulated in Cartan parameters to the physical K\"ahler parameters. Recovering known equations~\cite{Batyrev:1998kx} provides a check of the proposal, after which we apply the same construction to further examples.

\subsubsection{Grassmannians}\label{PFMGR}
We begin with Grassmannians and describe an algorithm for deriving their quantum differential equations. The procedure differs from that of \cite{Batyrev:1998kx}. We first treat $Gr(2,N)$ and then extend the construction to $Gr(k,N)$. According to our proposal, the mirror periods of $Gr(2,N)$ take the form
\begin{equation}\label{GSPG2N}
  \Pi\left(t\right)=\left.\left(\partial_{t^{1}}-\partial_{t^{2}}\right)\right|_{t^{1}=t^{2}=t} \Pi\left(t^{1}+i\pi,t^{2}+i\pi\right),
\end{equation}
where $\Pi\left(t^{1},t^{2}\right)$ satisfies the quantum differential equation of the mirror of $\mathbb{CP}^{N-1}\times\mathbb{CP}^{N-1}$:
\begin{align}\label{PFABG2N}
 \left(-\partial_{t^{1}}\right)^{N} \Pi\left(t^{1}+i\pi,t^{2}+i\pi\right) & =(-1)^{N+1}q^{1}  \Pi\left(t^{1}+i\pi,t^{2}+i\pi\right),\\\nonumber
    \left(-\partial_{t^{2}}\right)^{N} \Pi\left(t^{1}+i\pi,t^{2}+i\pi\right) & =(-1)^{N+1}q^{2}  \Pi\left(t^{1}+i\pi,t^{2}+i\pi\right).
\end{align}
Let us denote by $\Pi_{0}$, $\Pi_{1}, \cdots,\Pi_{N-1}$ a linearly independent set of solutions to the quantum differential equation of the mirror of $\mathbb{CP}^{N-1}$ on this shifted branch, so that
\begin{equation}\label{SPFP}
  \left(-\partial_{t^{a}}\right)^{N} \Pi_{i}  =(-1)^{N+1}q^{a}\Pi_{i}
\end{equation}
for $i=0,\ldots,N-1$. The derivation does not require explicit formulas for these solutions. For concreteness, they may be chosen as the coefficients of $\Sigma^i$ in the mirror generating function, equivalently the appropriately shifted $I$-function of $\mathbb{CP}^{N-1}$. Thus,
\begin{equation}\label{PPS}
  \Pi_{\mathbb{CP}^{N-1}}\left(\Sigma;t+i\pi\right)=\Pi_{0}\left(t\right)\cdot 1+\Pi_{1}\left(t\right)\cdot \Sigma+\cdots+\Pi_{N-1}\left(t\right)\cdot \Sigma^{N-1}, \qquad\rm{mod} \quad\Sigma^{N}=0.
\end{equation}

The general solution to Eq. (\ref{PFABG2N}) is of the form
\begin{equation}\label{SSG2N}
  \Pi\left(t^{1}+i\pi,t^{2}+i\pi\right)=\sum^{N-1}_{i,j=0}C_{ij}\Pi_{i}\left(t^{1}\right)\Pi_{j}\left(t^{2}\right),
\end{equation}
where the $C_{ij}$ are constants. To see this, use the first equation in Eq.~\eqref{PFABG2N} to write
\begin{equation}\nonumber
  \Pi\left(t^{1}+i\pi,t^{2}+i\pi\right)=\sum^{N-1}_{i=0}\Pi_{i}\left(t^{1}\right)g_{i}\left(t^{2}\right)
\end{equation}
for some functions $g_i(t^2)$. The second equation in Eq.~\eqref{PFABG2N} then implies
\begin{equation}\nonumber
  \sum^{N-1}_{i=0}\Pi_{i}\left(t^{1}\right)\left(\left(-\partial_{t^{2}}\right)^{N}+(-1)^N q^{2}\right)g_{i}\left(t^{2}\right)=0.
\end{equation}
Because $\Pi_{i}$ are linearly independent, the equation above implies
\begin{equation}\nonumber
  \left(\left(-\partial_{t^{2}}\right)^{N}+(-1)^N q^{2}\right)g_{i}\left(t^{2}\right)=0
\end{equation}
for all $i$ and $t^{2}$.  Then there exist constants $C_{ij}$ such that
\begin{equation}\nonumber
  g_{i}\left(t^{2}\right)=\sum^{N-1}_{j=0}C_{ij}\Pi_{j}\left(t^{2}\right).
\end{equation}

The image of the construction in Eq.~\eqref{GSPG2N} consists of functions of the form
\begin{equation}\label{GSPG2N2}
  \Pi\left(t\right)=\sum^{N-1}_{i,j=0}A_{ij}\left(\frac{d\Pi_{i}\left(t\right)}{dt}\Pi_{j}\left(t\right)-\Pi_{i}\left(t\right)\frac{d\Pi_{j}\left(t\right)}{dt}\right),
\end{equation}
where the constant matrix $A_{ij}$ satisfies
\begin{equation}\nonumber
  A_{ij}=-A_{ji}.
\end{equation}
Now we outline the proof of (\ref{GSPG2N2}). From \eqref{SSG2N}, we have
\begin{equation}\nonumber
  \Pi\left(t^{1}+i\pi,t^{2}+i\pi\right)=\sum^{N-1}_{i,j=0}C_{ij}\Pi_{i}\left(t^{1}\right)\Pi_{j}\left(t^{2}\right)
\end{equation}
for some $N\times N$ constant matrix $C_{ij}$. If we decompose $C_{ij}$ into symmetric and antisymmetric parts, then
\begin{equation}\nonumber
  \Pi\left(t^{1}+i\pi,t^{2}+i\pi\right)=\Pi_{S}\left(t^{1},t^{2}\right)+\Pi_{A}\left(t^{1},t^{2}\right),
\end{equation}
where
\begin{align}\nonumber
  \Pi_{S}\left(t^{1},t^{2}\right) &=\sum^{N-1}_{i,j=0}S_{ij}\Pi_{i}\left(t^{1}\right)\Pi_{j}\left(t^{2}\right), \\\nonumber
  \Pi_{A}\left(t^{1},t^{2}\right) &=\sum^{N-1}_{i,j=0}A_{ij}\Pi_{i}\left(t^{1}\right)\Pi_{j}\left(t^{2}\right), \\\nonumber
  C_{ij}=S_{ij}+A_{ij},&\qquad S_{ij}=S_{ji},\qquad A_{ij}=-A_{ji}.
\end{align}
Since
\begin{equation}\nonumber
  \left.\left(\partial_{t^{1}}-\partial_{t^{2}}\right)\right|_{t^{1}=t^{2}=t}  \Pi_{S}\left(t^{1},t^{2}\right)=0,
\end{equation}
one concludes
\begin{equation}\nonumber
  \Pi\left(t\right)= \left.\left(\partial_{t^{1}}-\partial_{t^{2}}\right)\right|_{t^{1}=t^{2}=t}\Pi_{A}\left(t^{1},t^{2}\right)=\sum^{N-1}_{i,j=0}A_{ij}\left(\frac{d\Pi_{i}\left(t\right)}{dt}\Pi_{j}\left(t\right)-\Pi_{i}\left(t\right)\frac{d\Pi_{j}\left(t\right)}{dt}\right).
\end{equation}

A basis of the vector space of antisymmetric $N\times N$ matrices therefore gives a spanning set for the image of Eq.~\eqref{GSPG2N}. Define
\begin{equation}\nonumber
  \Pi_{ij}\left(t\right)=\frac{d\Pi_{i}\left(t\right)}{dt}\Pi_{j}\left(t\right)-\Pi_{i}\left(t\right)\frac{d\Pi_{j}\left(t\right)}{dt},
\end{equation}
for $0\leq i<j\leq N-1$. Every period obtained from Eq.~\eqref{GSPG2N} is a linear combination of these functions.

The same construction extends to $Gr(k,N)$. We recall Eq.~\eqref{PGkN} from Section~\ref{VR2FM}:
\begingroup\small 
\begin{align}\label{Pi_GrCP}
\Pi_{Gr(k,N)}\left(\Sigma_{1},\cdots, \Sigma_{k};t\right)
=&\;
\left.
\prod_{1\leq a<b\leq k}
\frac{-\partial_{t^{a}}+\partial_{t^{b}}}{\Sigma_{a}-\Sigma_{b}}
\right |_{t^{a}=t}
\cdot \notag\\
&\;
\Pi_{\mathbb{CP}^{N-1}\times\cdots\times \mathbb{CP}^{N-1}}
\left(\Sigma_{1},\cdots ,\Sigma_{k};
t^{1}+i(k-1)\pi,\cdots, t^{k}+i(k-1)\pi\right).
\end{align}
\endgroup

For general $k$ we write $\Pi_i(\tau)$ for the unshifted projective-space coefficient functions and display the shift in their argument. Expanding each factor as in Eq.~\eqref{PPS}, we have
\begin{equation}
\begin{split}
&\Pi_{\mathbb{CP}^{N-1}\times\cdots\times \mathbb{CP}^{N-1}}
\left(\Sigma_{a}; t^{a}+i(k-1)\pi\right)\\
&=
\sum_{i_{1},\cdots,i_{k}=0}^{N-1}
\Sigma^{i_{1}}_{1}\cdots\Sigma^{i_{k}}_{k}
\Pi_{i_{1}}\left(t^{1}+i(k-1)\pi\right)\cdots
\Pi_{i_{k}}\left(t^{k}+i(k-1)\pi\right).
\end{split}
\end{equation}

The differential operator
\begin{equation}
\prod_{1\leq a<b\leq k}
\frac{-\partial_{t^{a}}+\partial_{t^{b}}}{\Sigma_{a}-\Sigma_{b}}
\end{equation}
is antisymmetric in $t^{a}$. Therefore, one obtains the Schur expansion
\begin{equation}\label{PGR}
\Pi_{Gr(k,N)}(\Sigma_1,\dots,\Sigma_k;t)
=
\sum_{\lambda}
S_\lambda(\Sigma_1,\dots,\Sigma_k)\,\Pi_\lambda(t).
\end{equation}

Here $S_\lambda$ is the Schur polynomial representing
the Grassmannian cohomology class described in
Section~\ref{VR2FM}. The partition
$\lambda=(\lambda_1,\ldots,\lambda_k)$ satisfies
\[
N-k\geq\lambda_1\geq\cdots\geq\lambda_k\geq0.
\]
The coefficient functions are
\begin{equation}\label{PGrCoeff}
\Pi_\lambda(t)
=
\det\left[
\Pi_{m_b}^{(a-1)}(\tau)
\right]_{a,b=1}^{k},
\qquad
m_b=\lambda_{k+1-b}+b-1,
\end{equation}
where
\[
\tau=t+i(k-1)\pi,
\qquad
\Pi_m^{(a)}(\tau)
=
\left(-\frac{d}{d\tau}\right)^a\Pi_m(\tau).
\]
The solution indices satisfy
$0\leq m_1<\cdots<m_k\leq N-1$.

For fixed solution indices $m_1<\cdots<m_k$, we denote
the antisymmetric combinations by
\[
\mathcal A_{\mathbf m}(a_1,\ldots,a_k)
=
\sum_{\alpha_1,\ldots,\alpha_k=1}^{k}
\epsilon_{\alpha_1\cdots\alpha_k}
\prod_{i=1}^{k}
\Pi_{m_{\alpha_i}}^{(a_i)}(\tau),
\]
where $\epsilon_{\alpha_1\cdots\alpha_k}$ is the
Levi-Civita symbol. In particular,
\[
\Pi_\lambda(t)
=
\mathcal A_{\mathbf m}(0,1,\ldots,k-1).
\]

Each $\Pi_i(t+i(k-1)\pi)$ satisfies
\begin{equation}\label{SPFP2}
\left(-\frac{d}{dt}\right)^N\Pi_i
=
(-1)^{N+k-1}q\,\Pi_i.
\end{equation}

Equations~\eqref{PGR} and \eqref{SPFP2} lead to an
algorithm for deriving differential equations in the
single physical K\"ahler parameter of $Gr(k,N)$.

$\bullet$ For each derivative $-\frac{d}{dt}$, we use
the product rule to increment the derivative orders
$a_1,\ldots,a_k$:
\[
-\frac{d}{dt}\mathcal A_{\mathbf m}(a_1,\ldots,a_k)
=
\sum_{i=1}^{k}
\mathcal A_{\mathbf m}
(a_1,\ldots,a_i+1,\ldots,a_k).
\]
Repeated derivative orders give zero after
antisymmetrization, while reordering distinct derivative
orders contributes the sign of the permutation.

$\bullet$ When $a_i=N$, Eq.~\eqref{SPFP2} gives
\[
\mathcal A_{\mathbf m}
(a_1,\ldots,N,\ldots,a_k)
=
(-1)^{N+k-1}q\,
\mathcal A_{\mathbf m}
(a_1,\ldots,0,\ldots,a_k).
\]

$\bullet$ Differentiate the period repeatedly until
it can be expressed as a linear combination of
lower-order derivatives. One can then read off the
quantum differential equation
\[
{\rm Poly}\!\left(-\frac{d}{dt}\right)
\cdot\Pi_\lambda(t)=0,
\]
where ${\rm Poly}$ is a polynomial in the differential
operator $-\frac{d}{dt}$, with coefficients possibly
depending on $q$. One should note that
$-\frac{d}{dt}$ acts nontrivially on $q$ as well.
In particular,
\[
-\frac{d}{dt}\bigl(qf(t)\bigr)
=
q\left(1-\frac{d}{dt}\right)f(t).
\]

Now let us look at a concrete example, namely the quantum
differential equation of $Gr(2,4)$. Consider
\begin{equation}\nonumber
  \Pi
  \equiv
  \varepsilon^{m_1,m_2}
  \Pi_{m_1}(t)\Pi_{m_2}^{(1)}(t),
  \qquad m_1,m_2\in\{0,\ldots,3\},
\end{equation}
where $\varepsilon^{m_1,m_2}=-\varepsilon^{m_2,m_1}$
are fixed antisymmetric coefficients, and repeated indices
are summed over. Equivalently,
\[
\Pi
=
\sum_{0\leq i<j\leq3}
\varepsilon^{i,j}\,
\mathcal A_{(i,j)}(0,1).
\]
Here $\Pi_i(t)$ denotes the shifted solution
$\Pi_i(t+i\pi)$ appearing in Eq.~\eqref{SPFP2}. Using $\left(-\frac{d}{dt}\right)^{4}\Pi_{l}\left(t\right)=-q\Pi_{l}\left(t\right)$, we can then compute
\begin{align}\nonumber
  \Pi^{(1)} &=  \varepsilon^{m_{1},m_{2}}\Pi_{m_{1}}\left(t\right)\Pi^{(2)}_{m_{2}}\left(t\right)\\\nonumber
   \Pi^{(2)} &=  \varepsilon^{m_{1},m_{2}}\left(\Pi^{(1)}_{m_{1}}\left(t\right)\Pi^{(2)}_{m_{2}}\left(t\right)+\Pi_{m_{1}}\left(t\right)\Pi^{(3)}_{m_{2}}\left(t\right)\right)\\\nonumber
     \Pi^{(3)} &= \varepsilon^{m_{1},m_{2}}\left(2\Pi^{(1)}_{m_{1}}\left(t\right)\Pi^{(3)}_{m_{2}}\left(t\right)\right)\\\nonumber
  \Pi^{(4)} &=  \varepsilon^{m_{1},m_{2}}\left(2\Pi^{(2)}_{m_{1}}\left(t\right)\Pi^{(3)}_{m_{2}}\left(t\right)+2q\Pi_{m_{1}}\left(t\right)\Pi^{(1)}_{m_{2}}\left(t\right)\right)\\\nonumber
  \Pi^{(5)} &=  \varepsilon^{m_{1},m_{2}}\left(4q\Pi_{m_{1}}\left(t\right)\Pi^{(2)}_{m_{2}}\left(t\right)+2q\Pi_{m_{1}}\left(t\right)\Pi^{(1)}_{m_{2}}\left(t\right)\right)\\\nonumber
  &= 4q\Pi^{(1)}+2q\Pi
\end{align}
thus, we have
\begin{equation}\label{GR24PF}
  \left(-\frac{d}{dt}\right)^{5}\Pi=2q\left(2\left(-\frac{d}{dt}\right)+1\right)\Pi.
\end{equation}
This agrees with \cite{Batyrev:1998kx}. Since $Gr(2,4)$ is the quadric hypersurface $\mathbb{CP}^{5}[2]$, Eq.~\eqref{GR24PF} also agrees with the differential equation for the mirror of that quadric. Further examples appear in Appendix~\ref{app:grass}.

\subsubsection{The algorithm}

We next consider GLSMs involving vector bundles over Grassmannians. Section~\ref{PFMGR} treated the Grassmannian itself; we now incorporate the additional bundle data.

The corresponding abelian periods need not factor into products of one-variable periods as they do in the Grassmannian case. The elimination procedure can nevertheless use the $S_k$ Weyl symmetry of $U(k)$ to organize the required antisymmetric combinations.

Motivated by this, we define
\begin{equation}
\Pi_M(S_\lambda;t)
=
\left.
\sum_{\sigma\in S_k}
\operatorname{sgn}(\sigma)
\prod_{i=1}^{k}
(-\partial_{t^i})^{\sigma(i)-1}
G_{m_1,\ldots,m_k}(t^1,\ldots,t^k)
\right|_{t^1=\cdots=t^k=t}
\cdot S_\lambda,
\end{equation}
where
\[
m_b=\lambda_{k+1-b}+b-1,
\qquad
0\leq m_1<\cdots<m_k\leq N-1.
\]
Here $G_{m_1,\ldots,m_k}(t^1,\ldots,t^k)$ denotes
the coefficient of
$\Sigma_1^{m_1}\cdots\Sigma_k^{m_k}$
in the cohomology expansion of the corresponding
abelian generating function.

More generally, for fixed solution indices
$0\leq m_1<\cdots<m_k\leq N-1$ and distinct
derivative orders
$0\leq a_1<\cdots<a_k\leq N-1$, we define
\begin{equation}
\begin{aligned}
\Pi_{\mathbf m;\mathbf a}(t)
={}&
\left.
\sum_{\sigma\in S_k}
\operatorname{sgn}(\sigma)
\prod_{i=1}^{k}
(-\partial_{t^i})^{a_{\sigma(i)}}
G_{m_1,\ldots,m_k}(t^1,\ldots,t^k)
\right|_{t^1=\cdots=t^k=t}.
\end{aligned}
\end{equation}
In particular, the periods associated with
the Schur expansion are
\[
\Pi_{m_1,\ldots,m_k}(t)
=
\Pi_{\mathbf m;(0,1,\ldots,k-1)}(t).
\]

In general, these periods may not be linearly independent. 
Nevertheless, they provide a convenient generating set for deriving the quantum differential equation via the following procedure:

\begin{itemize}

\item Acting with $-\partial_{t^i}$ increases the index $a_i$ to $a_i+1$ in
\begin{equation}\label{Per}
(-\partial_{t^1})^{a_{1}}\cdots(-\partial_{t^k})^{a_{k}}
G_{m_1,\cdots,m_k}(t^1,\cdots,t^k)\Big|_{t^{a}=t}.
\end{equation}
Due to the antisymmetry of the Levi-Civita symbol, if $a_i+1$ coincides with any other index, the corresponding antisymmetrized contribution vanishes; the individual expression \eqref{Per} need not vanish.

\item When $a_i=N$, the quantum differential equation satisfied by the corresponding abelian period is used to reduce the order of differentiation.

\item Define $\theta_t=\sum_i\partial_{t^i}$. Repeated application of $\theta_t$ to $\Pi_{m_1,\cdots,m_k}(t)$ produces relations of the form
\[
\theta_t^{q}\Pi_{m_1,\cdots,m_k}(t)
=
\sum_{p<q} c_p\,\theta_t^{p}\Pi_{m_1,\cdots,m_k}(t),
\]
which determine the quantum differential equation satisfied by $\Pi_{m_1,\cdots,m_k}(t)$.

\end{itemize}
We illustrate the procedure with the following examples.

\paragraph{Symplectic Grassmannian.}
Consider the symplectic Grassmannians $SG(k,2n)$. We recall their generating function from Eq.~\eqref{PSG}:
\begin{align}\nonumber
  \Pi_{SG(k,2n)}&\left(\Sigma_{a};t\right)=\prod_{1\leq a<b\leq k}\left.\frac{-\partial_{t^{a}}+\partial_{t^{b}}}{\Sigma_{a}-\Sigma_{b}}\right|_{t^{a}=t}\cdot \Pi^{Ab}\left(\Sigma_{a};t^{a}\right),
  \\\nonumber
 \Pi^{Ab}\left(\Sigma_{a};t^{a}\right)= &\sum_{d_{1},\ldots,d_{k}=0}^{\infty}\frac{\prod_{1\leq b< c\leq k}\prod^{d_{b}+d_{c}}_{j=1}\left(\Sigma_{b}+\Sigma_{c}+j\right)}{\prod^{k}_{a=1}\prod^{d_{a}}_{l=1}\left(\Sigma_{a}+l\right)^{2n}}\cdot\\\nonumber
&\left(-1\right)^{(k-1)\left(d_{1}+\cdots+d_{k}\right)}\exp\left(-t^{1}\left(d_{1}+\Sigma_{1}\right)-\cdots-t^{k}\left(d_{k}+\Sigma_{k}\right)\right).
\end{align}
Working modulo $\Sigma_a^{2n}=0$ for each $a$, the coefficients of the series show that $\Pi^{Ab}(\Sigma_a;t^a)$ satisfies
\begin{equation}\label{ABSG}
  \left(-\partial_{t^{a}}\right)^{2n}\Pi^{Ab}=(-1)^{k-1}e^{-t^{a}}\prod_{b\neq a}\left(-\partial_{t^{a}}-\partial_{t^{b}}+1\right)\Pi^{Ab}.
\end{equation}
Then from (\ref{PSG}), we have a result similar to that in the Grassmannian case, i.e.
\begin{equation}\nonumber
  \Pi_{SG(k,2n)}\left(t\right)=\varepsilon^{m_{1},\cdots,m_{k}}\left(-\partial_{t^{1}}\right)^{0}\left(-\partial_{t^{2}}\right)^{1}\cdots\left(-\partial_{t^{k}}\right)^{k-1}G_{m_{1},\cdots,m_{k}}\left(t^{1},\cdots,t^{k}\right)_{t^{a}=t}\cdot S_{\lambda},
\end{equation}
where $G_{m_{1},\cdots,m_{k}}\left(t^{1},\cdots,t^{k}\right)$ satisfies
\begin{equation}\nonumber
  \left(-\partial_{t^{a}}\right)^{2n}G_{m_{1},\cdots,m_{k}}\left(t^{1},\cdots,t^{k}\right)=(-1)^{k-1}e^{-t^{a}}\prod_{b\neq a}\left(-\partial_{t^{a}}-\partial_{t^{b}}+1\right)G_{m_{1},\cdots,m_{k}}\left(t^{1},\cdots,t^{k}\right).
\end{equation}

Now let us consider a special case, the Lagrangian Grassmannian $LG(2,4)$. We have
\begin{align}\nonumber
   \left(-\partial_{t^{1}}\right)^{4}G_{m_{1}m_{2}}\left(t^{1},t^{2}\right)&=-q^{1}\left(-\partial_{t^{1}}-\partial_{t^{2}}+1\right)G_{m_{1}m_{2}}\left(t^{1},t^{2}\right),  \\\nonumber
   \left(-\partial_{t^{2}}\right)^{4}G_{m_{1}m_{2}}\left(t^{1},t^{2}\right)&=-q^{2}\left(-\partial_{t^{1}}-\partial_{t^{2}}+1\right)G_{m_{1}m_{2}}\left(t^{1},t^{2}\right).
\end{align}

Set $-\theta_{t}=-\partial_{t^{1}}-\partial_{t^{2}}$; then we compute
\begingroup\footnotesize 
\begin{align}\nonumber
 \left(-\theta_{t}\right)\cdot\varepsilon^{m_{1}m_{2}} G^{0,1}_{m_{1}m_{2}}\left(t^{1},t^{2}\right)  &=\varepsilon^{m_{1}m_{2}}  G^{0,2}_{m_{1}m_{2}}\left(t^{1},t^{2}\right)\\\nonumber
   \left(-\theta_{t}\right)^{2}\cdot \varepsilon^{m_{1}m_{2}}G^{0,1}_{m_{1}m_{2}}\left(t^{1},t^{2}\right)  &=  \varepsilon^{m_{1}m_{2}}\left( G^{1,2}_{m_{1}m_{2}}\left(t^{1},t^{2}\right)+G^{0,3}_{m_{1}m_{2}}\left(t^{1},t^{2}\right)\right)\\\nonumber
 \left(-\theta_{t}\right)^{3}\cdot \varepsilon^{m_{1}m_{2}}G^{0,1}_{m_{1}m_{2}}\left(t^{1},t^{2}\right)  &=  \varepsilon^{m_{1}m_{2}}\left( 2G^{1,3}_{m_{1}m_{2}}\left(t^{1},t^{2}\right)\right)\\\nonumber
  \left(-\theta_{t}\right)^{4}\cdot \varepsilon^{m_{1}m_{2}}G^{0,1}_{m_{1}m_{2}}\left(t^{1},t^{2}\right)  &=  \varepsilon^{m_{1}m_{2}}\left( 2G^{2,3}_{m_{1}m_{2}}\left(t^{1},t^{2}\right)+2q^{2}\left(-\theta_{t}+1\right)G^{0,1}_{m_{1}m_{2}}\left(t^{1},t^{2}\right)\right)\\\nonumber
   \left(-\theta_{t}\right)^{5}\cdot \varepsilon^{m_{1}m_{2}}G^{0,1}_{m_{1}m_{2}}\left(t^{1},t^{2}\right)  &=  \varepsilon^{m_{1}m_{2}}\left( 4q^{2}\left(-\theta_{t}+1\right)G^{0,2}_{m_{1}m_{2}}\left(t^{1},t^{2}\right)+2q^{2}\left(-\theta_{t}+1\right)G^{0,1}_{m_{1}m_{2}}\left(t^{1},t^{2}\right)\right).
\end{align}
\endgroup
One can then obtain the quantum differential equation of the mirror of $LG(2,4)$:
\begin{equation}\nonumber
  \left(-\frac{d}{dt}\right)\left[\left(-\frac{d}{dt}\right)^{4}-2q\left(-2\frac{d}{dt}+1\right)\right]\Pi(t)=0.
\end{equation}
This is consistent with the identification of $LG(2,4)$ with the quadric hypersurface $\mathbb{CP}^{4}[2]$.

The same procedure applies to $SG(2,2n)$. For example, imposing the classical cohomology boundary conditions for $SG(2,6)$ gives the following annihilating differential operator:
\begin{align}\nonumber
  \left(-\frac{d}{dt}\right)^{8}&\left(-\frac{d}{dt}-1\right)^{5}-q\left(-\frac{d}{dt}\right)^{5}\left(-2\frac{d}{dt}+1\right)\left(13\frac{d^{2}}{dt^{2}}-13\frac{d}{dt}+4\right)\\\nonumber
  &-q^{2}\left(-3\frac{d}{dt}+2\right)\left(-3\frac{d}{dt}+3\right)\left(-3\frac{d}{dt}+4\right).
\end{align}
We leave the study of general cases to the interested reader.

\paragraph{Calabi-Yau threefolds in Grassmannians.}
For Calabi-Yau complete intersections in Grassmannians, the auxiliary periods generally do not factor as in the Grassmannian case. We therefore apply the elimination procedure described above. We begin with $Gr(2,4)[4]$ and denote the period obtained by antisymmetrization by $\Pi(t)$. Our proposal gives
\begin{equation}\label{GR244}
  \Pi\left(t\right)=\left(-\partial_{t^{1}}+\partial_{t^{2}}\right)_{t^{1}=t^{2}=t}\cdot f\left(t^{1},t^{2}\right),
\end{equation}
where $f(t^1,t^2)$ is the auxiliary abelian period. The equations below include the contribution from the fiber direction; the compact Calabi-Yau period is recovered by the insertion described in Section~\ref{VR2FM}.

\begin{align}\label{PFAbGR244}
  \left(-\partial_{t^{1}}\right)^{4} f\left(t^{1},t^{2}\right)& =-q^{1} D\cdot f\left(t^{1},t^{2}\right),\\\nonumber
   \left(-\partial_{t^{2}}\right)^{4} f\left(t^{1},t^{2}\right)& =-q^{2} D \cdot f\left(t^{1},t^{2}\right).
\end{align}
Here $q^a=e^{-t^a}$, and $D=\prod_{k=0}^{3}\bigl(-4(\partial_{t^1}+\partial_{t^2})+k\bigr)$ accounts for the fiber contribution. Equation~\eqref{GR244} selects the part of $f$ that is antisymmetric in $t^1,t^2$. Restriction to the diagonal commutes with the tangential derivative $\partial_{t^1}+\partial_{t^2}$, which becomes $d/dt$. We define
\begin{equation}\nonumber
  h\left(t^{1},t^{2}\right)=\left(-\partial_{t^{1}}+\partial_{t^{2}}\right)\cdot f\left(t^{1},t^{2}\right),
\end{equation}
then, from (\ref{PFAbGR244}), one computes
\begin{align}\nonumber
  \left(\partial_{t^{1}}+\partial_{t^{2}}\right)^{3}h\left(t^{1},t^{2}\right) & =-\left(\partial^{4}_{t^{1}}+2\partial^{3}_{t^{1}}\partial_{t^{2}}-2\partial^{3}_{t^{2}}\partial_{t^{1}}-\partial^{4}_{t^{2}}\right)\cdot f\left(t^{1},t^{2}\right). \\\nonumber
   &=\left(q^{1}-q^{2}\right)D\cdot f\left(t^{1},t^{2}\right)-2\left(\partial^{3}_{t^{1}}\partial_{t^{2}}-\partial^{3}_{t^{2}}\partial_{t^{1}}\right)\cdot f\left(t^{1},t^{2}\right).
\end{align}
\begin{align}\nonumber
  \left(\partial_{t^{1}}+\partial_{t^{2}}\right)^{5}h\left(t^{1},t^{2}\right) & =\left(q^{1}-q^{2}\right)\left(1-2\left(\partial_{t^{1}}+\partial_{t^{2}}\right)+\left(\partial_{t^{1}}+\partial_{t^{2}}\right)^{2}+2\left(\partial_{t^{1}}\partial_{t^{2}}\right)\right)\cdot D\cdot f\left(t^{1},t^{2}\right) \\\nonumber
   &+2\left(-q^{1}\partial_{t^{2}}+q^{2}\partial_{t^{1}}\right)D\cdot f\left(t^{1},t^{2}\right)+4\left(q^{1}\partial^{2}_{t^{2}}-q^{2}\partial^{2}_{t^{1}}\right)D\cdot f\left(t^{1},t^{2}\right).
\end{align}
In the limit $t^{1}\rightarrow t$, $t^{2}\rightarrow t$, $h\left(t^{1},t^{2}\right)\rightarrow h\left(t\right)$; therefore
\begin{equation}\nonumber
  \left(\frac{d}{dt}\right)^{5}h\left(t\right)=2q\left(2\frac{d}{dt}-1\right)D\cdot h\left(t\right),
\end{equation}
where $D=\prod^{3}_{k=0}\left(4\left(-\frac{d}{dt}\right)+k\right)$. Let $\theta=-\frac{d}{dt}$; then
\begin{equation}\nonumber
  \theta^{5}h\left(t\right)=2q\left(2\theta+1\right)D\cdot h\left(t\right).
\end{equation}
As explained in Section~\ref{VR2FM}, the compact Calabi-Yau period is $\widetilde h(t)=\theta h(t)$, up to a constant normalization. It satisfies
\begin{equation}\nonumber
  \left(\theta^{4}-16q\left(2\theta+1\right)^{2}\left(4\theta+1\right)\left(4\theta+3\right)\right)\widetilde{h}\left(t\right)=0.
\end{equation}
This is the Picard-Fuchs equation for the mirror of $Gr(2,4)[4]$. Similar calculations apply to more general examples, although the intermediate expressions become considerably longer.

A useful operator substitution streamlines these calculations. We first illustrate it for the quintic. The quantum differential equation for the mirror of $\mathbb{CP}^{4}$ is
\begin{equation}\nonumber
  \left(-\partial_{t}\right)^{5}\Pi_{\mathbb{CP}^{4}}=-q \Pi_{\mathbb{CP}^{4}}.
\end{equation}
Replace $q$ by $q\widehat D$, with $\widehat D=(-5\partial_t)(-5\partial_t+1)\cdots(-5\partial_t+4)$. This gives the differential equation for the mirror of ${\cal O}(-5)$:
\begin{equation}\nonumber
  \left(-\partial_{t}\right)^{5}\Pi_{{\cal O}(-5)}=-q \left(-5\partial_{t}\right)\cdots\left(-5\partial_{t}+4\right)\Pi_{{\cal O}(-5)}.
\end{equation}
Using $\Pi_{\rm quintic}=(-\partial_t)\Pi_{{\cal O}(-5)}$, up to normalization as in Section~\ref{VR2FM}, we recover
\begin{equation}\nonumber
  \left(-\partial_{t}\right)^{4}\Pi_{{\rm quintic}}=-5q\left(-5\partial_{t}+1\right)\cdots\left(-5\partial_{t}+4\right)\Pi_{{\rm quintic}}.
\end{equation}
One can interpret the negative sign in front of $q$ as a finite shift in the theta angle by $5\pi$ \cite{Morrison:1994fr}.

We now apply the same substitution to Calabi-Yau complete intersections in Grassmannians, beginning with $Gr(2,5)[1^2,3]$. With $\theta=-d/dt$, the Grassmannian operator obtained by the procedure of Section~\ref{PFMGR} is
\begin{equation}\nonumber
  \bigl[\theta^7(\theta-1)^3+q\theta^3(11\theta^2+11\theta+3)-q^2\bigr]\Pi_{Gr(2,5)}(t)=0.
\end{equation}
 Now we replace $q$ by $q\widehat{D}$, where
 \begin{equation}\nonumber
   \widehat D=\theta^2(3\theta)(3\theta+1)(3\theta+2).
 \end{equation}
This substitution yields
\begin{align}\label{PFTlGr25}
    \left(-\frac{d}{dt}\right)^{7}\left(-\frac{d}{dt}-1\right)^{3}\cdot\Pi\left(t\right)+&q\widehat{D}\left(11\left(-\frac{d}{dt}\right)^{5}+11\left(-\frac{d}{dt}\right)^{4}+3\left(-\frac{d}{dt}\right)^{3}\right)\cdot\Pi\left(t\right) \\\nonumber
     &-\left(q\widehat{D}\right)\left(q\widehat{D}\right)\cdot\Pi\left(t\right)=0.
\end{align}
The order of the factors is essential: $q$ does not commute with $\widehat D$, so $(q\widehat D)^2\neq q^2\widehat D^2$. Equation~\eqref{PFTlGr25} annihilates the period of the corresponding local model. Following Section~\ref{VR2FM}, set $\widetilde\Pi(t)=\Pi^{(3)}(t)$ to obtain the compact-mirror period, up to normalization. Equation~\eqref{PFTlGr25} then becomes
\begin{align}\label{PFTGr25CY}\nonumber
   &\left(-\frac{d}{dt}\right)^{4}\left(-\frac{d}{dt}-1\right)^{3}\cdot\widetilde{\Pi}\left(t\right)-3\left(q\widehat{D}\right)\left(q\left(-3\frac{d}{dt}+1\right)\left(-3\frac{d}{dt}+2\right)\right)\cdot\widetilde{\Pi}\left(t\right) \\
     &+3q\left(-\frac{d}{dt}\right)^{3}\left(-3\frac{d}{dt}+1\right)\left(-3\frac{d}{dt}+2\right)\left(11\left(-\frac{d}{dt}\right)^{2}+11\left(-\frac{d}{dt}\right)+3\right)\cdot\widetilde{\Pi}\left(t\right) =0.
\end{align}
For this one-parameter Calabi-Yau threefold, we expect a fourth-order Picard-Fuchs operator. The higher-order operator in Eq.~\eqref{PFTGr25CY} is reducible. To factor it, use
\begin{align}\nonumber
  q\cdot\left(-\frac{d}{dt}\right)^{3}\times\left(\cdots\right)&=\left(-\frac{d}{dt}-1\right)^{3}\left(q\times\cdots\right),\\\nonumber \left(-3\frac{d}{dt}+1\right)\left(-3\frac{d}{dt}+2\right)\cdot\left(q\cdots\right)&= q\left(-3\frac{d}{dt}+4\right)\left(-3\frac{d}{dt}+5\right)\times\left(\cdots\right).
\end{align}
Thus we obtain
\begin{align}\label{PFTGr25CY2}
  \left(\theta-1\right)^{3}\cdot\Bigg(&\Bigg.\theta^{4}+3q\left(3\theta+2\right)\left(3\theta+1\right)\left(11\theta^{2}+11\theta+3\right)
  \\\nonumber
  &-9q^{2}\left(3\theta+5\right)\left(3\theta+4\right)\left(3\theta+2\right)\left(3\theta+1\right)\Bigg.\Bigg)\cdot\widetilde{\Pi}\left(t\right)=0.
\end{align}
The factorization in Eq.~\eqref{PFTGr25CY2}
does not by itself imply the fourth-order equation,
since
\[
\ker(\theta-1)^3
=
\operatorname{span}
\{q,q\log q,q(\log q)^2\}.
\]
To remove this ambiguity, we examine the Frobenius
expansion at $q=0$.
For a holomorphic period
\[
\widetilde\Pi_0(q)=a_0+a_1q+O(q^2),
\]
the fourth-order operator in parentheses gives
\[
\mathcal L\widetilde\Pi_0
=
(a_1+18a_0)q+O(q^2).
\]
Thus the condition $a_1=-18a_0$ is sufficient
to eliminate the extra solution in the holomorphic
sector. For the logarithmic periods, the coefficients
of $q$, $q\log q$, and $q(\log q)^2$ in
$\mathcal L\widetilde\Pi$ must likewise vanish.

With this in mind, the expected Picard-Fuchs equation
for the mirror of $Gr(2,5)[1,1,3]$ is
\begin{align}\label{PFTGr25CY3}
 \Bigg(&\Bigg.\theta^{4}+3q\left(3\theta+2\right)\left(3\theta+1\right)\left(11\theta^{2}+11\theta+3\right)
  \\\nonumber
  &-9q^{2}\left(3\theta+5\right)\left(3\theta+4\right)\left(3\theta+2\right)\left(3\theta+1\right)\Bigg.\Bigg)\cdot\widetilde{\Pi}\left(t\right)=0.
\end{align}

After the change $q\mapsto-q$, this operator agrees with \cite{Batyrev:1998kx}. The same sign convention appeared in the Yukawa couplings in Section~\ref{YC}, as expected from their relation to the Picard-Fuchs system.

The same procedure applies to $Gr(2,5)[1,2^2]$, $Gr(2,6)[1^4,2]$, $Gr(2,7)[1^7]$, and $Gr(3,6)[1^6]$, with results agreeing with \cite{Batyrev:1998kx}. We expect analogous calculations to yield mirror Picard-Fuchs equations for more general Calabi-Yau models, including those considered in \cite{Jockers:2012zr}.

\subsubsection{Flag manifolds}
The method also applies to other targets with known GLSM descriptions. We illustrate this extension for flag manifolds, leaving the development of a general elimination algorithm to future work.

GLSMs for flag manifolds were studied in \cite{Donagi:2007hi,Guo:2018iyr}. Their mirror constructions provide a starting point for deriving quantum differential equations. We present explicit calculations for $Fl(1,2,3)$ and $Fl(1,2,4)$.

\Needspace{5\baselineskip}
\noindent{\textbf{$Fl(1,2,3)$}}. The period of the mirror of $Fl(1,2,3)$ can be expressed as
\begin{equation}\label{F123}
  \Pi\left(t,\widetilde{t}\right)=\left(-\partial_{t^{1}}+\partial_{t^{2}}\right)_{t^{1}=t^{2}=\widetilde{t}}\cdot f\left(t; t^{1},t^{2}\right),
\end{equation}
where $f\left(t; t^{1},t^{2}\right)$ obeys the partial differential equations
\begin{align}\label{APFF123}
   \left(-\partial_{t}+\partial_{t^{1}}\right)&\cdot\left(-\partial_{t}+\partial_{t^{2}}\right)f\left(t; t^{1},t^{2}\right)=e^{-t} f\left(t; t^{1},t^{2}\right) ,\\\nonumber
   \left(-\partial_{t^{1}}\right)^{3}&f\left(t; t^{1},t^{2}\right)=-\widetilde{q}^{1}\left(-\partial_{t}+\partial_{t^{1}}\right)f\left(t; t^{1},t^{2}\right),\\\nonumber
    \left(-\partial_{t^{2}}\right)^{3}&f\left(t; t^{1},t^{2}\right)=-\widetilde{q}^{2}\left(-\partial_{t}+\partial_{t^{2}}\right)f\left(t; t^{1},t^{2}\right).
\end{align}
As before, define $h(t;t^1,t^2)=(-\partial_{t^1}+\partial_{t^2})f(t;t^1,t^2)$. Equation~\eqref{APFF123} gives
\begin{align}\label{D2FL}
  \left(\partial_{t^{1}}+\partial_{t^{2}}\right)^{2}&h\left(t; t^{1},t^{2}\right)=\left(-\partial^{3}_{t^{1}}+\partial^{3}_{t^{2}}\right)\cdot  f\left(t; t^{1},t^{2}\right)-\partial_{t^{1}}\partial_{t^{2}}\left(\partial_{t^{1}}-\partial_{t^{2}}\right)\cdot f\left(t; t^{1},t^{2}\right)\\\nonumber
  &=\left(-\widetilde{q}^{1}\left(-\partial_{t}+\partial_{t^{1}}\right)+\widetilde{q}^{2}\left(-\partial_{t}+\partial_{t^{2}}\right)\right)\cdot  f\left(t; t^{1},t^{2}\right)+\partial_{t^{1}}\partial_{t^{2}}h\left(t; t^{1},t^{2}\right).
\end{align}
Now, we have
\begingroup\footnotesize 
\begin{align}\nonumber
  \left.\left(-\partial_{t}\right)^{2} h\left(t; t^{1},t^{2}\right)\right|_{t^{1}=t^{2}=\widetilde{t}}&=q \cdot \Pi\left(t; \widetilde{t}\right)+\left(-\partial_{t}\right)\left(-\partial_{\widetilde{t}}\right)\cdot \Pi\left(t; \widetilde{t}\right)-\left.\left(-\partial_{t^{1}}\right)\left(-\partial_{t^{2}}\right)\cdot h\left(t; t^{1},t^{2}\right)\right|_{t^{1}=t^{2}=\widetilde{t}},\\\nonumber
  &=\left(q+\widetilde{q}\right)\Pi\left(t; \widetilde{t}\right)+\left(-\partial_{t}\right)\left(-\partial_{\widetilde{t}}\right)\cdot \Pi\left(t; \widetilde{t}\right)-\left(-\partial_{\widetilde{t}}\right)^{2}\cdot \Pi\left(t; \widetilde{t}\right).
\end{align}
\endgroup
Here we used Eq.~\eqref{D2FL} and the chain-rule identity $\partial_{\widetilde t}=(\partial_{t^1}+\partial_{t^2})|_{t^1=t^2=\widetilde t}$. We therefore obtain the first quantum differential equation for $Fl(1,2,3)$:
\begin{equation}\label{1PFF23}
   \left(\left(-\partial_{t}\right)^{2}-\left(-\partial_{t}\right)\left(-\partial_{\widetilde{t}}\right)+ \left(-\partial_{\widetilde{t}}\right)^{2}\right)\cdot\Pi\left(t; \widetilde{t}\right)=\left(q+\widetilde{q}\right)\cdot \Pi\left(t; \widetilde{t}\right).
\end{equation}
A further differentiation gives
\begin{align}\nonumber
  \left.\left(-\partial_{t^{1}}-\partial_{t^{2}}\right)^{3}h\left(t; t^{1},t^{2}\right)\right|_{t^{1}=t^{2}=\widetilde{t}}=-\widetilde{q}\left(-1+\partial_{t}+\partial_{\widetilde{t}}\right)\Pi\left(t; \widetilde{t}\right).
\end{align}
This gives the second quantum differential equation for the mirror of $Fl(1,2,3)$,
\begin{equation}\label{2PFF23}
  \left(-\partial_{\widetilde{t}}\right)^{3}\Pi\left(t; \widetilde{t}\right)=-\widetilde{q}\left(-1+\partial_{t}+\partial_{\widetilde{t}}\right)\Pi\left(t; \widetilde{t}\right).
\end{equation}
Our method of getting quantum differential equations is not complicated; however, to the best of our knowledge, our computation is a new result.

\Needspace{5\baselineskip}
\noindent{\textbf{$Fl(1,2,4)$}}.  The period of the mirror of $Fl(1,2,4)$ satisfies
\begin{equation}\label{F124}
  \Pi\left(t,\widetilde{t}\right)=\left(-\partial_{t^{1}}+\partial_{t^{2}}\right)_{t^{1}=t^{2}=\widetilde{t}}\cdot g\left(t; t^{1},t^{2}\right),
\end{equation}
where $g\left(t; t^{1},t^{2}\right)$ obeys the partial differential equations
\begin{align}\label{APFF124}
   \left(-\partial_{t}+\partial_{t^{1}}\right)&\cdot\left(-\partial_{t}+\partial_{t^{2}}\right)g\left(t; t^{1},t^{2}\right)=e^{-t} g\left(t; t^{1},t^{2}\right) ,\\\nonumber
   \left(-\partial_{t^{1}}\right)^{4}&g\left(t; t^{1},t^{2}\right)=\widetilde{q}^{1}\left(-\partial_{t}+\partial_{t^{1}}\right)g\left(t; t^{1},t^{2}\right),\\\nonumber
    \left(-\partial_{t^{2}}\right)^{4}&g\left(t; t^{1},t^{2}\right)=\widetilde{q}^{2}\left(-\partial_{t}+\partial_{t^{2}}\right)g\left(t; t^{1},t^{2}\right).
\end{align}
Applying the same procedure to Eq.~\eqref{APFF124} gives the quantum differential equations for the mirror of $Fl(1,2,4)$:
\begin{align}\label{PF124}
  &\left(2\left(-\partial_{t}\right)^{2}\left(-\partial_{\widetilde{t}}\right)-2\left(-\partial_{t}\right)\left(-\partial_{\widetilde{t}}\right)^{2}+\left(-\partial_{\widetilde{t}}\right)^{3} \right)\cdot\Pi\left(t;\widetilde{t}\right) =\left(2q\left(-\partial_{\widetilde{t}}\right)-\widetilde{q}\right)\cdot\Pi\left(t;\widetilde{t}\right)
  \\\nonumber
 &\left(-\partial_{\widetilde{t}}\right)^{5}\cdot\Pi\left(t;\widetilde{t}\right) =-\widetilde{q}\left(\left(-\partial_{\widetilde{t}}+1\right)^{2}+\left(-2\partial_{t}\right)+2\left(-\partial_{t}\right)\left(-\partial_{\widetilde{t}}\right)+2\left(-\partial_{t}\right)^{2}-2q\right)\cdot\Pi\left(t;\widetilde{t}\right).
 \end{align}
As far as we know, our computation of $Fl(1,2,4)$ is also a new result.

The same construction can be applied to general flag manifolds. For the two-step flag manifold $Fl(k_1,k_2,N)$, the mirror period has the form
\begin{equation}\label{SPFL}
  \Pi\left(t; \widetilde{t}\right)=\left.\prod_{1\leq a<b\leq k_{1}}\left(-\partial_{t^{a}}+\partial_{t^{b}}\right)\right|_{t^{a}=t}\cdot\left.\prod_{1\leq a<b\leq k_{2}}\left(-\partial_{\widetilde{t}^{a}}+\partial_{\widetilde{t}^{b}}\right)\right|_{\widetilde{t}^{a}=\widetilde{t}}\cdot f\left(t^{1},\ldots,t^{k_{1}};\widetilde{t}^{1},\ldots,\widetilde{t}^{k_{2}}\right).
\end{equation}
Let $q^a=e^{-t^a}$ and $\widetilde q^b=e^{-\widetilde t^b}$. Before restricting to the diagonal, the auxiliary function $f$ satisfies
\begin{align*}
  \prod_{b=1}^{k_2}\left(-\partial_{t^a}+\partial_{\widetilde t^b}\right)f
     &=(-1)^{k_1-1+k_2}q^a f, \qquad 1\leq a\leq k_1,\\
  \left(-\partial_{\widetilde t^b}\right)^N f
     &=(-1)^{k_2-1+N+k_1}\widetilde q^b
       \prod_{a=1}^{k_1}\left(-\partial_{t^a}+\partial_{\widetilde t^b}\right)f,
       \qquad 1\leq b\leq k_2.
\end{align*}
Let $\mathcal D$ denote the product of the two alternating differential operators in Eq.~\eqref{SPFL} before restriction. Differentiating the period along the diagonal gives
\[
  \partial_t^m\partial_{\widetilde t}^n\Pi(t;\widetilde t)
  =\left.\left(\sum_{a=1}^{k_1}\partial_{t^a}\right)^m
          \left(\sum_{b=1}^{k_2}\partial_{\widetilde t^b}\right)^n
          \mathcal D f\right|_{t^a=t,\,\widetilde t^b=\widetilde t}.
\]
Appropriate choices of $m$ and $n$ give differential relations among the physical periods. It would be useful to organize these relations into a closed formula for a generating system of quantum differential equations for $Fl(k_1,k_2,N)$. The same approach extends to flags with more steps.

\subsubsection{Comments on periods and quantum differential equations}
The preceding derivations used relations among differential operators without requiring explicit period solutions. An alternative approach starts from the fundamental period and searches for operators that annihilate it:

$\bullet$ Compute the fundamental period $\Pi_0(t^a)$ from the proposed Cartan-parameter system and restrict to the physical parameters.

$\bullet$ Find differential equations in the physical parameters that annihilate the restricted period. For one physical parameter $t$, these take the form $\operatorname{Poly}(\partial_t;q)\,\Pi_0(t)=0$.

For example, the fundamental period for the mirror of $Gr(2,N)$ is
\begin{equation}\nonumber
  \Pi_0(t)=\sum_{d_1,d_2\geq0}\frac{\bigl((-1)^{N+1}q\bigr)^{d_1+d_2}}{(d_1!)^N(d_2!)^N}\left[1-\frac{N(d_1-d_2)}{2}\bigl(H_{d_1}-H_{d_2}\bigr)\right].
\end{equation}
Here $H_d=\sum_{j=1}^{d}j^{-1}$ and $H_0=0$, and the sum runs over all $d_1,d_2\geq0$. One may search for an annihilating operator ${\rm Poly}(\partial_t;q)$ using this series. A closed formula for such operators for general $Gr(2,N)$ would be useful. The same strategy can be applied to other examples; any operator inferred from finitely many coefficients must then be verified for the full series.

\section{Gromov-Witten invariants}\label{GWI}
In this section, we use the techniques developed above and the procedure outlined in Section~\ref{GWFM} to compute instanton numbers for Calabi-Yau manifolds described by nonabelian GLSMs.

\subsection{Complete intersections in Grassmannians}\label{GrGW}

We illustrate the procedure of Section~\ref{GWFM} with the Calabi-Yau hypersurface $Gr(2,4)[4]$. Further examples are presented in Appendix~\ref{app:grassCY}.

Following Section~\ref{VR2FM}, the generating function for the mirror of $Gr(2,4)[4]$ is
\begin{align}\label{PGr244}
  \Pi_{Gr(2,4)[4]}&(\Sigma_{1},\Sigma_{2}; t)=\left.\frac{-\partial_{t^{1}}+\partial_{t^{2}}}{\Sigma_{1}-\Sigma_{2}}\right|_{t^{1}=t^{2}=t}\sum^{\infty}_{d_{1},d_{2}=0}\frac{\prod^{4\left(d_{1}+d_{2}\right)}_{i=1}\left(4
  \left(\Sigma_{1}+\Sigma_{2}\right)+i\right)}{\prod^{d_{1}}_{l_{1}=1}\left(\Sigma_{1}+l_{1}\right)^{4}\cdot\prod^{d_{2}}_{l_{2}=1}\left(\Sigma_{2}+l_{2}\right)^{4}}
  \\\nonumber
  &\cdot\exp\left(-\left(\Sigma_{1}+d_{1}\right)\left(t^{1}+i\pi\right)\right)\exp\left(-\left(\Sigma_{2}+d_{2}\right)\left(t^{2}+i\pi\right)\right).
\end{align}

To compute the instanton numbers, we need only $\Pi_{Gr(2,4)[4],0}$ and $\Pi_{Gr(2,4)[4],1}$, which can be extracted directly from Eq.~\eqref{PGr244}. Acting on the exponential factors with the normalized operator $(-\partial_{t^{1}}+\partial_{t^{2}})/(\Sigma_{1}-\Sigma_{2})$ gives the factor
\begin{equation}\label{G24F}
  \frac{d_{1}-d_{2}}{\Sigma_{1}-\Sigma_{2}}+1,
\end{equation}
while the reciprocal denominator has the following expansion in $\Sigma_1$ and $\Sigma_2$:
\begingroup\footnotesize 
\begin{align}\label{DGR24}
   \prod^{d_{1}}_{l_1=1}\left(\Sigma_{1}+l_1\right)^{-4}\cdot\prod^{d_{2}}_{l_2=1}\left(\Sigma_{2}+l_2\right)^{-4}&=\left(d_{1}!\right)^{-4}\left(d_{2}!\right)^{-4}\Bigg(\Bigg.1-4\Sigma_{1}H_{d_{1}}-4\Sigma_{2}H_{d_{2}}
   \\\nonumber
   &+\left(\Sigma_{1}+\Sigma_{2}\right)\left(\Sigma_{1}-\Sigma_{2}\right)\left(4H^{2}_{d_{1}}+\sum^{d_{1}}_{k=1}\frac{1}{k^{2}}-4H^{2}_{d_{2}}-\sum^{d_{2}}_{k=1}\frac{1}{k^{2}}\right)+\ldots\Bigg.\Bigg),
   \end{align}
\endgroup
where $H_{d}$ is the harmonic number $H_{d}:=\sum^{d}_{k=1}\frac{1}{k}$ and $H_{0}:=0$. From Eqs. (\ref{G24F}) and (\ref{DGR24}), we then read off
\begin{align}\label{PGR24401}
  \Pi_{0}(t)&=\sum_{d_{1},d_{2}\geq 0}\frac{\left(4d_{1}+4d_{2}\right)!}{\left(d_{1}!\right)^4\left(d_{2}!\right)^4}\left(1-2\left(d_{1}-d_{2}\right)\left(H_{d_{1}}-H_{d_{2}}\right)\right)\exp\left(-\left(t+i\pi\right)\left(d_{1}+d_{2}\right)\right)
  \\\nonumber
  \Pi_{1}(t)&=-\left(t+i\pi\right)\Pi_{0}(t)\\\nonumber
  &+\sum_{d_{1},d_{2}\geq 0}\frac{\left(4d_{1}+4d_{2}\right)!}{\left(d_{1}!\right)^4\left(d_{2}!\right)^4}\Bigg(\Bigg.4H_{4(d_{1}+d_{2})}-2H_{d_{1}}-2H_{d_{2}}-8\left(d_{1}-d_{2}\right)H_{4(d_{1}+d_{2})}\left(H_{d_{1}}-H_{d_{2}}\right)
  \\\nonumber
   &+\left(d_{1}-d_{2}\right)\left(4H^{2}_{d_{1}}+\sum^{d_{1}}_{k=1}\frac{1}{k^{2}}-4H^{2}_{d_{2}}-\sum^{d_{2}}_{k=1}\frac{1}{k^{2}}\right)\Bigg.\Bigg)\exp\left(-(d_{1}+d_{2})\left(t+i\pi\right)\right).
\end{align}
The first few terms of Eq.~\eqref{PGR24401} are
\begin{align}\nonumber
  \Pi_{0}(t)&=1 + 48 q + 15120 q^{2} + 7392000 q^{3} + 4414410000 q^{4}+{\cal O}(e^{-5t}),
  \\\nonumber
  \Pi_{1}(t)&=-t - i\pi + 16 q (16 - 3 (t + i\pi)) +144 q^{2} (634 - 105 ( t + i\pi))
  \\\nonumber
   &+\frac{6400}{3} q^{3} \left(21874 - 3465 \left(t + i\pi\right)\right) + 98000 q^{4} \left(290929 - 45045 \left(t + i\pi\right)\right)+{\cal O}(e^{-5t}).
\end{align}
The flat coordinate $\tau$ is
\begin{equation}\label{FGR24}
  \tau=-\frac{\Pi_{1}(t)}{\Pi_{0}(t)}-2i\pi t_{0}=-\log q-256q-79008 q^{2}-\frac{117004288}{3} q^{3}-23552020432 q^{4} +{\cal O}(e^{-5t}),
\end{equation}
where we have chosen $t_{0}=1/2$. Expressing $q$ in terms of $z=e^{-\tau}$, we have
\begin{equation}\label{IFGR24}
  q=z-256z^{2}+19296 z^{3}-2836480 z^{4}+{\cal O}(z^{5}).
\end{equation}
The Yukawa coupling of $Gr(2,4)[4]$ is given in Eq.~\eqref{CLGR24}. Substituting Eq.~\eqref{IFGR24} into Eq.~\eqref{CLATB} gives
\begin{equation}\label{GRZS}
 K_{zzz} = 8+1280z+739584 z^2 + 422690816 z^3 + 248570501376 z^4+{\cal O}(z^{5}).
\end{equation}
The first four instanton numbers $n_d$ of $Gr(2,4)[4]$, extracted using Eq.~\eqref{AGW}, are
\begin{equation}\label{GWG244}
  1280,\quad 92288,\quad 15655168,\quad 3883902528.
\end{equation}
These results agree with the computations in \cite{Batyrev:1998kx,Jockers:2012dk}. Moreover, $Gr(2,4)[4]$ can also be realized as the complete intersection $\mathbb{CP}^{5}[2,4]$, whose instanton numbers can be computed using toric mirror symmetry. This provides a nontrivial consistency check of our method for extracting Gromov-Witten invariants from nonabelian mirrors. Further examples are presented in Appendix~\ref{app:grassCY}.

\subsection{The Pfaffian Calabi-Yau threefold}\label{PFCYGW}

In \cite{Rødland:1998}, R\o{}dland conjectured that the Grassmannian Calabi-Yau threefold $X=Gr(2,7)[1^{7}]$ and a Pfaffian Calabi-Yau threefold $Y$, obtained by intersecting the Pfaffian locus in $\mathbb{CP}^{20}$ with a generic $\mathbb{CP}^{6}$, share a common quantum K\"ahler moduli space. Their identical Hodge numbers provide one consistency check. R\o{}dland also found that the proposed mirrors of $X$ and $Y$ satisfy the same one-parameter Picard-Fuchs equation. Hori and Tong subsequently gave a field-theoretic explanation of this relation in \cite{Hori:2006dk}, identifying $X$ and $Y$ with the two large-volume phases of a single GLSM, at $r\gg0$ and $r\ll0$, respectively. In the latter phase, the nonabelian gauge dynamics leads to a nonlinear sigma model on $Y$. We therefore compute the instanton numbers in this phase and propose a generating $I$-function for the periods of the Pfaffian Calabi-Yau threefold. To the best of our knowledge, this generating function is new.

The Pfaffian Calabi-Yau threefold and $X=Gr(2,7)[1^{7}]$ arise from the same GLSM. In the Pfaffian phase, we assign vector R-charge 1 to the fundamental fields and vector R-charge 0 to the fields in the inverse determinant representation. The mirror Landau-Ginzburg superpotential is then
\begin{align}\label{MP-Pfaffian}
  W=&\Sigma_{1}\left(-2\sum^{7}_{i=1}\log X^{1}_{i}+\log X_{1}-\log X_{2}-\sum^{7}_{\beta=1}Y_{\beta}-t^{1}\right)
  \\\nonumber
  +&\Sigma_{2}\left(-2\sum^{7}_{i=1}\log X^{2}_{i}+\log X_{2}-\log X_{1}-\sum^{7}_{\beta=1}Y_{\beta}-t^{2}\right)
  \\\nonumber
  +&\sum^{7}_{i=1} \left(X^{1}_{i}\right)^{2}+\sum^{7}_{i=1} \left(X^{2}_{i}\right)^{2}+X_{1}+X_{2}+\sum^{7}_{\beta=1}e^{-Y_{\beta}}.
  \end{align}
The fundamental variables are related to the logarithmic variables by $X^{1}_{i}=e^{-Y^{1}_{i}/2}$ and $X^{2}_{i}=e^{-Y^{2}_{i}/2}$. The critical-point equations are
\begin{equation}\nonumber
  \frac{\partial W}{\partial \Sigma_{a}}=\frac{\partial W}{\partial X^{a}_{i}}=\frac{\partial W}{\partial Y_{\beta}}=\frac{\partial W}{\partial X_1}=\frac{\partial W}{\partial X_2}=0.
\end{equation}
We then have
\begin{align}\nonumber
  X_{1}=-\Sigma_{1}&+\Sigma_{2}=-X_{2}  \\\nonumber
   e^{-Y^{a}_{i}}=\Sigma_{a},\qquad& e^{-Y_{\beta}}=-\Sigma_{1}-\Sigma_{2} ,
\end{align}
and
\begin{align}\label{PCYCL}
  \Sigma^{7}_{1} &=q^{1}\left(\Sigma_{1}+\Sigma_{2}\right)^{7}  \\\nonumber
   \Sigma^{7}_{2} &=q^{2}\left(\Sigma_{1}+\Sigma_{2}\right)^{7} ,\qquad \Sigma_{1}\neq\Sigma_{2}.
\end{align}
For $\Sigma_a\neq0$, Eq.~\eqref{PCYCL} gives $\bigl((\Sigma_1+\Sigma_2)/\Sigma_a\bigr)^7=1/q^a$. This ratio tends to zero along any family of such solutions with $|q^a|\to\infty$. The equations leave the overall scale of the $\Sigma_a$ undetermined.

The period in this phase is
\begin{equation}\nonumber
  \Pi=\int_{\mathcal C_{\Sigma}}d\Sigma_1\,d\Sigma_2
  \int_{\Gamma}\prod_{i=1}^{7}\prod_{a=1}^{2}dX_i^a
  \prod_{\beta=1}^{7}dY_{\beta}\,dX_1\,dX_2\,
  \exp\left[-W+\frac{t^1+t^2}{2}\right].
\end{equation}
The extra factor $e^{(t^1+t^2)/2}$ is inserted for the same reason
as in \cite{Rødland:1998}: it gives the appropriate normalization of
the holomorphic top form near $q=\infty$.

We first integrate out the root variables $X_1$ and $X_2$.
The Pfaffian phase is reached as $q^1,q^2\to\infty$, and
from Eq.~(\ref{PCYCL}) we have, for nonzero Coulomb-branch variables,
\begin{equation}\nonumber
  \left(\frac{\Sigma_1}{\Sigma_2}\right)^7=\frac{q^1}{q^2},
  \qquad
  \Sigma_1=\left(\frac{q^1}{q^2}\right)^{1/7}\omega^j\Sigma_2,
  \qquad \omega=e^{2\pi i/7},\quad j=0,\ldots,6.
\end{equation}
The constraint $\Sigma_1\neq\Sigma_2$ restricts the allowed
branches. This relation, however, does not by itself imply that
$\Sigma_1$ and $\Sigma_2$ both approach zero. We instead use
the root-cycle prescription of Section~\ref{NGQS}. Write
\[
\Delta=\Sigma_1-\Sigma_2,
\qquad
W_{\rm root}=X_1+X_2+\Delta\log X_1-\Delta\log X_2.
\]
For fixed real $\Sigma_a$ with $\Delta>0$, the critical point is
$X_1=-\Delta$, $X_2=\Delta$. We take $X_2$ along the positive
real axis and $X_1$ along the clockwise Hankel contour
$\mathcal H$ specified in Section~\ref{NGQS}, with
$\log(-\Delta)=\log\Delta+i\pi$.
With the product orientation $dX_1\wedge dX_2$, the paired
root integral is
\begin{align*}
\mathcal I_{\rm root}
&=
\left(\int_{\mathcal H}dX_1\,X_1^{-\Delta}e^{-X_1}\right)
\left(\int_0^\infty dX_2\,X_2^{\Delta}e^{-X_2}\right)\\
&=
\bigl(1-e^{-2\pi i\Delta}\bigr)
\Gamma(1-\Delta)\Gamma(1+\Delta)\\
&=2\pi i\Delta e^{-i\pi\Delta}.
\end{align*}
The calculation is initially performed for $0<\Delta<1$;
the last expression gives its analytic continuation.
Thus the Weyl factor and the accompanying phase both follow
from the specified root cycle.

We use the Weyl-symmetric theta-angle branch below.
It is obtained from this expression by continuing
$t^2$ to $t^2-2\pi i$, with the remaining integration cycles
continued accordingly. Since the integrand contains
$e^{t^1\Sigma_1+t^2\Sigma_2}$, the root phase becomes
\[
e^{-i\pi(\Sigma_1-\Sigma_2)}e^{-2\pi i\Sigma_2}
=e^{-i\pi(\Sigma_1+\Sigma_2)}.
\]
This continuation leaves $q^a=e^{-t^a}$ unchanged but fixes
the branch of the periods. The normalization factor
$e^{(t^1+t^2)/2}$ acquires only an overall minus sign.
Absorbing this sign and the constant $2\pi i$ into the
normalization, we obtain
\begin{equation}\nonumber
  \Pi\propto\int_{\mathcal C_{\Sigma}}d\Sigma_1\,d\Sigma_2
  \int_{\Gamma'}\prod_{i=1}^{7}\prod_{a=1}^{2}dX_i^a
  \prod_{\beta=1}^{7}dY_{\beta}\,
  (\Sigma_1-\Sigma_2)
  \exp\left[-W_{\rm eff}+\frac{t^1+t^2}{2}\right],
\end{equation}
where
\begin{align}\nonumber
  W_{\rm eff}={}&
  \Sigma_1\left(-2\sum_{i=1}^{7}\log X_i^1
        -\sum_{\beta=1}^{7}Y_\beta-t^1+i\pi\right)\\\nonumber
  &+\Sigma_2\left(-2\sum_{i=1}^{7}\log X_i^2
        -\sum_{\beta=1}^{7}Y_\beta-t^2+i\pi\right)
  +\sum_{i=1}^{7}\sum_{a=1}^{2}(X_i^a)^2
  +\sum_{\beta=1}^{7}e^{-Y_\beta}.
\end{align}

Define the scalar period of the associated abelian theory by
\begin{equation}\nonumber
  \Pi^{Ab}(t^1,t^2)=
  \int_{\mathcal C_{\Sigma}}d\Sigma_1\,d\Sigma_2
  \int_{\Gamma'}\prod_{i=1}^{7}\prod_{a=1}^{2}dX_i^a
  \prod_{\beta=1}^{7}dY_\beta\,
  \exp\left[-W_{\rm eff}+\frac{t^1+t^2}{2}\right].
\end{equation}
As expected, this abelian period satisfies the Picard-Fuchs equations
\begin{align}\label{ABPFF}
  \left(-\frac{\partial}{\partial t^1}\right)^7\Pi^{Ab}
  &=q^1\left(-\frac{\partial}{\partial t^1}
             -\frac{\partial}{\partial t^2}+1\right)^7\Pi^{Ab},
  \\\nonumber
  \left(-\frac{\partial}{\partial t^2}\right)^7\Pi^{Ab}
  &=q^2\left(-\frac{\partial}{\partial t^1}
             -\frac{\partial}{\partial t^2}+1\right)^7\Pi^{Ab},
\end{align}
where $q^a=e^{-t^a}$.

Integrating out $X_i^a$ and $Y_\beta$, we obtain the Mellin--Barnes
representation
\begin{align}\label{PABP}
  \Pi^{Ab}\propto{}&
  \int_{\mathcal C_\Sigma}d\Sigma_1\,d\Sigma_2\,
  \Gamma^7\left(\Sigma_1+\frac12\right)
  \Gamma^7\left(\Sigma_2+\frac12\right)
  \Gamma^7(-\Sigma_1-\Sigma_2)
  \\\nonumber
  &\times\exp\left[
  t^1\Sigma_1+t^2\Sigma_2
  +\frac{t^1+t^2}{2}
  -i\pi(\Sigma_1+\Sigma_2)\right].
\end{align}
The minus sign in front of $i\pi(\Sigma_1+\Sigma_2)$ follows directly from the $+i\pi$ terms in $W_{\rm eff}$.
The Mellin--Barnes contours are chosen in an initial convergence
region and then deformed according to the Pfaffian-phase residue
prescription.  In particular, the relevant residues arise from
intersections of the poles described below; it is not necessary
to assume that both $\Sigma_1$ and $\Sigma_2$ have positive real
parts at every pole.

The factor $\Gamma(-\Sigma_1-\Sigma_2)$ has poles at
$\Sigma_1+\Sigma_2=n$, with $n\geq0$.  Writing
\begin{equation}\nonumber
  \Sigma_1+\Sigma_2=n+\epsilon_1+\epsilon_2,
\end{equation}
we have
\begin{equation}\nonumber
  \Gamma(-n-\epsilon_1-\epsilon_2)
  =\frac{\pi(-1)^{n+1}}
  {\sin[\pi(\epsilon_1+\epsilon_2)]
   \Gamma(1+n+\epsilon_1+\epsilon_2)}.
\end{equation}
At fixed $\Sigma_1+\Sigma_2$ the other two Gamma functions can
be written as
\begin{align}\nonumber
  &\Gamma^7\left(\Sigma_1+\frac12\right)
   \Gamma^7\left(n+\frac12+\epsilon_1+\epsilon_2-\Sigma_1\right),
  \\\nonumber
  \text{or}\qquad
  &\Gamma^7\left(\Sigma_2+\frac12\right)
   \Gamma^7\left(n+\frac12+\epsilon_1+\epsilon_2-\Sigma_2\right).
\end{align}
The two Weyl-related pole families are therefore specified by
\begin{align}\nonumber
  n+\frac12+\epsilon_1+\epsilon_2-\Sigma_1
    &=-m+\epsilon_2,
  \\\nonumber
  n+\frac12+\epsilon_1+\epsilon_2-\Sigma_2
    &=-m+\epsilon_1,
  \qquad m\geq0.
\end{align}
Equivalently, the two branches are
\begin{align}\nonumber
  &(\Sigma_1,\Sigma_2)
   =\left(n+m+\frac12+\epsilon_1,
           -m-\frac12+\epsilon_2\right),
  \\\nonumber
  \text{or}\qquad
  &(\Sigma_1,\Sigma_2)
   =\left(-m-\frac12+\epsilon_1,
           n+m+\frac12+\epsilon_2\right).
\end{align}
They are exchanged by $1\leftrightarrow2$.

We now use the reflection formula
\begin{equation}\nonumber
  \Gamma(-m+\epsilon_a)
  =\frac{\pi(-1)^m}
  {\sin(\pi\epsilon_a)\,\Gamma(1+m-\epsilon_a)},
  \qquad a=1,2.
\end{equation}
For the first pole family we thus find, up to an overall constant,
\begin{align}\label{PfAbResidues}
  \Pi^{Ab}\propto{}&
  \sum_{m,n\geq0}\int d\epsilon_1\,d\epsilon_2\,
  \frac{(-1)^{m+n}}
  {\sin^7[\pi(\epsilon_1+\epsilon_2)]
   \sin^7(\pi\epsilon_2)}
  \\\nonumber
  &\times
  \frac{\Gamma^7(1+m+n+\epsilon_1)}
  {\Gamma^7(1+m-\epsilon_2)
   \Gamma^7(1+n+\epsilon_1+\epsilon_2)}
  \\\nonumber
  &\times\exp\Big[
  (t^1-i\pi)(n+m+1+\epsilon_1)
  +(t^2-i\pi)(-m+\epsilon_2)\Big]
  \\\nonumber
  &+(1\leftrightarrow2).
\end{align}
Here $1\leftrightarrow2$ exchanges
$(t^1,\epsilon_1)$ with $(t^2,\epsilon_2)$, leaving $m,n$
unchanged.  In particular, the second term contains
$\Gamma^7(1+m+n+\epsilon_2)$ and
$\Gamma^7(1+m-\epsilon_1)$, together with the exponential
\begin{equation}\nonumber
  \exp\Big[(t^1-i\pi)(-m+\epsilon_1)
           +(t^2-i\pi)(n+m+1+\epsilon_2)\Big].
\end{equation}
The two residue expansions are understood by analytic continuation
to the Weyl-invariant locus $t^1=t^2=t$.

The factorial structure of these residues motivates a
cohomology-valued abelian generating series.  In passing to this
series, we use the conventional normalization in which the
$m,n$-independent Gamma-class factor and the common exponential
phase in the formal cohomology variables are removed.  The
Weyl projection is then implemented by
$(-\partial_{t^1}+\partial_{t^2})/(\Sigma_1-\Sigma_2)$.
Using the second residue family as the displayed term, we propose
\begin{align}\label{PfGF}
  \Pi_{\rm Pfaffian}(\Sigma_1,\Sigma_2,t)
  ={}&-\left.\frac{-\partial_{t^1}+\partial_{t^2}}
  {\Sigma_1-\Sigma_2}\right|_{t^1=t^2=t}
  \Bigg[\sum_{m,n\geq0}
  \frac{(-1)^m\displaystyle\prod_{k=1}^{m+n}(\Sigma_2+k)^7}
  {\displaystyle\prod_{k=1}^{n}(\Sigma_1+\Sigma_2+k)^7
   \displaystyle\prod_{k=1}^{m}(-\Sigma_1+k)^7}
  \\\nonumber
  &\quad\times\exp\Bigg[
    \frac{t^1+t^2}{2}(\Sigma_1+\Sigma_2)
   +\frac{t^1-t^2}{2}(\Sigma_1-\Sigma_2)
  \\\nonumber
  &\hspace{4.7cm}
   +t^2(n+1)+(t^2-t^1)m\Bigg]
   +(1\leftrightarrow2)\Bigg].
\end{align}
The $1\leftrightarrow2$ operation in this last formula exchanges
$(t^1,\Sigma_1)$ with $(t^2,\Sigma_2)$.  The overall
minus sign in $\Pi_{\rm Pfaffian}$ is chosen to agree
with the large-radius fundamental period.

To determine the mirror map for the Pfaffian Calabi-Yau threefold, we expand
\begin{equation}\nonumber
   \Pi_{\rm{Pfaffian}}=\Pi_{0}(t)+\Pi_{1}(t)\left(-\Sigma_{1}-\Sigma_{2}\right)+\ldots.
\end{equation}
The operator $(-\partial_{t^{1}}+\partial_{t^{2}})/(\Sigma_{1}-\Sigma_{2})$ produces the factors
\begin{equation}\nonumber
  -1+\frac{2m+n+1}{\Sigma_{1}-\Sigma_{2}},\qquad -1-\frac{2m+n+1}{\Sigma_{1}-\Sigma_{2}},
\end{equation}
for the displayed term and its Weyl image, respectively. The required expansions are
\begin{align}\nonumber
  \prod^{m}_{k=1}&\left(\frac{k+\Sigma_{2}}{k-\Sigma_{1}}\right)^{7}=\prod^{m}_{k=1}\left(\left(1+\frac{\Sigma_{2}}{k}\right)\cdot\left(1+\frac{\Sigma_{1}}{k}+\frac{\Sigma^{2}_{1}}{k^{2}}+\ldots\right)\right)^{7}
\\\nonumber
 &=\prod^{m}_{k=1}\left(1+\frac{\Sigma_{1}+\Sigma_{2}}{k}+\frac{\Sigma_{1}\cdot\Sigma_{2}}{k^{2}}+\frac{\left(\frac{\Sigma_{1}+\Sigma_{2}}{2}+\frac{\Sigma_{1}-\Sigma_{2}}{2}\right)^{2}}{k^{2}}+\ldots\right)^{7}
  \\\nonumber
  &=1+7(\Sigma_1+\Sigma_2)H_m+\frac{49}{2}(\Sigma_1+\Sigma_2)^2H_m^2\\\nonumber
  &\quad+\frac{7}{2}(\Sigma_1+\Sigma_2)(\Sigma_1-\Sigma_2)\sum_{k=1}^m\frac{1}{k^2}+\cdots,
\end{align}
\begin{align}\nonumber
  \frac{1}{\prod^{n}_{k=1}\left(k+\Sigma_{1}+\Sigma_{2}\right)^{7}}=&\frac{1}{\left(n!\right)^{7}}\prod^{n}_{k=1}\left[\frac{1}{1+\frac{\Sigma_{1}+\Sigma_{2}}{k}}\right]^{7}
  \\\nonumber
  &=\frac{1}{\left(n!\right)^{7}}\left(1-7H_{n}\left(\Sigma_{1}+\Sigma_{2}\right)+\ldots\right)
\end{align}
and
\begin{equation}\nonumber
  \prod^{m+n}_{k=m+1}\left(k+\Sigma_{2}\right)^{7} = \left(\frac{\left(m+n\right)!}{m!}\right)^{7}\left(1+7\left(H_{m+n}-H_{m}\right)\left(\frac{\Sigma_{1}+\Sigma_{2}}{2}-\frac{\Sigma_{1}-\Sigma_{2}}{2}\right)+\ldots\right).
\end{equation}
Equation~\eqref{PfGF} then gives
\begin{align}\label{PfP0}
  \Pi_{0}(t)&=\sum_{n,m=0}^{\infty}\frac{(-1)^{m}\left(\left(m+n\right)!\right)^{7}}{\left(n!\right)^{7}\left(m!\right)^{7}}\left(1+\frac{7}{2}\left(2m+n+1\right)\left(H_{m+n}-H_{m}\right)\right)
  \\\nonumber
  &\cdot\left.\left(e^{t^{2}(n+1)}e^{\left(t^{2}-t^{1}\right)m}+e^{t^{1}(n+1)}e^{\left(t^{1}-t^{2}\right)m}\right)\right|_{t^{1}=t^{2}=t},
\end{align}
and
\begin{align}\label{PfP1}
  \Pi_{1}(t)&=-t\Pi_{0}(t)-\sum_{n,m=0}^{\infty}\frac{(-1)^{m}\left(\left(m+n\right)!\right)^{7}}{\left(n!\right)^{7}\left(m!\right)^{7}}\Bigg(\Bigg.\left(7H_{m}-7H_{n}+\frac{7\left(H_{n+m}-H_{m}\right)}{2}\right)
  \\\nonumber
  &-\frac{7}{2}\left(2m+n+1\right)\left(\sum^{m}_{k=1}\frac{1}{k^{2}}\right)+\frac{49\left(2m+n+1\right)\left(H_{m}-H_{n}\right)\left(H_{m+n}-H_{m}\right)}{2}\\\nonumber &+\frac{7}{4}\left(2m+n+1\right)\left(7\left(H_{m+n}-H_{m}\right)^{2}-\sum^{m+n}_{k=m+1}\frac{1}{k^{2}}\right)\Bigg.\Bigg)
  \\\nonumber
  &\cdot\left.\left(e^{(n+1)t^{2}}e^{\left(t^{2}-t^{1}\right)m}+e^{(n+1)t^{1}}e^{\left(t^{1}-t^{2}\right)m}\right)\right|_{t^{1}=t^{2}=t}.
\end{align}
At each fixed degree $n$, we evaluate the alternating sum over $m$ on the diagonal by inserting a factor $e^{m\delta}$ in each Weyl term separately, with real $\delta<0$ and taking $\delta\to0^-$ after summation. We use the identities
\begin{align}\nonumber
   \frac{-e^{\delta}}{1+e^{\delta}}&=\sum^{\infty}_{m=1}(-1)^m\exp\left(m\delta\right),
   \\\nonumber
    \left(\frac{-e^{\delta}}{1+e^{\delta}}\right)^{(k)}&=\sum^{\infty}_{m=1}m^{k}(-1)^m\exp\left(m\delta\right),
   \\\nonumber
    \frac{d \mathrm{Li}_{2}(z)}{dz}&=-\frac{\ln\left(1-z\right)}{z}.
\end{align}
Here $\mathrm{Li}_{2}(z)=\sum_{k=1}^{\infty}z^{k}/k^{2}$ is the dilogarithm, initially defined by this series for $|z|<1$, and the superscript $(k)$ denotes the $k$th derivative with respect to $\delta$.

The first several terms of $\Pi_{0}(t)$ and $\Pi_{1}(t)$ are
\begin{align}\label{PfP01}
  \Pi_{0}\left(t\right)&=q^{-1}-17q^{-2}+1549q^{-3}- 215585q^{-4}+36505501q^{-5}+\ldots,
  \\\nonumber
  \Pi_{1}\left(t\right)&=-t\Pi_{0}+70q^{-2}-7413q^{-3}+ \frac{3268573}{3}q^{-4}-\frac{1138372375}{6}q^{-5}+\ldots.
\end{align}
The mirror map is
\begin{equation}\nonumber
  z^{-1}=e^{\tau}=e^{-\frac{\Pi_{1}\left(t\right)}{\Pi_{0}\left(t\right)}-2i\pi t_{0}}.
\end{equation}
Choosing $t_0=1/2$ and inverting the mirror map gives
\begin{equation}\label{MPPF}
  q^{-1}=-z^{-1}+70z^{-2}-1127z^{-3}+47530z^{-4}+{\cal O}(z^{-5}).
\end{equation}
Recall that the Yukawa coupling is
\begin{equation}\nonumber
\langle(-\Sigma_1-\Sigma_2)^3\rangle
=
\frac{14q^{-2}(1+3q^{-1})}
{1+289q^{-1}-57q^{-2}-q^{-3}}.
\end{equation}
This follows from Eq.~\eqref{CLGR27} by analytically
continuing to the Pfaffian phase and taking into account
the sign of the insertion $(-\Sigma_1-\Sigma_2)^3$.
Dividing by $\Pi_0(t)^2$ and including the
logarithmic-coordinate Jacobian in Eq.~\eqref{CLATB},
we obtain
\begin{equation}\label{AGWPf}
K_{z^{-1}z^{-1}z^{-1}}
=
14+588z^{-1}+97412z^{-2}
+15765456z^{-3}
+2647082116z^{-4}+\cdots.
\end{equation}
Consequently, the first four instanton numbers of the Pfaffian Calabi-Yau threefold are
 \begin{equation}\label{GWPf}
  588,\quad 12103,\quad 583884,\quad 41359136.
\end{equation}

These instanton numbers agree with \cite{Rødland:1998,Hosono:2007vf}. The generating function proposed in Eq.~\eqref{PfGF} appears to be new. It would therefore be interesting to give it a precise mathematical interpretation.

\subsection{The determinantal Gulliksen-Neg{\aa}rd Calabi-Yau threefold}

Our final example is the Gulliksen-Neg{\aa}rd Calabi-Yau threefold \cite{Gulliksen:1971}, described by a GLSM with two physical K\"ahler parameters. The GLSM was studied in \cite{Jockers:2012zr}. Its gauge group is $U(1)\times U(2)$, with twisted chiral field-strength superfields denoted by $\Sigma_0$ and $\Sigma$, respectively. We write $\Sigma_1$ and $\Sigma_2$ for the Cartan components of $\Sigma$. The matter content consists of eight chiral superfields $\Phi_\mu$ of $U(1)$ charge $+1$, four chiral superfields $P^i$ in a bifundamental representation of $U(1)\times U(2)$, and four chiral superfields $X_i$ in the antifundamental representation of $U(2)$. The $U(1)$ charges of $P^i$ and $X_i$ are $-1$ and $0$, respectively. The superpotential is
\begin{equation}\nonumber
  W={\rm tr} \left(PA(\Phi)X\right).
\end{equation}
Here $A(\Phi)=\sum_{\mu=1}^{8}A^{\mu}\Phi_{\mu}$, where the $A^{\mu}$ are constant $4\times4$ matrices. This superpotential gives the PAX model its name \cite{Jockers:2012zr}. We denote the auxiliary Cartan FI parameters by $r^0,r^1,r^2$ and recover the two physical parameters by setting $r^1=r^2=r$. We focus on geometric phase I, in the region $r^0+r^1+r^2>0$, $r^1>0$, and $r^2>0$.

The mirror Landau-Ginzburg model has the superpotential
\begin{align}\label{GNMLG}
  W=&\Sigma_{0}\left(\sum^{8}_{\mu=1}Y_{\mu}-\sum^{4}_{i=1}\left(Y^{1}_{P_{i}}+Y^{2}_{P_{i}}\right)-t^{0}\right)
  \\\nonumber
  +&\Sigma_{1}\left(\sum^{4}_{i=1}Y^{1}_{P_{i}}+\sum^{4}_{i=1}\log X^{1}_{i}+\log X_{1}-\log X_{2}-t^{1}\right)
  \\\nonumber
  +&\Sigma_{2}\left(\sum^{4}_{i=1}Y^{2}_{P_{i}}+\sum^{4}_{i=1}\log X^{2}_{i}+\log X_{2}-\log X_{1}-t^{2}\right)
  \\\nonumber
  +&\sum^{8}_{\mu=1}e^{-Y_{\mu}}+\sum^{4}_{i=1}\left(e^{-Y^{1}_{P_{i}}}+e^{-Y^{2}_{P_{i}}}\right)+\sum^{4}_{i=1}\left(X^{1}_{i}+X^{2}_{i}\right)+X_{1}+X_{2}.
\end{align}

The same procedure gives the following mirror generating series:
 \begin{align}\label{GNMP}
   \Pi_{GN} & \left(\Sigma_0,\Sigma_{1}, \Sigma_{2}, t^{0},t\right)
    \\\nonumber
    &=\left.\frac{-\partial_{t^{1}}+\partial_{t^{2}}}{\Sigma_{1}-\Sigma_{2}}\right|_{t^{1}=t^{2}=t}\sum^{\infty}_{d,d_{1},d_{2}=0}
    \frac{\prod^{d_{1}+d}_{k=1}\left(\Sigma_{1}+k\right)^{4}\prod^{d_{2}+d}_{k=1}\left(\Sigma_{2}+k\right)^{4}}{\prod^{d}_{k=1}\left(\Sigma_{0}+k\right)^{8}\prod^{d_{1}}_{k=1}\left(\Sigma_{1}-\Sigma_{0}+k\right)^{4}\prod^{d_{2}}_{k=1}\left(\Sigma_{2}-\Sigma_{0}+k\right)^{4}}
    \\\nonumber
    &\cdot \exp\left(-\left(\Sigma_{0}+d\right)t^{0}-\left(\Sigma_{1}+d+d_{1}\right)\left(t^{1}+i\pi\right)-\left(\Sigma_{2}+d+d_{2}\right)\left(t^{2}+i\pi\right)\right).
    \end{align}

As discussed in \cite{Jockers:2012zr}, the divisor classes represented by $\Sigma_0$ and $\Sigma_1+\Sigma_2-2\Sigma_0$ form a basis of $H^{1,1}$. Introduce the variables
  \begin{equation}\nonumber
    \widetilde{\Sigma}_{1}\equiv\Sigma_{1}-\Sigma_{0},\qquad \widetilde{\Sigma}_{2}\equiv\Sigma_{2}-\Sigma_{0}.
  \end{equation}
  With these variables, Eq.~\eqref{GNMP} becomes
  \begin{align}\label{GNMP2}
   \Pi_{GN}&\left(\Sigma_0,\Sigma_{1}, \Sigma_{2}, t^{0},t\right)
   \\\nonumber
    &=\left.\frac{-\partial_{t^{1}}+\partial_{t^{2}}}{\widetilde{\Sigma}_{1}-\widetilde{\Sigma}_{2}}\right|_{t^{1}=t^{2}=t}\sum^{\infty}_{d,d_{1},d_{2}=0}
    \frac{\prod^{d_{1}+d}_{k=1}\left(\widetilde{\Sigma}_{1}+\Sigma_{0}+k\right)^{4}\prod^{d_{2}+d}_{k=1}\left(\widetilde{\Sigma}_{2}+\Sigma_{0}+k\right)^{4}}{\prod^{d}_{k=1}\left(\Sigma_{0}+k\right)^{8}\prod^{d_{1}}_{k=1}\left(\widetilde{\Sigma}_{1}+k\right)^{4}\prod^{d_{2}}_{k=1}\left(\widetilde{\Sigma}_{2}+k\right)^{4}}
    \\\nonumber
    &\cdot \exp\left(-\left(\Sigma_{0}+d\right)\left(t^{0}+t^{1}+t^{2}+2i\pi\right)-\left(\widetilde{\Sigma}_{1}+d_{1}\right)\left(t^{1}+i\pi\right)-\left(\widetilde{\Sigma}_{2}+d_{2}\right)\left(t^{2}+i\pi\right)\right).
    \end{align}
Here the summation variables correspond to the Cartan fluxes $(d,d+d_1,d+d_2)$.

We expand $\Pi_{GN}(\Sigma_0,\Sigma_1,\Sigma_2,t^0,t)$ in the divisor basis:
    \begin{equation}\nonumber
      \Pi_{GN}\left(\Sigma_0,\Sigma_{1}, \Sigma_{2}, t^{0},t\right)=\Pi_{0}\left(t^{0},t\right)+\Pi_{11}\left(t^{0},t\right)\Sigma_{0}+\Pi_{12}\left(t^{0},t\right)\left(\Sigma_{1}+\Sigma_{2}-2\Sigma_{0}\right)+\cdots.
    \end{equation}

  Note the following expansions, with $a=1,2$,
  \begin{align*}\nonumber
     \prod^{d}_{k=1}&\left(\Sigma_{0}+k\right)^{-8}=\frac{1}{\left(d!\right)^{8}}\left(1-8\Sigma_{0}H_{d}+\cdots\right),  \\\nonumber
     \prod^{d_{a}}_{k=1}&\left(\widetilde{\Sigma}_{a}+k\right)^{-4}=\frac{1}{\left(d_{a}!\right)^{4}}\left(1-4\widetilde{\Sigma}_{a}H_{d_{a}}+\widetilde{\Sigma}^{2}_{a}\left(8H^{2}_{d_{a}}+2\sum^{d_{a}}_{k=1}\frac{1}{k^{2}}\right)+\cdots\right),\\\nonumber
      \prod^{d_{a}+d}_{k=1}&\left(\widetilde{\Sigma}_{a}+\Sigma_{0}+k\right)^{4}=\left(\left(d_{a}+d\right)!\right)^{4}\Bigl(1+4\left(\widetilde{\Sigma}_{a}+\Sigma_{0}\right)H_{d_{a}+d}\\\nonumber &\qquad+\left(\widetilde{\Sigma}_{a}+\Sigma_{0}\right)^{2}\left(8H^{2}_{d_{a}+d}-2\sum^{d_{a}+d}_{k=1}\frac{1}{k^{2}}\right)+\cdots\Bigr).
  \end{align*}

  We obtain the following three periods.
  \begin{align}\label{GNP00}
    \Pi_{0} & =\sum^{\infty}_{d,d_{1},d_{2}=0}\frac{\left(\left(d_{1}+d\right)!\right)^{4}\left(\left(d_{2}+d\right)!\right)^{4}}{\left(d!\right)^{8}\left(\left(d_{1}\right)!\right)^{4}\left(\left(d_{2}\right)!\right)^{4}}\cdot e^{-d(t^{0}+2t)}e^{-\left(d_{1}+d_{2}\right)\left(t+i\pi\right)} \\\nonumber
     &\cdot\left(1+2\left(d_{1}-d_{2}\right)\left(H_{d_{1}+d}-H_{d_{2}+d}-H_{d_{1}}+H_{d_{2}}\right)\right),
  \end{align}

  \begin{align}\label{GNP11}
\Pi_{11}
={}&-(t^{0}+2t+2i\pi)\Pi_{0}
+\sum_{d,d_{1},d_{2}=0}^{\infty}
\frac{[(d_{1}+d)!]^{4}[(d_{2}+d)!]^{4}}
{(d!)^{8}(d_{1}!)^{4}(d_{2}!)^{4}}
e^{-d(t^{0}+2t)}e^{-(d_{1}+d_{2})(t+i\pi)}
\nonumber\\
&\times\Bigg[
4H_{d_{1}+d}+4H_{d_{2}+d}-8H_d
+(d_{1}-d_{2})\Bigg(
8\left(H_{d_{1}+d}^{2}-H_{d_{2}+d}^{2}\right)
-2\left(
\sum_{k=1}^{d_{1}+d}\frac{1}{k^{2}}
-\sum_{k=1}^{d_{2}+d}\frac{1}{k^{2}}
\right)
\nonumber\\
&\qquad
+16H_d\left(
H_{d_{1}}-H_{d_{2}}
-H_{d_{1}+d}+H_{d_{2}+d}
\right)
+8\left(H_{d_{1}+d}+H_{d_{2}+d}\right)
\left(H_{d_{2}}-H_{d_{1}}\right)
\Bigg)
\Bigg],
\end{align}

\begingroup\footnotesize 
\begin{align}\label{GNP12}
\Pi_{12}
={}&-\left(t+i\pi\right)\Pi_{0}
+\sum_{d,d_{1},d_{2}=0}^{\infty}
\frac{[(d_{1}+d)!]^{4}[(d_{2}+d)!]^{4}}
{(d!)^{8}(d_{1}!)^{4}(d_{2}!)^{4}}
e^{-d(t^{0}+2t)}e^{-(d_{1}+d_{2})(t+i\pi)}
\nonumber\\
&\times\Bigg[
2H_{d_{1}+d}+2H_{d_{2}+d}
-2H_{d_{1}}-2H_{d_{2}}+(d_{1}-d_{2})\Bigg(
4\left(H_{d_{1}+d}^{2}-H_{d_{2}+d}^{2}\right)
-\left(
\sum_{k=1}^{d_{1}+d}\frac{1}{k^{2}}
-\sum_{k=1}^{d_{2}+d}\frac{1}{k^{2}}
\right)
\nonumber\\
&\qquad
+4\left(H_{d_{1}}^{2}-H_{d_{2}}^{2}\right)
+\left(
\sum_{k=1}^{d_{1}}\frac{1}{k^{2}}
-\sum_{k=1}^{d_{2}}\frac{1}{k^{2}}
\right)
-8\left(
H_{d_{1}+d}H_{d_{1}}
-H_{d_{2}+d}H_{d_{2}}
\right)
\Bigg)
\Bigg].
\end{align}
\endgroup

Following \cite{Jockers:2012zr}, set $w=-q$ and $z=q^0w^{2}$. We choose the branch $\log z=-(t^0+2t+2i\pi)$ for the logarithmic periods. The period expansions are
 \begingroup\small 
\begin{align}\nonumber
   \Pi_{0}\left(w, z\right)&=1+2w+z+3{w}^{2}+z^{2}+4{w}^{3}-14{w}^{2}z-54{w} z^{2}+z^{3}+\ldots,
   \\\nonumber
   \Pi_{11}\left(w, z\right) &=-\left(t^{0}+2t+2i\pi\right)\Pi_{0}+4 {w} + 10 {w}^{2} + \frac{52}{3} {w}^{3} - 16 {w}z - 110 {w}^{2} z - 108 {w}z^{2}+\ldots,
   \\\nonumber
   \Pi_{12}\left(w, z\right) &=-\left(t+i\pi\right)\Pi_{0}+4 z + 6 z^{2} + 24 {w}z + \frac{22}{3} z^{3} + 80 {w}^{2}z - 108 {w}z^{2}+\ldots.
 \end{align}
\endgroup

With $\log(-w)=-t$, we take $\tau^0=-\Pi_{11}/\Pi_0$ and $\tau^1=-\Pi_{12}/\Pi_0-i\pi$, corresponding to $t_0^0=0$ and $t_0^1=1/2$ in Eq.~\eqref{FCFP}. The resulting mirror maps are
  \begin{align}\label{GNMM}
     \tau^{0}&=-\log z-4 {w} - 2 {w}^{2} - \frac{4}{3} {w}^{3} + 20 {w}z + 72 {w}^{2} z + 92 {w}z^{2}+\ldots,  \\
     \tau^{1}&=-\log\left( {-w}\right)-4 z - 2 z^{2} - \frac{4}{3} z^{3} - 16 {w}z - 36 {w}^{2} z +128 {w}z^{2}+\ldots.
  \end{align}

These flat coordinates agree with those obtained from the two-sphere partition function in \cite{Jockers:2012dk}. Combining them with the Yukawa couplings computed in Section~8.3 of \cite{Closset:2015rna}, using the period normalization and coordinate transformation in Eq.~\eqref{CLATB}, gives the following couplings in the flat variables $z^0=e^{-\tau^0}$ and $z^1=e^{-\tau^1}$
  \begin{align*}
 K_{z^0 z^0 z^0}={}&20+56z^0+56(z^0)^2+192z^0z^1
+56(z^0)^3+7168(z^0)^2z^1+56z^0(z^1)^2+\cdots,
  \\[4pt]
  K_{z^0 z^0 z^1}={}&20+192z^0z^1             +3584(z^0)^2z^1+112z^0(z^1)^2+\cdots,
  \\[4pt]
  K_{z^0 z^1 z^1}={}&16+192z^0z^1             +1792(z^0)^2z^1+224z^0(z^1)^2+\cdots,
  \\[4pt]
  K_{z^1 z^1 z^1}={}&8+192z^0z^1            +896(z^0)^2z^1+448z^0(z^1)^2+\cdots,
  \end{align*}
  from which the instanton numbers can be extracted:
 \begin{equation}
n_{1,0}=56,\qquad n_{1,1}=192,\qquad
  n_{2,1}=896,\qquad n_{1,2}=56, \quad \cdots.
 \end{equation}

These instanton numbers agree with the two-sphere partition function computation of \cite{Jockers:2012dk}.

\section{Conclusions and future directions}\label{cfd}
 
In this paper, we have proposed quantum differential equations for the periods~\cite{Hori:2000ck} of the nonabelian mirror theories of \cite{Gu:2018fpm}. The equations are initially formulated in auxiliary Cartan parameters, whose number equals the rank of the GLSM gauge group, and are then reduced to the physical K\"ahler parameters. We have checked the proposal by recovering the known quantum differential equations of Grassmannians and the Picard-Fuchs equations of Calabi-Yau complete intersections in Grassmannians \cite{Batyrev:1998kx}. We have also applied the construction to symplectic Grassmannians $SG(k,2n)$ and flag manifolds $Fl(k_1,\ldots,k_n,N)$. The resulting equations determine cohomology-valued generating functions for the mirrors of Fano varieties such as $Gr(k,N)$ and $SG(k,2n)$, in agreement with \cite{Bertram:2004}. For Calabi-Yau targets, we have proposed generating functions for complete intersections in Grassmannians and for the Pfaffian and Gulliksen-Neg{\aa}rd threefolds. The resulting instanton numbers agree with \cite{Batyrev:1998kx,Rødland:1998,Hosono:2007vf,Jockers:2012dk}. To the best of our knowledge, the generating functions proposed here for the two non-complete-intersection examples are new.

As reviewed in Section~\ref{LSMMS}, the nonabelian GLSMs considered here admit an effective description in terms of the associated Cartan theory \cite{Halverson:2013eua}. This perspective is related to the abelian/nonabelian correspondence. At the level of cohomology, see \cite{Martin:2000}; the genus-zero Gromov-Witten correspondence was conjectured in \cite{Bertram:2004}. The $I$-functions of nonabelian GIT quotients have also been studied using this correspondence \cite{Rachel:2018}. Our non-complete-intersection examples suggest that it would be useful to formulate a precise mathematical conjecture for their generating functions and to establish an appropriate nonabelian mirror theorem. Another direction concerns higher genus. Mathematical constructions of higher-genus GLSM invariants are available \cite{Ciocan:2016,Fan:2017,Tian:2018,Zhou:2019}, whereas a general physical treatment remains to be developed. Combining our approach with tools such as those of \cite{Bershadsky:1993cx} may provide a route to higher-genus invariants.

Our periods are defined in mirror Landau-Ginzburg models \cite{Hori:2000ck}. Although the corresponding A-cycles depend on the K\"ahler moduli, the periods are independent of those moduli. They may therefore help identify geometric mirrors of Calabi-Yau targets described by nonabelian GLSMs. Such a realization would provide further geometric insight into nonabelian mirror symmetry \cite{Strominger:1996it}. If the mirror Calabi-Yau manifold also admits a GLSM description, exact computations in the two gauge theories may reveal additional dualities.

Finally, heterotic compactifications provide a further motivation for extending these methods \cite{Candelas:1985en}. Progress on heterotic mirror symmetry includes \cite{Melnikov:2010sa,Melnikov:2012hk,Gu:2017nye,Gu:2019byn}, but a general framework remains incomplete. It would be interesting to determine whether the period methods developed here can contribute to a suitable heterotic analogue of Gromov-Witten theory.

\section*{Acknowledgements}
We thank Mauricio Romo for useful discussions and comments. WG would like to thank Sheldon Katz, Dave Morrison, and Shing-Tung Yau for sharing the early history of mirror symmetry with him. WG is supported by the National Natural Science Foundation
of China (Grant No.~12575077).   JG acknowledges support from the National Natural Science Foundation of China (Grant No.~12475005), the Natural Science Foundation of Shanghai Municipality (Grant No.~24ZR1468600), and the Fundamental Research Funds for the Central Universities. YW is supported by a start-up grant from Zhejiang University. 

\section*{Statement of AI usage}

We carried out the original research presented in this paper during
the COVID-19 pandemic, before modern generative AI tools became
widely available. The pandemic limited our opportunities to discuss
the work with experts in related areas of mathematics and physics.
This may have affected the clarity of our original presentation.

In the years that followed, we received valuable feedback from
several experts. Their comments helped us recognize where our
terminology and exposition needed clarification. In particular,
we needed to explain more precisely how we distinguish quantum
differential operators from Picard--Fuchs operators and how the
two are related. We therefore began planning a revision well before
we started using AI tools for this work. Around 2024, one of us,
Jirui Guo, suggested submitting the paper to a journal. This gave
us a further reason to revise the manuscript systematically.
Although progress was slow, we carried out most of the planned
revisions ourselves.

We developed the original ideas, obtained the principal results,
and reached the main conclusions of this work without AI assistance.
These ideas, results, and conclusions already appear in the first
version of the manuscript posted on arXiv.

During the later stages of revision, we used AI tools to proofread
the manuscript and to check selected formulas and sign conventions.
These tools included GPT-6 Astra, accessed through Codex.
This helped us identify typographical errors and inconsistencies
that we had previously overlooked.

We also acknowledge a technical suggestion from AI concerning the
choice of integration cycle for the root variables in the A-brane
period integral. For each positive root $\alpha$, our prescription
in the chosen real chamber is to integrate $X_\alpha$ along the
positive real axis and $X_{-\alpha}$ along a Hankel contour. We
examined this choice independently before adopting it in place of
steps in our earlier derivation that we had not adequately justified.

The authors take full responsibility for all arguments, calculations,
and conclusions in the revised manuscript.

\appendix

\section{Quantum differential equations of mirrors of toric varieties}\label{APFMT}

In this appendix, we review the QDEs satisfied by periods of mirrors of toric varieties. An abelian GLSM has $k$ twisted chiral field-strength superfields $\Sigma_a$ and $N$ matter fields $\Phi_i$, with charge matrix $Q_i^a$. Its mirror is a Landau-Ginzburg theory on $(\mathbb{C}^{\star})^{N}\times\mathbb{C}^{k}$ with superpotential
\begin{equation} \nonumber
  W=\Sigma_{a}\left(Q^{a}_{i}Y_{i}-t^{a}\right)+\sum_{i}e^{-Y_{i}}.
\end{equation}
Following Eq.~\eqref{0PER}, the period for a cycle $\gamma_\ell$ is
\begin{equation}\nonumber
  \Pi_{1,\ell}=\int_{\gamma_{\ell}}\prod_{a}d\Sigma_{a}\prod_{i}dY_{i}e^{-\Sigma_{a}\left(Q^{a}_{i}Y_{i}-t^{a}\right)-\sum_{i}e^{-Y_{i}}}.
\end{equation}
Integrating out the $\Sigma$ fields, we have
\begin{equation}\nonumber
   \Pi_{1,\ell}=\int\prod_{i}dY_{i}\prod_{a}\delta\left(\sum_i Q^{a}_{i}Y_{i}-t^{a}\right)\exp\left(-\sum_{i}e^{-Y_{i}}\right).
\end{equation}
We have suppressed the dependence of the integration cycles on the A-brane data discussed in Section~\ref{AILGM}.

Using the global symmetry of the theory, we introduce auxiliary parameters as in \cite{Hori:2000kt}\footnote{This construction can also be understood through T-duality for GLSMs with an $\Omega$-deformation \cite{Nekrasov:2002qd,Losev:1999tu}.}:
\begin{equation} \nonumber
  \Pi(\mu_i,t^a)\equiv\Pi_{1,\ell}(\mu_i,t^a)=\int\prod_{i}dY_{i}\prod_{a}\delta\left(\sum_i Q^{a}_{i}Y_{i}-t^{a}\right)\exp\left(-\sum_{i}\mu_{i}e^{-Y_{i}}\right).
\end{equation}
For each fixed $a$, differentiating the exponential gives
\begin{equation}\label{PFMT}
  \left[\prod_{Q_i^a>0}\left(\frac{\partial}{\partial\mu_i}\right)^{Q_i^a}\right]\Pi(\mu_j,t^a)
  =(-1)^{\sum_iQ_i^a}e^{-t^a}
  \left[\prod_{Q_i^a<0}\left(\frac{\partial}{\partial\mu_i}\right)^{-Q_i^a}\right]\Pi(\mu_j,t^a).
\end{equation}
The relative sign comes from the minus sign contributed by each $\mu_i$ derivative. The index $a$ labels the equation and is not summed.
A shift of each $Y_i$ by $\log\mu_i$ removes the $\mu_i$ dependence from the exponential and shifts all the delta-function constraints. Thus,
\[
  \Pi(\mu_i,t^a)=\Pi(1,T^a),
  \qquad T^b=t^b-\sum_iQ_i^b\log\mu_i.
\]
Define
\[
  q_b=e^{-t^b},\qquad
  \theta_b=-\frac{\partial}{\partial t^b},\qquad
  D_i=\sum_{b=1}^{k}Q_i^b\theta_b,
  \qquad s_a=\sum_{Q_i^a>0}Q_i^a.
\]
The chain rule gives
\[
  \left(\frac{\partial}{\partial\mu_i}\right)^n\Pi(\mu_j,t^a)
  =\mu_i^{-n}\prod_{m=0}^{n-1}(D_i-m)\Pi(\mu_j,t^a).
\]
In particular, a field charged under several $U(1)$ factors contributes the full operator $D_i$.
Setting $\mu_j=1$, Eq.~\eqref{PFMT} becomes
\begin{equation}\label{PFMT2}
  \left[\prod_{Q_i^a>0}\prod_{m=0}^{Q_i^a-1}(D_i-m)\right]\Pi(t^a)
  =(-1)^{s_a}q_a
  \left[\prod_{Q_j^a<0}\prod_{\ell=0}^{-Q_j^a-1}(-D_j+\ell)\right]\Pi(t^a).
\end{equation}
These are the quantum differential equations satisfied by the mirror periods in the coordinate $q_a=e^{-t^a}$ used in the displayed integral. The phase-shifted coordinate $\widehat q_a=(-1)^{s_a}q_a$ absorbs the explicit sign, while $\widehat q_a\partial_{\widehat q_a}=\theta_a$.
For the charges $(-3,1,1,1)$ in \cite[Section~6.1]{Hori:2000kt}, Eq.~(6.8) of that reference has a positive coefficient. At $\mu_i=1$, its third derivative with respect to $\mu_1$ gives $(-3\theta)(-3\theta-1)(-3\theta-2)$. Thus its Eq.~(6.9), in the coordinate $q=e^{-t}$, reads $\theta^3\Pi=-q(3\theta+2)(3\theta+1)(3\theta)\Pi$.

\Needspace{5\baselineskip}
\noindent \textbf{Example: $\mathbb{CP}^{4}$}

The GLSM for $\mathbb{CP}^{4}$ contains one twisted chiral field-strength superfield $\Sigma$ and five chiral superfields with gauge charges $(1,1,1,1,1)$ and vector R-charge 0. Equation~\eqref{PFMT} gives
\begin{equation}\nonumber
 \prod^{5}_{i=1} \frac{\partial}{\partial \mu_{i}}\Pi=-e^{-t}\Pi.
\end{equation}
With $\theta=-d/dt$, the delta-function constraint implies that the period depends on the $\mu_i$ through $t-\log\bigl(\prod_{i=1}^{5}\mu_i\bigr)$. The quantum differential equation is therefore
\begin{equation}\nonumber
  \theta^{5}\Pi=-e^{-t}\Pi.
\end{equation}

\Needspace{5\baselineskip}
\noindent{\textbf{Example:} ${\cal O}(-5)\quad \mathrm{over} \quad \mathbb{CP}^{4}$}

The GLSM for the total space of $\mathcal{O}(-5)$ over $\mathbb{CP}^{4}$ contains one twisted chiral field-strength superfield $\Sigma$ and six chiral superfields with gauge charges $(1,1,1,1,1,-5)$ and vector R-charge 0. Thus,

\begin{equation}\nonumber
 \prod^{5}_{i=1} \frac{\partial}{\partial \mu_{i}}\Pi=e^{-t}\frac{\partial^{5}}{\partial\mu^{5}_{6} }\Pi.
\end{equation}
With $\theta=-d/dt$, the period depends on the $\mu_i$ through $t-\log\bigl((\prod_{i=1}^{5}\mu_i)/\mu_6^{5}\bigr)$. The quantum differential equation is

\begin{equation}\nonumber
  \theta^{5}\Pi=-e^{-t}\left(5\theta+4\right)\left(5\theta+3\right)\left(5\theta+2\right)\left(5\theta+1\right)\left(5\theta\right)\Pi.
\end{equation}

\Needspace{5\baselineskip}
\noindent{\textbf{Example:} ${\cal O}(-4,-4)\quad \mathrm{over} \quad \mathbb{CP}^{3}\times\mathbb{CP}^{3}$}

The GLSM for the total space of $\mathcal{O}(-4,-4)$ over $\mathbb{CP}^{3}\times\mathbb{CP}^{3}$ contains two twisted chiral field-strength superfields $\Sigma_1,\Sigma_2$ and nine chiral superfields with charge matrix
\begin{equation}\nonumber
\left(
  \begin{array}{ccccccccc}
    1 & 1 & 1 & 1 &  &  &  &  & -4\\
     &  &  &  &  1& 1 & 1 & 1 & -4 \\
  \end{array}
\right),
\end{equation}
where $\mu_p$ denotes the auxiliary coefficient of the field with charges $(-4,-4)$. Thus,
\begin{equation}\nonumber
 \prod^{4}_{i=1} \frac{\partial}{\partial \mu_{1i}}\Pi=e^{-t^{1}}\frac{\partial^{4}}{\partial\mu^{4}_{p} }\Pi,\qquad \prod^{4}_{i=1} \frac{\partial}{\partial \mu_{2i}}\Pi=e^{-t^{2}}\frac{\partial^{4}}{\partial\mu^{4}_{p} }\Pi.
\end{equation}
Define $\theta_1=-\partial/\partial t^1$ and $\theta_2=-\partial/\partial t^2$. The period depends on the auxiliary parameters through $t^1-\log\bigl((\prod_{i=1}^{4}\mu_{1i})/\mu_p^4\bigr)$ and $t^2-\log\bigl((\prod_{i=1}^{4}\mu_{2i})/\mu_p^4\bigr)$. The mixed-charge operator is $D_p=-4(\theta_1+\theta_2)$. The quantum differential equations are

\begin{equation}\nonumber
  \theta^{4}_{1}\Pi=e^{-t^{1}}\left(4\left(\theta_{1}+\theta_{2}\right)+3\right)\left(4\left(\theta_{1}+\theta_{2}\right)+2\right)\left(4\left(\theta_{1}+\theta_{2}\right)+1\right)4\left(\theta_{1}+\theta_{2}\right)\Pi,
\end{equation}
\begin{equation}\nonumber
  \theta^{4}_{2}\Pi=e^{-t^{2}}\left(4\left(\theta_{1}+\theta_{2}\right)+3\right)\left(4\left(\theta_{1}+\theta_{2}\right)+2\right)\left(4\left(\theta_{1}+\theta_{2}\right)+1\right)4\left(\theta_{1}+\theta_{2}\right)\Pi.
\end{equation}

\section{Picard-Fuchs equations of abelian mirrors of complete intersections}\label{PFEMS}

As reviewed in Section~\ref{MC}, the fundamental mirror variable for a matter field $\phi_i$ of nonzero vector R-charge $q_i$ is $X_i=\exp(-q_iY_i/2)$. The GLSM for a complete intersection has $k$ twisted chiral field-strength superfields $\Sigma_a$ ($a=1,\ldots,k$), $M$ chiral superfields $\phi_i$ ($1\leq i\leq M$) with charge matrix $Q_i^a$, and $N-M$ chiral superfields $P_j$ ($M<j\leq N$) with charge matrix $\widetilde Q_j^a$. Here $Q_i^a\geq0$ and $\widetilde Q_j^a\leq0$. The superpotential is $W=\sum_{j=M+1}^N P_jG_j$.
In the geometric phase, we assign vector R-charge 2 to the $P_j$ and vector R-charge 0 to the $\phi_i$. The mirror of a GLSM for a complete intersection is an LG model defined on a noncompact Calabi-Yau manifold $\left(\mathbb{C}^{\star}\right)^{M}\times\mathbb{C}^{N-M+k}$ with a superpotential
\begin{equation}\label{SMLG}
  W=\Sigma_{a}\left(Q^{a}_{i}Y_{i}-\widetilde{Q}^{a}_{j}\log X_{j}-t^{a} \right)+\sum^{M}_{i=1}\exp\left(-Y_{i}\right)+\sum^{N}_{j=M+1}X_{j}.
\end{equation}
The period is defined as
\begin{equation}\label{PMLGS}
  \widetilde{\Pi}=\int \prod_{a}d\Sigma_{a}\prod_{i}dY_{i}\prod_{j}dX_{j}e^{-\Sigma_{a}\left(Q^{a}_{i}Y_{i}-\widetilde{Q}^{a}_{j}\log X_{j}-t^{a} \right)-\sum^{M}_{i=1}\exp\left(-Y_{i}\right)-\sum^{N}_{j=M+1}X_{j}}.
\end{equation}
Integrating out the $\Sigma_{a}$ fields, we have
\begin{align}\label{PMLGS2}
  \widetilde{\Pi}=\int \prod_{i}dY_{i}\prod_{j}dX_{j}\prod_{a}&\delta\left(Q^{a}_{i}Y_{i}-\widetilde{Q}^{a}_{j}\log X_{j}-t^{a} \right)
  \\\nonumber
  &\exp\left(-\sum^{M}_{i=1}\exp\left(-Y_{i}\right)-\sum^{N}_{j=M+1}X_{j}\right).
\end{align}
We perform the following transformation:
\begin{equation*}
  Y_i \mapsto Y_i-\log\mu_{i}, \qquad
  X_j \mapsto \mu_{j}X_j , \qquad
  t^a \mapsto t^a- \log\left(\prod_i \mu_{i}^{Q^a_i}\prod_j \mu_{j}^{\widetilde{Q}^{a}_{j}}\right).
\end{equation*}
We then consider
\begin{align}\label{PMLGS3}
 \widetilde{\Pi}(\mu_{i}, t^{a})
 =\int &\prod^{M}_{i=1}dY_{i}
 \prod^{N}_{j=M+1}\left(\mu_{j}dX_{j}\right)
 \prod_{a}\delta\left(
 \sum Q^{a}_{i}Y_{i}
 -\widetilde{Q}^{a}_{j}\log X_{j}
 -t^{a}
 \right)
 \nonumber\\
 &\times
 \exp\left(
 -\sum^{M}_{i=1}\mu_{i}e^{-Y_{i}}
 -\sum^{N}_{j=M+1}\mu_{j}X_{j}
 \right),
\end{align}
where, with a slight abuse of notation, we use the same symbol $t^a$ for
\[
 t^a- \log\left(\prod_i \mu_{i}^{Q^a_i}\prod_j \mu_{j}^{\widetilde{Q}^{a}_{j}}\right).
\]

Let
\[
q_a=e^{-t^a},
\qquad
\varepsilon_a=(-1)^{\sum_{i=1}^{M}Q_i^a},
\qquad
\theta_a=q_a\frac{\partial}{\partial q_a}
=-\frac{\partial}{\partial t^a},
\]
and define
\[
\mathcal Q_i
=\sum_b Q_i^b\theta_b,
\qquad
D_j=\sum_b\widetilde Q_j^b\theta_b,
\qquad
d_j^a=-\widetilde Q_j^a\geq 0.
\]
Here the parameters $t^a$ denote the effective
invariant coordinates, with the coefficients
$\mu_i$ fixed in the chosen coefficient gauge. The sign $\varepsilon_a$ is the factor $(-1)^{s_a}$ in Eq.~\eqref{PFMT2}; we keep $q_a=e^{-t^a}$ without absorbing this phase.

One then finds that
\begin{equation}\label{PMLGS4}
\begin{aligned}
\widetilde{\Pi}(\mu_i,t^a)
&=
\prod_{j=M+1}^{N}
\left(
\mu_j\frac{\partial}{\partial\mu_j}
\right)\Pi(\mu_i,t^a)\\
&=
\left(\prod_{j=M+1}^{N}D_j\right)
\Pi(\mu_i,t^a).
\end{aligned}
\end{equation}

The period $\Pi(\mu_i,t^a)$ is the
toric-mirror period defined above and
satisfies Eq.~\eqref{PFMT}. To express
the corresponding equations for the
complete-intersection periods, define
\[
\mathcal A_a(\theta)
=
\prod_{i=1}^{M}
\prod_{k=0}^{Q_i^a-1}
(\mathcal Q_i-k),
\]
and
\[
\mathcal B_a(\theta)
=
\prod_{j=M+1}^{N}
\prod_{\ell=0}^{d_j^a-1}
(-D_j+\ell).
\]
The original GKZ equation can then be
written as
\[
\mathcal A_a(\theta)\Pi
=
\varepsilon_a q_a\mathcal B_a(\theta)\Pi.
\]

Rather than introducing inverse
differential operators, we act with
$\prod_jD_j$ on this equation. Since
the operators $D_j$ commute with
$\mathcal A_a(\theta)$, we obtain
\begin{equation}\label{PFMT2-CI}
\mathcal A_a(\theta)\widetilde{\Pi}
=
\varepsilon_a\left(\prod_{j=M+1}^{N}D_j\right)
\left[
q_a\mathcal B_a(\theta)\Pi
\right].
\end{equation}

Using the exact product rule
\[
D_j(q_af)
=
q_a(D_j+\widetilde Q_j^a)f,
\]
we find
\begin{align}\label{PFMT22}
\mathcal A_a(\theta)\widetilde{\Pi}
&=
\varepsilon_a q_a
\left[
\prod_{j=M+1}^{N}
(D_j+\widetilde Q_j^a)
\right]
\mathcal B_a(\theta)\Pi
\nonumber\\
&=
\varepsilon_a q_a
\left[
\prod_{j=M+1}^{N}
(D_j-d_j^a)
\right]
\mathcal B_a(\theta)\Pi .
\end{align}

For each $P$ field, the identity
\[
(D_j-d_j^a)
\prod_{\ell=0}^{d_j^a-1}
(-D_j+\ell)
=
D_j
\prod_{\ell=1}^{d_j^a}
(-D_j+\ell)
\]
allows the product to be reorganized.
Consequently, the differential
equations for the complete-intersection
periods take the closed form
\begin{align}\label{PFMT3}
\mathcal A_a(\theta)\widetilde{\Pi}
&=
\varepsilon_a q_a
\left[
\prod_{j=M+1}^{N}
\prod_{\ell=1}^{d_j^a}
(-D_j+\ell)
\right]
\left(\prod_{j=M+1}^{N}D_j\right)
\Pi
\nonumber\\
&=
\varepsilon_a q_a
\left[
\prod_{j=M+1}^{N}
\prod_{\ell=1}^{d_j^a}
(-D_j+\ell)
\right]\widetilde{\Pi}.
\end{align}

Here empty products are understood to be
one. This derivation includes all mixed
terms generated by the product rule for
multiple $P$ fields, without introducing
inverse differential operators. The resulting closed GKZ-type equations
need not form the minimal Picard--Fuchs
system; further factorization may be
required in a general multiparameter
model \cite{Hosono:1994ax}. The index $a$ labels the equation and is not summed. Equations of this type belong to the GKZ framework \cite{Gel'fand:1989,Gel'fand:1990}. If a GKZ operator factors as $\mathcal D_{\rm GKZ}=\mathcal A\,\mathcal L_{\rm PF}$, every solution of $\mathcal L_{\rm PF}f=0$ also solves $\mathcal D_{\rm GKZ}f=0$. The higher-order GKZ system may also admit additional solutions that do not satisfy the Picard-Fuchs equation \cite{Cox:1999}. We illustrate this distinction below.

Finally, our argument for Eq.~\eqref{PMLGS4} is restricted to the large-volume limit. To outline it, recall the period pairing $\Pi=\langle\gamma_i\mid\mathcal{O}_a(X)\rangle$. The relevant ground states in the complete-intersection mirror and in the mirror of the corresponding toric GLSM are related by
\begin{equation}\nonumber
  \mid{\cal O}_{a}(X)\rangle_{\rm CI}\equiv\prod_{j}\left(-X_{j}\right)\mid{\cal O}_{a}(X)\rangle_{\rm variety}.
\end{equation}

Here $\mathrm{CI}$ denotes the complete intersection. The discussion in Section~\ref{AILGM} suggests that the relevant A-branes can be identified in the large-volume limit, leading to Eq.~\eqref{PMLGS4} in that regime. Extending this relation over the full K\"ahler moduli space requires control of the continued cycles; we leave this question for future work.

\Needspace{5\baselineskip}
\noindent \textbf{Example: Quintic}

The GLSM for the quintic is a $U(1)$ gauge theory with five matter fields $\phi_i$, $i=1,\ldots,5$, and a field $P$, with gauge charges $(1,1,1,1,1,-5)$ and superpotential $W=PG$. We assign vector R-charge 2 to $P$ and vector R-charge 0 to the $\phi_i$. In the geometric phase, the $\phi_i$ may acquire nonzero expectation values. The mirror period in the normalization of Eq.~\eqref{PMLGS4} is
\begin{equation}\nonumber
  \widetilde{\Pi}_{\rm raw}=\int\prod^{5}_{i=1}dY_{i}dX_{p}\delta\left(\sum^{5}_{i=1}Y_{i}+5\log X_{p}-t\right)\exp\left(-\sum^{5}_{i=1}e^{-Y_{i}}-X_{p}\right).
\end{equation}
For a single $P$ field, Eq.~\eqref{PMLGS4} gives $\widetilde{\Pi}_{\rm raw}=-5\theta\Pi$. We choose the overall normalization used in Section~\ref{VR2FM}, so that
\begin{equation}\nonumber
  \widetilde{\Pi}=-\widetilde{\Pi}_{\rm raw}=5\theta\Pi.
\end{equation}
Here $\Pi$ is the period of the mirror of the total space of $\mathcal{O}_{\mathbb{CP}^{4}}(-5)$. Using this relation in the local-model equation of Appendix~\ref{APFMT} gives the Picard-Fuchs equation for the quintic mirror:
\begin{equation}\label{PFQ1}
  \theta^{4}\widetilde{\Pi}=-5q\left(5\theta+4\right)\left(5\theta+3\right)\left(5\theta+2\right)\left(5\theta+1\right)\widetilde{\Pi}.
\end{equation}
Equation~\eqref{PFMT3} instead gives the higher-order equation
\begin{equation}\label{PFQ2}
  \theta^{5}\widetilde{\Pi}=-q\left(5\theta+5\right)\left(5\theta+4\right)\left(5\theta+3\right)\left(5\theta+2\right)\left(5\theta+1\right)\widetilde{\Pi}.
\end{equation}
Applying $\theta$ to Eq.~\eqref{PFQ1} gives Eq.~\eqref{PFQ2}, using $\theta(qf)=q(\theta+1)f$.

In the coordinate $\widehat q=-q$, these equations take the familiar positive-sign form. The logarithmic derivative is unchanged. For the parameter $\psi$ of Section~\ref{VR2FM}, we have $\psi^5=q^{-1}=-\widehat q^{-1}$, so the finite singularity is at $q=-5^{-5}$, consistently with $\psi^5=-5^5$.

To distinguish the ambient Euler-twisted series $I_{\rm amb}$ from the local-model series $\Pi_{{\cal O}(-5)}$ in Eq.~\eqref{PGNQ}, write
\[
 I_{\rm amb}(\Sigma,t)=e^{-\Sigma t}\sum_{d\geq0}(-q)^d
    \frac{\prod_{\ell=1}^{5d}(5\Sigma+\ell)}
         {\prod_{\ell=1}^{d}(\Sigma+\ell)^5}
 \quad \bmod\ \Sigma^5 .
\]
With the corresponding Frobenius normalization of the local series, $\theta\Pi_{{\cal O}(-5)}=\Sigma I_{\rm amb}$ term by term, including the degree-zero term. We expand the ambient series in $H^*(\mathbb{CP}^4)$ \cite{Givental:1996,Cox:1999} as
\begin{equation}\label{QUGF}
  I_{\rm amb}\left(\Sigma, t\right)=\Pi_{0}\left(t\right)+\Sigma\cdot\Pi_{1}\left(t\right)+\Sigma^{2}\cdot\Pi_{2}\left(t\right)+\Sigma^{3}\cdot\Pi_{3}\left(t\right)+\Sigma^{4}\cdot\Pi_{4}\left(t\right).
\end{equation}

The coefficients $\Pi_0,\ldots,\Pi_4$ solve Eq.~\eqref{PFQ2}, whereas the Picard-Fuchs equation~\eqref{PFQ1} has only four linearly independent solutions. Multiplication by the Euler class $5\Sigma$ gives
\begin{equation}\label{QUGF2}
 \widetilde{ \Pi}\left(\Sigma, t\right)= 5\Sigma\, I_{\rm amb}\left(\Sigma, t\right)=5\Sigma\cdot\Pi_{0}\left(t\right)+5\Sigma^{2}\cdot\Pi_{1}\left(t\right)+5\Sigma^{3}\cdot\Pi_{2}\left(t\right)+5\Sigma^{4}\cdot\Pi_{3}\left(t\right).
\end{equation}
Here $\Sigma^5=0$ in $H^*(\mathbb{CP}^4)$. This removes the coefficient $\Pi_4(t)$, leaving the four Picard-Fuchs solutions $\Pi_0(t),\ldots,\Pi_3(t)$. See \cite[Sections~5.5.2 and~6.3.4]{Cox:1999} for details and Section~\ref{VR2FM} for the physical interpretation. The generating function for the quintic mirror is then written as
 \begin{equation}\label{QUGF3}
 \Pi_{\rm{quintic}}\left(\Sigma, t\right)=\frac{\widetilde{ \Pi}\left(\Sigma, t\right)}{5\Sigma} =\Pi_{0}\left(t\right)+\Sigma\cdot\Pi_{1}\left(t\right)+\Sigma^{2}\cdot\Pi_{2}\left(t\right)+\Sigma^{3}\cdot\Pi_{3}\left(t\right).
\end{equation}
Here $\Sigma^4=0$ after restriction to the quintic; the quotient by $5\Sigma$ is understood through the displayed polynomial expression, not as inversion of $\Sigma$ in cohomology.

\section{Examples of quantum differential equations of mirrors of Grassmannians}\label{app:grass}

As discussed in Section~\ref{PFEIPP}, we derive the one-parameter Grassmannian equations from Eq.~\eqref{MPGR}. To compare them with \cite{Batyrev:1998kx}, we use $q_{\rm lit}=(-1)^Nq$, with $q_{\rm lit}\partial_{q_{\rm lit}}=\theta$.
Throughout this appendix, $\Pi^{(r)}=(-d/dt)^r\Pi$, and
$\varepsilon^{m_1,\ldots,m_k}$ denotes a fixed totally antisymmetric
array of coefficients, with repeated solution indices summed over,
as in the $Gr(2,4)$ example in Section~\ref{PFMGR}.

\Needspace{5\baselineskip}
\noindent{$\textbf{Gr(3,4)}$}. We have
\begin{equation}\nonumber
  \Pi\equiv\varepsilon^{m_{1},m_{2},m_{3}}\Pi_{m_{1}}\left(t\right)\Pi^{(1)}_{m_{2}}\left(t\right)\Pi^{(2)}_{m_{3}}\left(t\right),\qquad m_1, m_2,m_3\in\{0,\cdots,3\}
\end{equation}
and
\begin{equation}\nonumber
  \left(-\frac{d}{dt}\right)^{4}\Pi_{j}=q\cdot\Pi_{j}.
\end{equation}
We then compute
\begin{align}\nonumber
   \Pi^{(1)}&=\varepsilon^{m_{1},m_{2},m_{3}}\Pi_{m_{1}}\left(t\right)\Pi^{(1)}_{m_{2}}\left(t\right)\Pi^{(3)}_{m_{3}}\left(t\right),  \\\nonumber
   \Pi^{(2)}&=\varepsilon^{m_{1},m_{2},m_{3}}\Pi_{m_{1}}\left(t\right)\Pi^{(2)}_{m_{2}}\left(t\right)\Pi^{(3)}_{m_{3}}\left(t\right) , \\\nonumber
   \Pi^{(3)}&=\varepsilon^{m_{1},m_{2},m_{3}}\Pi^{(1)}_{m_{1}}\left(t\right)\Pi^{(2)}_{m_{2}}\left(t\right)\Pi^{(3)}_{m_{3}}\left(t\right) , \\\nonumber
     \Pi^{(4)}&=\varepsilon^{m_{1},m_{2},m_{3}}q\Pi_{m_{1}}\left(t\right)\Pi^{(1)}_{m_{2}}\left(t\right)\Pi^{(2)}_{m_{3}}\left(t\right),
\end{align}
thus we get the quantum differential equation
\begin{equation}\nonumber
  \left(-\frac{d}{dt}\right)^{4}\Pi=q\Pi.
\end{equation}
This agrees with the isomorphism $Gr(3,4)\cong Gr(1,4)$.

\Needspace{5\baselineskip}
\noindent{$\textbf{Gr(2,5)}$}. We have
\begin{equation}\nonumber
  \Pi\equiv\varepsilon^{m,n}\Pi_{m}\left(t\right)\Pi^{(1)}_{n}\left(t\right), \qquad m,n\in\{0,\cdots,4\}
\end{equation}
and
\begin{equation}\nonumber
  \left(-\frac{d}{dt}\right)^{5}\Pi_{l}=q\cdot\Pi_{l}.
\end{equation}
Then we can compute
\begin{align}\nonumber
   \Pi^{(1)}&=\varepsilon^{m,n}\Pi_{m}\Pi^{(2)}_{n},  \\\nonumber
    \Pi^{(2)}&=\varepsilon^{m,n}\left(\Pi^{(1)}_{m}\Pi^{(2)}_{n}+\Pi_{m}\Pi^{(3)}_{n}\right),  \\\nonumber
    \Pi^{(3)}&=\varepsilon^{m,n}\left(2\Pi^{(1)}_{m}\Pi^{(3)}_{n}+\Pi_{m}\Pi^{(4)}_{n}\right),  \\\nonumber
      \Pi^{(4)}&=\varepsilon^{m,n}\left(2\Pi^{(2)}_{m}\Pi^{(3)}_{n}+3\Pi^{(1)}_{m}\Pi^{(4)}_{n}\right),  \\\nonumber
      \Pi^{(5)}&=\varepsilon^{m,n}\left(5\Pi^{(2)}_{m}\Pi^{(4)}_{n}-3q\Pi_{m}\Pi^{(1)}_{n}\right),  \\\nonumber
     \Pi^{(6)}&=\varepsilon^{m,n}\left(5\Pi^{(3)}_{m}\Pi^{(4)}_{n}-3q\Pi_{m}\Pi^{(1)}_{n}-8q\Pi_{m}\Pi^{(2)}_{n}\right),  \\\nonumber
       \Pi^{(7)}&=\varepsilon^{m,n}\left(-3q\Pi_{m}\Pi^{(1)}_{n}-11q\Pi_{m}\Pi^{(2)}_{n}-8q\Pi^{(1)}_{m}\Pi^{(2)}_{n}-13q\Pi_{m}\Pi^{(3)}_{n}\right),  \\\nonumber
      \Pi^{(8)}&=\varepsilon^{m,n}\Bigg(\Bigg.-3q\Pi_{m}\Pi^{(1)}_{n}-14q\Pi_{m}\Pi^{(2)}_{n}-19q\Pi^{(1)}_{m}\Pi^{(2)}_{n}-24q\Pi_{m}\Pi^{(3)}_{n}\\\nonumber
      &-21q\Pi^{(1)}_{m}\Pi^{(3)}_{n}-13q\Pi_{m}\Pi^{(4)}_{n}\Bigg.\Bigg),  \\\nonumber
       \Pi^{(9)}&=\varepsilon^{m,n}\Bigg(\Bigg.-3q\Pi_{m}\Pi^{(1)}_{n}-17q\Pi_{m}\Pi^{(2)}_{n}-33q\Pi^{(1)}_{m}\Pi^{(2)}_{n}-38q\Pi_{m}\Pi^{(3)}_{n}\\\nonumber
      &-64q\Pi^{(1)}_{m}\Pi^{(3)}_{n}-37q\Pi_{m}\Pi^{(4)}_{n}-21q\Pi^{(2)}_{m}\Pi^{(3)}_{n}-34q\Pi^{(1)}_{m}\Pi^{(4)}_{n}\Bigg.\Bigg),  \\\nonumber
        \Pi^{(10)}&=\varepsilon^{m,n}\Bigg(\Bigg.-3q\Pi_{m}\Pi^{(1)}_{n}-20q\Pi_{m}\Pi^{(2)}_{n}-50q\Pi^{(1)}_{m}\Pi^{(2)}_{n}-55q\Pi_{m}\Pi^{(3)}_{n}\\\nonumber
      &-135q\Pi^{(1)}_{m}\Pi^{(3)}_{n}-75q\Pi_{m}\Pi^{(4)}_{n}-85q\Pi^{(2)}_{m}\Pi^{(3)}_{n}-135q\Pi^{(1)}_{m}\Pi^{(4)}_{n}\\\nonumber
      &-55q\Pi^{(2)}_{m}\Pi^{(4)}_{n}+34q^{2}\Pi_{m}\Pi^{(1)}_{n}\Bigg.\Bigg).
\end{align}
The tenth derivative is a linear combination of lower-order derivatives:
\begin{equation}\nonumber
  \Pi^{(10)}=3\Pi^{(9)}-3\Pi^{(8)}+\Pi^{(7)}-q\left(11\Pi^{(5)}+11\Pi^{(4)}+3\Pi^{(3)}\right)+q^{2}\Pi.
\end{equation}
The quantum differential equation for the mirror of $Gr(2,5)$ is therefore
\begin{equation}\label{PFGr25}
  \bigl[\theta^7(\theta-1)^3+q\theta^3(11\theta^2+11\theta+3)-q^2\bigr]\Pi(t)=0,\qquad\theta=-d/dt.
\end{equation}
With $q_{\rm lit}=-q$, this agrees with the operator in \cite[Section~5.2]{Batyrev:1998kx}. The same procedure reproduces the quantum differential equations for $Gr(2,6)$, $Gr(3,6)$, and $Gr(2,7)$ obtained there. We omit these calculations. A uniform closed formula for Grassmannian quantum differential operators would be useful.

\section{Examples of the computation of instanton numbers of Calabi-Yau manifolds in Grassmannians}\label{app:grassCY}

\Needspace{5\baselineskip}
$\bullet$ $Gr(2, 5)\left[2^{2},1\right]$.

Following Section~\ref{VR2FM}, the cohomology-valued generating function for the mirror of $Gr(2,5)[2^2,1]$ is
\begingroup\small 
\begin{align}\label{PGr25221}
  &\Pi_{Gr(2,5)[2^{2},1]}(\Sigma_{1},\Sigma_{2}; t)=\left.\frac{-\partial_{t^{1}}+\partial_{t^{2}}}{\Sigma_{1}-\Sigma_{2}}\right|_{t^{1}=t^{2}=t}\sum^{\infty}_{d_{1},d_{2}=0}
   \\\nonumber
  &\frac{\prod^{2\left(d_{1}+d_{2}\right)}_{k=1}\left(2
  \left(\Sigma_{1}+\Sigma_{2}\right)+k\right)^{2}\prod^{d_{1}+d_{2}}_{k=1}\left(\left(\Sigma_{1}+\Sigma_{2}\right)+k\right)}{\prod^{d_{1}}_{k=1}\left(\Sigma_{1}+k\right)^{5}\cdot\prod^{d_{2}}_{k=1}\left(\Sigma_{2}+k\right)^{5}}
 \cdot\exp\left(-\left(\Sigma_{1}+d_{1}\right)t^{1}\right)\exp\left(-\left(\Sigma_{2}+d_{2}\right)t^{2}\right).
\end{align}
\endgroup
The coefficients $\Pi_{Gr(2,5)[2^2,1],0}$ and $\Pi_{Gr(2,5)[2^2,1],1}$ are
\begingroup\small 
\begin{align}\label{PGR2522101}
  \Pi_{0}(t)&=\sum_{d_{1},d_{2}=0}^{\infty}\frac{\left[\left(2d_{1}+2d_{2}\right)!\right]^{2}\left(d_{1}+d_{2}\right)!}{\left(d_{1}!\right)^{5}\left(d_{2}!\right)^{5}}\left(1-\frac{5}{2}\left(d_{1}-d_{2}\right)\left(H_{d_{1}}-H_{d_{2}}\right)\right)\exp\left(-\left(d_{1}+d_{2}\right)t\right),
  \\\nonumber
  \Pi_{1}(t)&=-t\Pi_{0} +
  \\\nonumber &\sum_{d_{1},d_{2}=0}^{\infty}\frac{\left[\left(2d_{1}+2d_{2}\right)!\right]^{2}\left(d_{1}+d_{2}\right)!}{\left(d_{1}!\right)^{5}\left(d_{2}!\right)^{5}}\Bigg(\Bigg.4H_{2d_{1}+2d_{2}}+H_{d_{1}+d_{2}}-\frac{5}{2}H_{d_{1}}-\frac{5}{2}H_{d_{2}}-\left(d_{1}-d_{2}\right)
  \\\nonumber
   &\cdot\left(\left(10H_{2d_{1}+2d_{2}}+\frac{5}{2}H_{d_{1}+d_{2}}\right)\left(H_{d_{1}}-H_{d_{2}}\right)-\left(\frac{25}{4}\left(H^{2}_{d_{1}}-H^{2}_{d_{2}}\right)+\frac{5}{4}\left(\sum^{d_{1}}_{k=1}\frac{1}{k^{2}}-\sum^{d_{2}}_{k=1}\frac{1}{k^{2}}\right)\right)\right)\Bigg.\Bigg)
   \\\nonumber
   &\cdot\exp\left(-\left(d_{1}+d_{2}\right)t\right).
\end{align}
\endgroup
The series expansion of (\ref{PGR2522101}) gives
\begin{align}\nonumber
  \Pi_{0}(t)&=1 - 12q + 684 q^{2} - 58800 q^{3} + 6129900 q^{4}+{\cal O}(q^{5}),
  \\\nonumber
  \Pi_{1}(t)&=-t  - 4 q (11 - 3t) +6 q^{2} (491 - 114 t)
  \\\nonumber
   &-\frac{280}{3} q^{3} \left(2879 - 630t\right) + 315 q^{4} \left(91683 - 19460t\right)+{\cal O}(e^{-5t}).
\end{align}
The flat coordinate $\tau$ is
\begin{equation}\label{FGR25221}
  \tau=-\frac{\Pi_{1}(t)}{\Pi_{0}(t)}-2i\pi t_{0}=-i\pi-\log q+44q-2418 q^{2}+\frac{628784}{3} q^{3}-22123897 q^{4} +{\cal O}(e^{-5t}),
\end{equation}
where we have chosen $t_{0}=1/2$. In terms of the variable $z=e^{-\tau}$, we have
\begin{equation}\label{IFGR25221}
  q=-z+44z^{2}-486 z^{3}+11184 z^{4}+{\cal O}(z^{5}).
\end{equation}
Substituting these expressions into Eq.~\eqref{CLATB} gives
\begin{equation}\label{GRZS25221}
 K_{zzz}= 20+400z+44720 z^2 + 4439200 z^3 + 451877040 z^4+{\cal O}(z^{5}).
\end{equation}
This implies that the first four instanton numbers of $Gr(2,5)[2^{2},1]$ are
\begin{equation}\label{GWG25221}
  400,\quad 5540,\quad 164400,\quad 7059880.
\end{equation}
These instanton numbers agree with \cite{Batyrev:1998kx}. Although our period formula~\eqref{PGR2522101} differs in appearance from the formula on page~31 of that reference, their agreement suggests a nontrivial combinatorial identity. A direct computation also gives identical data for $Gr(3,5)[2^2,1]$, as expected from Grassmannian duality.

\Needspace{5\baselineskip}
$\bullet$ $Gr(2, 5)\left[1^{2},3\right]$.

The generating function for the mirror of $Gr(2,5)[1^2,3]$ is
\begin{align}\label{PGr25113}
  &\Pi_{Gr(2,5)[1^{2},3]}(\Sigma_{1},\Sigma_{2}; t)=
   \\\nonumber
  &\left.\frac{-\partial_{t^{1}}+\partial_{t^{2}}}{\Sigma_{1}-\Sigma_{2}}\right|_{t^{1}=t^{2}=t}
 \sum^{\infty}_{d_{1},d_{2}=0}\frac{\prod^{3\left(d_{1}+d_{2}\right)}_{k=1}\left(3
  \left(\Sigma_{1}+\Sigma_{2}\right)+k\right)\prod^{d_{1}+d_{2}}_{k=1}\left(\left(\Sigma_{1}+\Sigma_{2}\right)+k\right)^{2}}{\prod^{d_{1}}_{k=1}\left(\Sigma_{1}+k\right)^{5}\cdot\prod^{d_{2}}_{k=1}\left(\Sigma_{2}+k\right)^{5}}
   \\\nonumber
  &\cdot\exp\left(-\left(\Sigma_{1}+d_{1}\right)t^{1}\right)\exp\left(-\left(\Sigma_{2}+d_{2}\right)t^{2}\right).
\end{align}

The coefficients $\Pi_{Gr(2,5)[1^2,3],0}$ and $\Pi_{Gr(2,5)[1^2,3],1}$ are
\begingroup\small 
\begin{align}\label{PGR2511301}
  \Pi_{0}(t)&=\sum_{d_{1},d_{2}=0}^{\infty}\frac{\left(3d_{1}+3d_{2}\right)!\cdot \left(\left(d_{1}+d_{2}\right)!\right)^{2}}{\left(d_{1}!\right)^{5}\cdot\left(d_{2}!\right)^{5}}\left(1-\frac{5}{2}\left(d_{1}-d_{2}\right)\left(H_{d_{1}}-H_{d_{2}}\right)\right)\exp\left(-\left(d_{1}+d_{2}\right)t\right),
  \\\nonumber
  \Pi_{1}(t)&=-t\cdot\Pi_{0} +
  \\\nonumber
   &\sum_{d_{1},d_{2}=0}^{\infty}\frac{\left(3d_{1}+3d_{2}\right)!\cdot \left(\left(d_{1}+d_{2}\right)!\right)^{2}}{\left(d_{1}!\right)^{5}\cdot\left(d_{2}!\right)^{5}}\Bigg(\Bigg.3H_{3d_{1}+3d_{2}}+2H_{d_{1}+d_{2}}-\frac{5}{2}H_{d_{1}}-\frac{5}{2}H_{d_{2}}-\left(d_{1}-d_{2}\right)
  \\\nonumber
   &\cdot\left(\left(5H_{d_{1}+d_{2}}+\frac{15}{2}H_{3d_{1}+3d_{2}}\right)\left(H_{d_{1}}-H_{d_{2}}\right)-\left(\frac{25}{4}\left(H^{2}_{d_{1}}-H^{2}_{d_{2}}\right)+\frac{5}{4}\left(\sum^{d_{1}}_{k=1}\frac{1}{k^{2}}-\sum^{d_{2}}_{k=1}\frac{1}{k^{2}}\right)\right)\right)\Bigg.\Bigg)
   \\\nonumber
   &\cdot\exp\left(-\left(d_{1}+d_{2}\right)t\right),
\end{align}
\endgroup
which have series expansions
\begin{align}\nonumber
  \Pi_{0}(t)&=1 - 18 q + 1710 q^{2} - 246960 q^{3} + 43347150 q^{4}+{\cal O}(q^5),
  \\\nonumber
  \Pi_{1}(t)&=-t - 3 q (25 - 6t) +\frac{9}{2} q^{2} (1833 - 380t)
  \\\nonumber
   &-98 q^{3} \left(12827 - 2520t\right) + \frac{135 }{4}q^{4} \left(6720919 - 1284360t\right)+{\cal O}(e^{-5t}).
\end{align}
So the flat coordinate $\tau$ is
\begin{equation}\label{FGR25113}
  \tau=-\frac{\Pi_{1}(t)}{\Pi_{0}(t)}-2i\pi t_{0}=-i\pi-\log q+75q- \frac{13797}{2} q^{2}+1004623 q^{3}-\frac{713717469}{4}  q^{4} +{\cal O}(e^{-5t}).
\end{equation}
Here we choose $t_0=1/2$. Inverting the mirror map, with $z=e^{-\tau}$, gives
\begin{equation}\label{IFGR25113}
  q=-z+75z^{2}-1539 z^{3}+60073 z^{4}+{\cal O}(z^{5}).
\end{equation}
Equation~\eqref{CLATB} gives
\begin{equation}\label{GRZS25113}
  K_{zzz}=15+540z+100980 z^2 + 16776045 z^3 + 2873237940 z^4+{\cal O}(z^{5}).
\end{equation}
Consequently, the first four instanton numbers of $Gr(2,5)[1^{2},3]$ are
\begin{equation}\label{GWG25113}
  540,\quad 12555,\quad 621315,\quad 44892765.
\end{equation}

\Needspace{5\baselineskip}
$\bullet$ $Gr(2, 6)\left[1^{4},2\right]$.

The generating function for the mirror of $Gr(2,6)[1^4,2]$ is
\begin{align}\label{PGr2612}
  &\Pi_{Gr(2,6)[1^{4},2]}(\Sigma_{1},\Sigma_{2}; t)=\left.\frac{-\partial_{t^{1}}+\partial_{t^{2}}}{\Sigma_{1}-\Sigma_{2}}\right|_{t^{1}=t^{2}=t}
  \\\nonumber
  &\sum^{\infty}_{d_{1},d_{2}=0}\frac{\prod^{2\left(d_{1}+d_{2}\right)}_{k=1}\left(2
  \left(\Sigma_{1}+\Sigma_{2}\right)+k\right)\prod^{d_{1}+d_{2}}_{k=1}\left(\left(\Sigma_{1}+\Sigma_{2}\right)+k\right)^{4}}{\prod^{d_{1}}_{k=1}\left(\Sigma_{1}+k\right)^{6}\cdot\prod^{d_{2}}_{k=1}\left(\Sigma_{2}+k\right)^{6}} \\\nonumber
  &\cdot\exp\left(-\left(\Sigma_{1}+d_{1}\right)\left(t^{1}+i\pi\right)\right)\exp\left(-\left(\Sigma_{2}+d_{2}\right)\left(t^{2}+i\pi\right)\right),
\end{align}
which implies
\begingroup\small 
\begin{align}\label{PGr26121}
\Pi_{0}(t)&=\sum_{d_{1},d_{2}=0}^{\infty}\frac{\left(2d_{1}+2d_{2}\right)!\left[\left(d_{1}+d_{2}\right)!\right]^{4}}{\left(d_{1}!\right)^{6}\left(d_{2}!\right)^{6}}\left(1-3\left(d_{1}-d_{2}\right)\left(H_{d_{1}}-H_{d_{2}}\right)\right)\cdot\exp\left(-\left(d_{1}+d_{2}\right)\left(t+i\pi\right)\right),
\\\nonumber
\Pi_{1}(t)&=-(t + i\pi)\Pi_{0} +
  \\\nonumber &\sum_{d_{1},d_{2}=0}^{\infty}\frac{\left(2d_{1}+2d_{2}\right)!\left[\left(d_{1}+d_{2}\right)!\right]^{4}}{\left(d_{1}!\right)^{6}\left(d_{2}!\right)^{6}}\Bigg(\Bigg.2H_{2d_{1}+2d_{2}}+4H_{d_{1}+d_{2}}-3H_{d_{1}}-3H_{d_{2}}-\left(d_{1}-d_{2}\right)
  \\\nonumber
   &\cdot\left(\left(6H_{2d_{1}+2d_{2}}+12H_{d_{1}+d_{2}}\right)\left(H_{d_{1}}-H_{d_{2}}\right)-\left(9\left(H^{2}_{d_{1}}-H^{2}_{d_{2}}\right)+\frac{3}{2}\left(\sum^{d_{1}}_{k=1}\frac{1}{k^{2}}-\sum^{d_{2}}_{k=1}\frac{1}{k^{2}}\right)\right)\right)\Bigg.\Bigg)
   \\\nonumber
   &\cdot\exp\left(-\left(d_{1}+d_{2}\right)\left(t+i\pi\right)\right),
  \end{align}
\endgroup
whose series expansions read
\begin{align}\nonumber
  \Pi_{0}(t)&=1 + 8 q + 288 q^{2} + 15200 q^{3} + 968800 q^{4}+{\cal O}(q^{5}),
  \\\nonumber
  \Pi_{1}(t)&=-t - i\pi + 2 q (13 - 4(t + i\pi)) +3 q^{2} (367 - 96 ( t + i\pi))
  \\\nonumber
   &+\frac{20}{3} q^{3} \left(9301 - 2280 \left(t + i\pi\right)\right) + q^{4} \left(\frac{24519335}{6} - 968800  \left(t + i\pi\right)\right)+{\cal O}(e^{-5t}).
\end{align}
Thus, the flat coordinate $\tau$ is
\begin{equation}\label{FGR2612}
  \tau=-\frac{\Pi_{1}(t)}{\Pi_{0}(t)}-2i\pi t_{0}=-\log q-26q-893 q^{2}-\frac{142124 }{3} q^{3}-\frac{6110349}{2}q^{4} +{\cal O}(e^{-5t}),
\end{equation}
where we have chosen $t_{0}=1/2$. Then we have
\begin{equation}\label{IFGR2612}
  q=z-26z^{2}+121 z^{3}-1372z^{4}+{\cal O}(z^{5}).
\end{equation}
Equation~\eqref{CLATB} then gives
\begin{equation}\label{GRZS2612}
  K_{zzz}=28+280z+ 21672 z^2 + 1303624 z^3 + 81880232 z^4+{\cal O}(z^{5}).
\end{equation}
Therefore, the first four instanton numbers of $Gr(2,6)[1^{4},2]$ are
\begin{equation}\label{GWG2612}
  280,\quad 2674,\quad 48272,\quad 1279040.
\end{equation}

\Needspace{5\baselineskip}
$\bullet$ $Gr(3, 6)\left[1^{6}\right]$.

The generating function of the mirror of $Gr(3, 6)\left[1^{6}\right]$ is
\begin{align}\label{PGr361}
  \Pi_{Gr(3,6)[1^{6}]}&(\Sigma_{1},\Sigma_{2},\Sigma_3; t)=\left.\left(\frac{-\partial_{t^{1}}+\partial_{t^{2}}}{\Sigma_{1}-\Sigma_{2}}\right)\left(\frac{-\partial_{t^{1}}+\partial_{t^{3}}}{\Sigma_{1}-\Sigma_{3}}\right)\left(\frac{-\partial_{t^{2}}+\partial_{t^{3}}}{\Sigma_{2}-\Sigma_{3}}\right)\right|_{t^{1}=t^{2}=t^{3}=t}\\\nonumber
  &\cdot\sum^{\infty}_{d_{1},d_{2},d_{3}=0}\frac{\prod^{d_{1}+d_{2}+d_{3}}_{k=1}\left(\Sigma_{1}+\Sigma_{2}+\Sigma_3+k\right)^{6}}{\prod^{d_{1}}_{k=1}\left(\Sigma_{1}+k\right)^{6}\cdot\prod^{d_{2}}_{k=1}\left(\Sigma_{2}+k\right)^{6}\cdot\prod^{d_{3}}_{k=1}\left(\Sigma_{3}+k\right)^{6}} \\\nonumber
  &\cdot\exp\left(-\left(\Sigma_{1}+d_{1}\right)t^{1}\right)\exp\left(-\left(\Sigma_{2}+d_{2}\right)t^{2}\right)\exp\left(-\left(\Sigma_{3}+d_{3}\right)t^{3}\right).
\end{align}
Applying the normalized differential operator in Eq.~\eqref{PGr361} to the exponential factors gives
\begin{equation}\label{Gr36f}
  \frac{\left(\Sigma_{1}-\Sigma_{2}+d_{1}-d_{2}\right)\left(\Sigma_{1}-\Sigma_{3}+d_{1}-d_{3}\right)\left(\Sigma_{2}-\Sigma_{3}+d_{2}-d_{3}\right)}{\left(\Sigma_{1}-\Sigma_{2}\right)\left(\Sigma_{1}-\Sigma_{3}\right)\left(\Sigma_{2}-\Sigma_{3}\right)}.
\end{equation}
It is therefore convenient to express $\Sigma_1,\Sigma_2,\Sigma_3$ in terms of their sum and pairwise differences:
\begin{align}\nonumber
  \Sigma_{1}&=\frac{\Sigma_{1}+\Sigma_{2}+\Sigma_{3}}{3}+\frac{\Sigma_{1}-\Sigma_{2}}{3}+\frac{\Sigma_{1}-\Sigma_{3}}{3},\\\nonumber
  \Sigma_{2}&=\frac{\Sigma_{1}+\Sigma_{2}+\Sigma_{3}}{3}+\frac{\Sigma_{2}-\Sigma_{1}}{3}+\frac{\Sigma_{2}-\Sigma_{3}}{3},\\\nonumber
  \Sigma_{3}&=\frac{\Sigma_{1}+\Sigma_{2}+\Sigma_{3}}{3}+\frac{\Sigma_{3}-\Sigma_{1}}{3}+\frac{\Sigma_{3}-\Sigma_{2}}{3}.
\end{align}
This gives explicit expressions for $\Pi_0(t)$ and $\Pi_1(t)$. We display only their first few terms:
\begin{align}\nonumber
  \Pi_{0}(t)&=1 + 6 q + 126 q^{2} + 3948 q^{3} + 149310 q^{4}+{\cal O}(q^{5}),
  \\\nonumber
  \Pi_{1}(t)&=-t + 2 q (8 - 3t) +2 q^{2} (208 - 63t)
  \\\nonumber
   &+q^{3} \left(\frac{42136}{3}  - 3948 t \right) + 6q^{4} \left(91976 - 24885 t\right)+{\cal O}(e^{-5t}).
\end{align}
Then the flat coordinate is
\begin{equation}\label{FGR361}
  \tau=-\frac{\Pi_{1}(t)}{\Pi_{0}(t)}-2i\pi t_{0}=-\log q-16q-320 q^{2}-\frac{30328 }{3} q^{3}-387712q^{4} +{\cal O}(e^{-5t}),
\end{equation}
where we have taken $t_{0}=0$. Then,
\begin{equation}\label{IFGR361}
  q=z-16z^{2}+64 z^{3}-552z^{4}+{\cal O}(z^{5}).
\end{equation}
Equation~\eqref{CLATB} then gives
\begin{equation}\label{GRZS361}
  K_{zzz}=42+210z+ 9618  z^2 + 354018  z^3 + 12933522 z^4+{\cal O}(z^{5}).
\end{equation}
Thus the first four instanton numbers of $Gr(3,6)[1^{6}]$ are
\begin{equation}\label{GWG361}
  210,\quad 1176,\quad 13104,\quad 201936.
\end{equation}

\Needspace{5\baselineskip}
$\bullet$ $Gr(2, 7)\left[1^{7}\right]$.

The generating function for the mirror of $Gr(2,7)[1^7]$ is
\begin{align}\label{PGr27}
  \Pi_{Gr(2,7)[1^{7}]}&(\Sigma_{1},\Sigma_{2}; t)=\left.\frac{-\partial_{t^{1}}+\partial_{t^{2}}}{\Sigma_{1}-\Sigma_{2}}\right|_{t^{1}=t^{2}=t}\sum^{\infty}_{d_{1},d_{2}=0}\frac{\prod^{d_{1}+d_{2}}_{k=1}\left(\left(\Sigma_{1}+\Sigma_{2}\right)+k\right)^{7}}{\prod^{d_{1}}_{k=1}\left(\Sigma_{1}+k\right)^{7}\cdot\prod^{d_{2}}_{k=1}\left(\Sigma_{2}+k\right)^{7}}\cdot
  \\\nonumber
  &\exp\left(-\left(\Sigma_{1}+d_{1}\right)t^{1}\right)\exp\left(-\left(\Sigma_{2}+d_{2}\right)t^{2}\right),
\end{align}
which implies
\begin{align}\label{PGR2701}
   \Pi_{0}(t)&=\sum_{d_{1},d_{2}=0}^{\infty}\frac{\left[\left(d_{1}+d_{2}\right)!\right]^{7}}{\left(d_{1}!\right)^{7}\cdot\left(d_{2}!\right)^{7}}\left(1-\frac{7}{2}\left(d_{1}-d_{2}\right)\left(H_{d_{1}}-H_{d_{2}}\right)\right)\exp\left(-\left(d_{1}+d_{2}\right)t\right),
  \\\nonumber
  \Pi_{1}(t)&=-t\cdot\Pi_{0} +
  \\\nonumber
   &\sum_{d_{1},d_{2}=0}^{\infty}\frac{\left[\left(d_{1}+d_{2}\right)!\right]^{7}}{\left(d_{1}!\right)^{7}\cdot\left(d_{2}!\right)^{7}}\Bigg(\Bigg.7H_{d_{1}+d_{2}}-\frac{7}{2}H_{d_{1}}-\frac{7}{2}H_{d_{2}}-\left(d_{1}-d_{2}\right)
  \\\nonumber
   &\cdot\left(\frac{49}{2}H_{d_{1}+d_{2}}\left(H_{d_{1}}-H_{d_{2}}\right)-\left(\frac{49}{4}\left(H^{2}_{d_{1}}-H^{2}_{d_{2}}\right)+\frac{7}{4}\left(\sum^{d_{1}}_{k=1}\frac{1}{k^{2}}-\sum^{d_{2}}_{k=1}\frac{1}{k^{2}}\right)\right)\right)\Bigg.\Bigg)
   \\\nonumber
   &\cdot\exp\left(-\left(d_{1}+d_{2}\right)t\right),
\end{align}
whose series expansions read
\begin{align}\label{PGR2701-series}
  \Pi_{0}(t)&=1 - 5 q + 109 q^{2} - 3317 q^{3} + 121501 q^{4}+{\cal O}(q^{5}),
  \\\nonumber
  \Pi_{1}(t)&=-t  -  q (14 - 5t) +q^{2} (357 - 109 t )
  \\\nonumber
   &-\frac{1}{3} q^{3} \left(35105 - 9951t\right) + q^{4} \left(\frac{2669975}{6} - 121501 t\right)+{\cal O}(e^{-5t}).
\end{align}
Then the flat coordinate $\tau$ is
\begin{equation}\label{FGR27}
  \tau=-\frac{\Pi_{1}(t)}{\Pi_{0}(t)}-2i\pi t_{0}=-i\pi-\log q+14q- 287 q^{2}+\frac{26222}{3} q^{3}-\frac{647143}{2} q^{4} +{\cal O}(e^{-5t}),
\end{equation}
where we have chosen $t_{0}=1/2$. In terms of $z=e^{-\tau}$, we have
\begin{equation}\label{IFGR27}
  q=-z+14z^{2}-7 z^{3}-14z^{4}+{\cal O}(z^{5}).
\end{equation}
Equation~\eqref{CLATB} gives
\begin{equation}\label{GRZS27}
  K_{zzz}=42+196 z+ 9996 z^2 + 344176 z^3 + 12685708 z^4+{\cal O}(z^{5}).
\end{equation}
Hence the first four instanton numbers of $Gr(2,7)[1^{7}]$ are
\begin{equation}\label{GWG27}
  196,\quad 1225,\quad 12740,\quad 198058.
\end{equation}

\end{document}